\documentclass[amssymb,amsmath,prd,twocolumn,preprintnumbers,superscriptaddress,nofootinbib]{revtex4-1}

\usepackage{graphicx}
\usepackage{bm}
\usepackage{amssymb,amsmath}
\usepackage{mathrsfs}
\usepackage{latexsym}
\usepackage{color}
\usepackage[normalem]{ulem} 
\usepackage{dcolumn}
\usepackage[colorlinks=true,citecolor=blue,urlcolor=blue]{hyperref}
\usepackage[usenames,dvipsnames]{xcolor}
\usepackage{xspace}
\usepackage{graphicx}
\usepackage{diagbox}

\usepackage{multirow}
\usepackage{xfp} 
\usepackage{comment}
\usepackage{nameref}
\usepackage{chngcntr}

\newcommand{\typ}{\mathrm{typ}}

\newcommand{\rhoc}{\rho_{\rm c}}
\newcommand{\rhonuc}{\rho_{\mathrm{nuc}}}

\newcommand{\stitch}{\text{stitch}}

\newcommand{\Mmax}{\ensuremath{M_{\rm max}}}

\newcommand{\eps}{\epsilon}

\newcommand{\e}[1]{\times 10^{#1}}
\newcommand{\deriv}[2]{\frac{d #1}{d #2}}
\newcommand{\pderiv}[2]{\frac{\partial #1}{\partial #2}}

\DeclareMathOperator{\nuc}{nuc}

\usepackage[english]{babel}
\usepackage{xspace}
\usepackage{color,soul}

\newcommand{\smallchange}{\delta}
\newcommand{\neweos}{enthalpy }
\newcommand{\spectre}{\texttt{SpECTRE}}

\newcommand{\quant}[1]{\left(#1\right)}
\newcommand{\quantb}[1]{\left[#1\right]}

\newcommand{\red}[1]{{#1}}

\newcommand{\enthalpyptadordlevfourlabel}{enthalpy-pt-ppao-70}
\newcommand{\enthalpyptadordlevfourelements}{$\left(48\right)^3$}
\newcommand{\enthalpyptadordlevfourfdgridspacing}{67}
\newcommand{\enthalpyptadordlevfourrhoc}{0.0021}

\newcommand{\enthalpyptadordlevfourradius}{11.8}
\newcommand{\enthalpyptadordlevfourcost}{24.6}
\newcommand{\enthalpyptadordlevfourcperd}{32}
\newcommand{\enthalpyptmonotlevfourlabel}{enthalpy-pt-mc-70}
\newcommand{\enthalpyptmonotlevfourelements}{$\left(48\right)^3$}
\newcommand{\enthalpyptmonotlevfourfdgridspacing}{67}
\newcommand{\enthalpyptmonotlevfourrhoc}{0.0021}

\newcommand{\enthalpyptmonotlevfourradius}{11.8}
\newcommand{\enthalpyptmonotlevfourcost}{20.7}
\newcommand{\enthalpyptmonotlevfourcperd}{32}

\newcommand{\enthalpypolytropelevthreelabel}{enthalpy-polytrope-mc-130}
\newcommand{\enthalpypolytropelevthreeelements}{$\left(24\right)^3$}
\newcommand{\enthalpypolytropelevthreefdgridspacing}{134}
\newcommand{\enthalpypolytropelevthreerhoc}{0.00128}

\newcommand{\enthalpypolytropelevthreeradius}{14.1}
\newcommand{\enthalpypolytropelevthreecost}{2.07}
\newcommand{\enthalpypolytropelevthreecperd}{19}

\newcommand{\polytropicpolytropelevthreelabel}{polytropic-polytrope-mc-130}
\newcommand{\polytropicpolytropelevthreeelements}{$\left(24\right)^3$}
\newcommand{\polytropicpolytropelevthreefdgridspacing}{134}
\newcommand{\polytropicpolytropelevthreerhoc}{0.00128}

\newcommand{\polytropicpolytropelevthreeradius}{14.1}
\newcommand{\polytropicpolytropelevthreecost}{2.1}
\newcommand{\polytropicpolytropelevthreecperd}{19}

\newcommand{\spectralslylevthreelabel}{spectral-sly-mc-220}
\newcommand{\spectralslylevthreeelements}{$\left(24\right)^3$}
\newcommand{\spectralslylevthreefdgridspacing}{224}
\newcommand{\spectralslylevthreerhoc}{0.00138}

\newcommand{\spectralslylevthreeradius}{11.5}
\newcommand{\spectralslylevthreecost}{3.5}
\newcommand{\spectralslylevthreecperd}{9}

\newcommand{\enthalpyslylevthreelabel}{enthalpy-sly-mc-220}
\newcommand{\enthalpyslylevthreeelements}{$\left(24\right)^3$}
\newcommand{\enthalpyslylevthreefdgridspacing}{224}
\newcommand{\enthalpyslylevthreerhoc}{0.00138}

\newcommand{\enthalpyslylevthreeradius}{11.5}
\newcommand{\enthalpyslylevthreecost}{4.1}
\newcommand{\enthalpyslylevthreecperd}{9}

\newcommand{\enthalpyptlevthreelabel}{enthalpy-pt-mc-130}
\newcommand{\enthalpyptlevthreeelements}{$\left(24\right)^3$}
\newcommand{\enthalpyptlevthreefdgridspacing}{134}
\newcommand{\enthalpyptlevthreerhoc}{0.0021}

\newcommand{\enthalpyptlevthreeradius}{11.8}
\newcommand{\enthalpyptlevthreecost}{5.1}
\newcommand{\enthalpyptlevthreecperd}{16}
\newcommand{\spectraldbhflevthreelabel}{spectral-dbhf-mc-130}
\newcommand{\spectraldbhflevthreeelements}{$\left(24\right)^3$}
\newcommand{\spectraldbhflevthreefdgridspacing}{134}
\newcommand{\spectraldbhflevthreerhoc}{0.001}

\newcommand{\spectraldbhflevthreeradius}{13.4}
\newcommand{\spectraldbhflevthreecost}{4.1}
\newcommand{\spectraldbhflevthreecperd}{18}

\newcommand{\enthalpydbhflevthreelabel}{enthalpy-dbhf-mc-130}
\newcommand{\enthalpydbhflevthreeelements}{$\left(24\right)^3$}
\newcommand{\enthalpydbhflevthreefdgridspacing}{134}
\newcommand{\enthalpydbhflevthreerhoc}{0.001}

\newcommand{\enthalpydbhflevthreeradius}{13.5}
\newcommand{\enthalpydbhflevthreecost}{3.7}
\newcommand{\enthalpydbhflevthreecperd}{18}

\newcommand{\enthalpysmoothptlevthreelabel}{enthalpy-smoothpt-170}
\newcommand{\enthalpysmoothptlevthreeelements}{$\left(24\right)^3$}
\newcommand{\enthalpysmoothptlevthreefdgridspacing}{168}
\newcommand{\enthalpysmoothptlevthreerhoc}{0.0021}

\newcommand{\enthalpysmoothptlevthreefigs}{NA}
\newcommand{\enthalpysmoothptlevthreecost}{3.5}
\newcommand{\enthalpysmoothptlevthreeradius}{11.9}
\newcommand{\enthalpysmoothptlevthreeinfo}{\ref{sec:dbhf-smoothed}}
\newcommand{\enthalpysmoothptlevthreecperd}{12}

\newcommand{\enthalpyptadordlevfourfigs}{\ref{fig:dbhf-2507-sim}}
\newcommand{\enthalpyptadordlevfourinfo}{\ref{sec:dbhf-2507-grmhd}}
\newcommand{\enthalpyptmonotlevfourfigs}{\ref{fig:dbhf-2507-sim}}
\newcommand{\enthalpyptmonotlevfourinfo}{\ref{sec:dbhf-2507-grmhd}}
\newcommand{\enthalpypolytropelevthreefigs}{\ref{fig:polytrope-comparison}}
\newcommand{\enthalpypolytropelevthreeinfo}{\ref{sec:polytrope appendix}}
\newcommand{\polytropicpolytropelevthreefigs}{\ref{fig:polytrope-comparison}}
\newcommand{\polytropicpolytropelevthreeinfo}{\ref{sec:polytrope appendix}}

\newcommand{\enthalpyslylevthreefigs}{\ref{fig:sly-compare-oscillation}}
\newcommand{\enthalpyslylevthreeinfo}{\ref{sec:sly-grmhd}}
\newcommand{\spectralslylevthreefigs}{\ref{fig:sly-compare-oscillation}}
\newcommand{\spectralslylevthreeinfo}{\ref{sec:sly-grmhd}}
\newcommand{\enthalpyptlevthreefigs}{N/A}
\newcommand{\enthalpyptlevthreeinfo}{\ref{sec:dbhf-2507-grmhd}}
\newcommand{\spectraldbhflevthreefigs}{\ref{fig:dbhf-grmhd-comparison}}
\newcommand{\spectraldbhflevthreeinfo}{\ref{sec:dbhf-grmhd}}
\newcommand{\enthalpydbhflevthreefigs}{\ref{fig:dbhf-grmhd-comparison}}
\newcommand{\enthalpydbhflevthreeinfo}{\ref{sec:dbhf-grmhd}}

\newcommand{\benchenthalpyslypressurefromdensity }{120}
\newcommand{\benchenthalpyslyinternalenergyfromdensity}{120}
\newcommand{\benchenthalpyslymorepressurefromdensity }{224}
\newcommand{\benchenthalpyslymoreinternalenergyfromdensity }{225}
\newcommand{\benchenthalpypressurefromdensity}{43}
\newcommand{\benchenthalpyinternalenergyfromdensity}{44}
\newcommand{\benchspectralpressurefromdensity}{62}
\newcommand{\benchspectralinternalenergyfromdensity}{315}
\newcommand{\benchpolytropepressurefromdensity}{57}
\newcommand{\benchpolytropeinternalenergyfromdensity}{58}

\newcommand{\CIT}{\affiliation{Department of Physics, California Institute of Technology, Pasadena, California 91125, USA}}
\newcommand{\CITLab}{\affiliation{LIGO Laboratory, California Institute of Technology, Pasadena, California 91125, USA}}
\newcommand{\UNH}{\affiliation{Department of Physics \& Astronomy, University of New Hampshire, 9 Library Way, Durham NH 03824, USA}}
 \newcommand{\Cornell}{\affiliation{Cornell Center for Astrophysics and Planetary Science, Cornell University, Ithaca, New York 14853, USA}}
 \newcommand{\Tapir}{\affiliation{Theoretical Astrophysics 350-17, California Institute of Technology, Pasadena, CA 91125, USA}}

\begin{document}

\title{
Simulating neutron stars with a flexible enthalpy-based equation of state parametrization in SpECTRE
}

\author{Isaac Legred}\CIT \CITLab
\author{Yoonsoo Kim}\CIT \Tapir
\author{Nils Deppe} \Tapir
\author{Katerina Chatziioannou} \CIT \CITLab 
\author{Francois Foucart}\UNH
\author{Fran\c{c}ois H\'{e}bert}\Tapir
\author{Lawrence E.~Kidder}\Cornell

\begin{abstract}
    Numerical simulations of neutron star mergers represent an essential step toward interpreting the full complexity of multimessenger observations and constraining the properties of supranuclear matter.  Currently, simulations are limited by an array of factors, including computational performance and input physics uncertainties, such as the neutron star equation of state.
    In this work, we expand the range of nuclear phenomenology efficiently available to simulations by introducing a new analytic parametrization of cold, beta-equilibrated matter that is based on the relativistic enthalpy.  We show that the new \emph{enthalpy parametrization} can capture a range of nuclear behavior, including strong phase transitions.
    We implement the enthalpy parametrization in the \spectre\, code, simulate isolated neutron stars, and compare performance to the commonly used spectral and polytropic parametrizations.  
    We find comparable computational performance for nuclear models that are well represented by either parametrization, such as simple hadronic EoSs. We show that the enthalpy parametrization further allows us to simulate more complicated hadronic models or models with phase transitions that are inaccessible to current parametrizations.  
\end{abstract}

\maketitle

\section{Introduction}

Multimessenger observations of the gravitational wave event GW170817~\cite{TheLIGOScientific:2017qsa,LIGOScientific:2017ync} have highlighted the role of neutron star binaries (BNS) in probing the physics of dense matter, e.g.,~\cite{Dietrich:2020eud,Chatziioannou:2020pqz,Piekarewicz:2022ycz}. In addition, further astronomical observations~\cite{TheLIGOScientific:2017qsa, Abbott:2020uma, Cromartie:2019kug, Fonseca:2021wxt, Miller:2019nzo, Miller:2021qha, Riley:2019yda, Riley:2021pdl, Antoniadis:2013pzd} and terrestrial nuclear experiments~\cite{Adhikari:2021phr, CREX:2022kgg} have facilitated new insights into the equation of state (EoS) of NS matter~\cite{Abbott:2018exr, Essick:2021ezv, Pang:2021jta,Pang:2021jta, Raaijmakers:2019dks, Raaijmakers:2021uju, Landry:2020vaw, Legred:2021, Miller:2019nzo,Miller:2021qha}. Nonetheless significant uncertainty exists about the properties of dense matter above nuclear saturation density\footnote{The saturation density of atomic nuclei is determined via theory and experiments~\cite{DrischlerCarbone2016};  here we fix a value for convenience.}, $\rhonuc \equiv 2.8\e{14}\mathrm{g}/\mathrm{cm}^3$, which translates to uncertainty in the properties of astrophysical NSs whose densities can reach $\sim7\rhonuc~$\cite{Pang:2021jta, Legred:2021}.

The merger phase of a BNS coalescence carries the largest imprint of nuclear matter and strong gravity and it can only be studied numerically. Numerical relativity (NR) simulations of BNS coalescences through merger require solving the equations of general relativistic magnetohydrodynamics (GRMHD) simultaneously with the Einstein field equations and, possibly, the Boltzmann equations for neutrino radiation transport~\cite{2006ApJ...637..296A, Font:2008fka, baumgarte_shapiro_2010}. The system of equations is closed with a nuclear EoS. See e.g.~\cite{Baiotti:2016qnr, Radice:2020ddv, Foucart:2020ats, Kyutoku:2021icp, 2013rehy.book.....R} for reviews of the field.  Such simulations have been used to interpret  existing signals, e.g.~\cite{Margalit:2017dij,Radice:2018ozg,Shibata:2019ctb,Koppel:2019pys,Annala:2021gom,Camilletti:2022jms}, and targeted simulations will likely be an essential tool for understanding future observations.

The most generic strategy for representing the nuclear EoS numerically is piecewise, i.e., using independent expressions in different density or pressure intervals. For example, interpolated tables of thermodynamic quantities such as pressure and internal energy at every value of the density and composition offer access to the widest range of nuclear behavior. However, the temperature- and composition-dependent tables currently used, e.g.~\cite{Typel:2022lcx}, have a significant memory footprint and evaluation requires computationally expensive operations such as constant access to the table and interpolation~\cite{Siegel:2017sav}. The latter may also be inaccurate (at low order) or prone to unphysical oscillations for EoSs with discontinuities or underresolved features (at high order). 
A related approach makes use of piecewise parametrizations such as a \emph{piecewise-polytrope}~\cite{Read:2008iy}, which is effectively a sparsely sampled table for the polytropic exponent. Though it can capture a range of high-density behavior, discontinuities in derivatives  of thermodynamic quantities can degrade simulation accuracy~\cite{Foucart:2019yzo,Raithel:2022san}.

A different strategy is based on functional representations of the EoS that stay smooth across density scales, such as a \emph{single-polytropic} or \emph{spectral} parametrization~\cite{Lindblom:2010bb,Greif:2018njt, Foucart:2019yzo,Raithel:2022san}.  Such parametrizations typically cannot fully represent nuclear EoS models, as they are restricted to a finite number of parameters in the density range of interest~\cite{Greif:2018njt,Foucart:2019yzo}. On top of this, smoothness across density scales fails to capture nuclear models that contain nuclear transitions to exotic degrees of freedom.

In this study, we propose a new parametrization of the nuclear EoS that bridges smooth and discontinuous models while balancing accuracy and computational efficiency.\footnote{We use the term ``model" to refer to a nuclear-theoretic prediction and ``parametrization" for a functional form for the EoS.} We parametrize the relativistic enthalpy~\cite{Lindblom:1992} via a combination of analytic polynomials and trigonometric functions.  Unlike pressure, the difference in enthalpy at densities $[\rhonuc, 3\rhonuc]$ for two EoSs is typically small compared to the enthalpy of either.  The enthalpy can thus be effectively written as a ``baseline" part  plus small corrections. We capitalize on this in order to write the enthalpy as a polynomial, typically capturing $\sim 99\%$ of the EoS, plus small trigonometric corrections, bringing the fit accuracy to 1 in $10^5$. Such a decomposition can capture a wide range of phenomenology with modest changes to the relevant parameters. In addition, further thermodynamic quantities such as the pressure can be evaluated efficiently and analytically.  

We implement this parametrization in \spectre~\cite{spectrecode, Kidder:2016hev}, a scalable next-generation multiphysics computational astrophysics code that uses task-based parallelism~\cite{laxmikant_kale_2020_3972617}.  A primary science target for \spectre\, is fast and accurate GRMHD simulations of BNS coalescences. We use \spectre\ to test the \emph{enthalpy parametrization} on isolated NSs in the Cowling approximation, i.e. we do not evolve the spacetime~\cite{Cowling:1941}, while evolving the ideal GRMHD equations~\cite{baumgarte_shapiro_2010} with a discontinuous Galerkin-finite difference (DG-FD) hybrid scheme~\cite{Deppe2022method, Deppe:2021bhi}. Though these simulations assume a static spacetime, they still allow us to evaluate the role of the enthalpy parametrization in questions of convergence, efficiency, and resolvability of nuclear physics in simulations.

We show that the \neweos parametrization is able to effectively represent a wide range of nuclear behavior, while incurring small additional computational costs relative to simpler parametrizations.  After reviewing the general requirements a parametrization must meet in Sec.~\ref{sec:flexible-eos}, we introduce the enthalpy parametrization in Sec.~\ref{sec:neweos}. 
\red{We demonstrate that it can faithfully fit various nuclear models ranging from smooth EoSs to phase transitions in Sec.~\ref{sec:verification}.
We perform numerical simulations with \spectre\, and find that for resolutions of at least $130$\,m, the EoS evaluation cost is subdominant to other simulation components. We also simulate hybrid stars with quark cores and find that such simulations can be carried out stably with better-than-expected runtime scaling properties under increasing resolution. 
We conclude with discussions in Sec.~\ref{sec:discussion}.
}

\section{EoS Parametrizations for Relativistic Simulations}
\label{sec:flexible-eos}

\subsection{General requirements}

We begin with a general discussion of the requirements phenomenological parametrizations of the nuclear EoS must meet for efficient use in numerical simulations. These include (i) faithful representation of target nuclear models, (ii) parametric extensibility, and (iii) computational performance related to smoothness (to the extent allowed by the underlying nuclear physics) and/or a fully analytic formalism.

The first requirement is that the parametrization is generic enough that it can faithfully represent the target nuclear physics.  While no standard faithfulness metrics exist, a common test is the $L^2$ difference of quantities of interest~\cite{Lindblom:2010bb, Read:2008iy}.  Nonetheless it is unclear how different metrics relate, for example the $L^2$ difference of the local polytropic indices and that of the mass-radius curve~\cite{Lindblom:2013kra, Foucart:2019yzo}.  One particular challenge to smooth parametrizations is modeling strong phase transitions~\cite{Han:2018mtj, Pang:2020ilf}. In general we would like a parametrization where, whatever the metric, we can improve the fit via iterative approximation.  In principle this is available to any parametrization by adding more parameters and smoothly changing parameter values. In practice, however, the functional form of the parametrization may limit the accessible parameter space, as shown in~\cite{Wysocki:2020myz} for the spectral parametrization.

The second requirement is that the parametrization allows us to parametrically explore a wide range of possible high-density behavior.  This entails continuously, and without significant fine-tuning, extending the parametrization to produce EoSs that might differ from existing nuclear models.  An example of such an extension would be a parameter which controls the pressure at a particular density and thus allows us to isolate the effect of this density scale on macroscopic observables. Another benefit of such continuous extensibility is that it allows us to construct a map from the EoS to observables, \red{e.g.~\cite{Ozel:2009}}.  This approach has already been successfully used in the case of binary black hole mergers to produce accurate \emph{surrogates} of the map of binary configurations to gravitational waves~\cite{Blackman:2015pia,Varma:2018mmi}.  A similar methodology could be used to construct a surrogate for the post-merger gravitational-wave signature of BNS mergers, whose EoS dependence is not well captured by a small number of parameters~\cite{Wijngaarden:2022sah,Breschi:2022xnc}.

At the same time, we consider practical requirements in terms of computational performance: speed and accuracy of the relevant evaluations, and smoothness of the thermodynamic quantities where possible.
A fully analytic form for the EoS and all the relevant thermodynamic quantities is a sufficient (but perhaps not necessary) condition.
Tabulated EoSs, while guaranteeing maximal flexibility, fail in this regard. Consider, for example, primitive variable recovery. Numerical simulations evolve the components of the stress-energy tensor which are nonlinear functions of primitive variables such the rest-mass baryon density $\rho$, pressure $p$, and specific internal energy $\epsilon$. This process involves inverting the relation between the stress-energy tensor and the primitive variables with root-finding routines during which the EoS, for example $p(\epsilon)$, is evaluated repeatedly. For tabulated EoSs this includes computing the temperature $T$ from $\epsilon$ via another root-find and then computing $p(T)$ via a table lookup and interpolation.  The EoS tables are typically too large to store in the CPU caches and so the nested root-finding routines require repeated loading of data from main memory, causing significant overhead that dominates simulation cost~\cite{Siegel:2017sav}.

Another advantage of fully analytical parametrizations is that they enable efficient computation of all necessary thermodynamic quantities in a consistent way. Besides tabulated EoSs, this also applies to certain parametrizations that require interpolation or numerical integration. For example, the spectral parametrization allows for analytic evaluation but not integration of $d\eps/d\rho$. Then, $\eps(\rho)$ is computed via a computationally expensive numerical integral as high accuracy is required to avoid thermodynamic inconsistency during primitive variable recovery.  \red{Even if tables are used in simulations, ensuring smoothness and consistency requires building higher-order interpolants (or sampling very densely).  This effectively amounts to constructing local parametrizations of the EoS which satisfy some stitching constraints.  Therefore, even the use of tables in NR simulations stands to gain from understanding fully analytic representations of the nuclear EoS.}

\subsection{Existing parametrizations of the EoS}

The simplest parametrization of cold, beta-equilibrated, dense matter is a single polytrope that prescribes a relationship between the rest-mass baryon density $\rho$ and the pressure $p$
\begin{equation}
\label{eq:polytrope}
    p(\rho) = K \rho ^{\Gamma}\,,
\end{equation}
where $\Gamma$ is the the polytropic exponent and $K$ is the polytropic constant; both are independent of $\rho$. For example, a degenerate neutron gas would obey a polytropic relation with $\Gamma=5/3$.  Polytropes have a long history in NS simulations, e.g.,~\cite{Shibata:1999wm,Etienne:2007jg,Baiotti:2008ra,Duez:2008rb}, and more recent code tests, e.g.,~\cite{Radice:2013hxh,DeBuhr:2015jqk,Deppe:2021bhi}, due to their simplicity, low computational cost, and the fact that they allow for analytic evaluation of pressure, internal energy, specific enthalpy, and rest-mass density. Nonetheless, their simplicity makes polytropes incompatible with realistic EoS nuclear models, either hadronic (for example, polytropes do not satisfy the same universal relations as hadronic models~\cite{Yagi:2013awa}) or hybrid ones that include multiple degrees of freedom.

Piecewise-polytropes~\cite{Read:2009yp} extend single-polytropes to multiple polytropic segments at different densities, thereby decoupling low- and high-density behavior.  With enough piecewise segments, piecewise-polytropes can also fit EoSs with strong phase transitions~\cite{Ujevic:2022nkr}.
While piecewise-polytropes retain some of the computational simplicity of the single-polytrope and have been employed in BNS mergers~\cite{Hotokezaka:2011dh,Lackey:2013axa,Dietrich:2018upm, Dietrich:2017aum}, the lack of smoothness across stitching boundaries tends to increase the computational cost and decrease the accuracy~\cite{Foucart:2019yzo,Raithel:2022san}. Extensions to continuous polytropic indices~\cite{OBoyle:2020qvf,Raithel:2022san} \red{guarantee differentiability of the pressure; however,  it is unclear how to extend the parametrization to guarantee further derivatives of the pressure exist at the stitching point.  Generically stitching two $\mathcal{C}^n$ functions to form a globally $\mathcal{C}^{n}$ function requires matching $n+1$ derivatives, which may require the introduction of functions to the parametrization of $p(\rho)$ for example, which make it difficult to solve for $e(\rho)$ analytically.   }

Finally, the spectral parametrization~\cite{Lindblom:2010bb}  accurately reflects a broad range of nuclear models while maintaining smoothness across density scales. 
The parametrization has a similar form to a polytrope
\begin{equation}
    p(\rho) = K \rho ^{\Gamma(\rho)}\,,
\end{equation}
but now $\Gamma(\rho)$ is expanded in a basis of smooth functions, typically a polynomial.
The spectral parametrization can successfully reproduce hadronic nuclear models with a comparable number of parameters as polytropes, though it cannot capture sharp changes in the speed of sound that are associated with phase transitions~\cite{Lindblom:2010bb, Foucart:2019yzo}.
Compared to piecewise polytropes and other EoS with discontinuities, the spectral parametrization can lead to reduced computational cost in simulations ~\cite{Foucart:2019yzo} for a given accuracy requirement, while remaining more computationally intensive than pure polytropes. 
Our current implementation of the spectral EoS balances faithfulness to nuclear models and computational efficiency by expressing $\Gamma(\rho)$ as a a polynomial in $\log \rho$~\cite{Foucart:2019yzo}. More complex basis functions could improve faithfulness, but they would come at the cost of computational efficiency since computation of the internal energy requires a numeric integral whose accuracy depends on how rapidly $\Gamma(\rho)$ varies.

The above discussion highlights the role of balancing faithfulness and computational efficiency in selecting EoS parametrizations for numerical simulations.  While the single-polytrope is  computationally efficient, it is too restrictive in terms of nuclear physics.  Piecewise-polytropes expand the range of nuclear models accessible, but at the cost of longer runtimes and loss of accuracy due to non-smoothness at the stitching boundaries. The spectral parametrization strikes some balance, but performs optimally when few parameters are used; it is therefore restricted to simple nuclear models.  Ultimately, we would prefer an EoS parametrization which is able to fit to a problem-specific precision,  matching the level of other errors in simulations at the lowest possible cost. This motivates the introduction of a new parametrization with increased flexibility to model a wider range of nuclear EoSs without considerable performance losses.

\section{\neweos parametrization of the EoS}
\label{sec:neweos}

In this section we introduce a new \emph{enthalpy parametrization} with a flexible number of degrees of freedom that expands the range of microscopic physics we are able to represent in numerical simulations.  
In the following, we work in geometric units: $c=1$, $G=1$.

\subsection{Parametrizing the enthalpy}

The specific enthalpy of a system $h$ is defined as the enthalpy per unit mass. In relativistic contexts it represents the energy required to inject a unit of rest mass into the system while remaining in thermodynamic equilibrium. The first law of thermodynamics requires that at zero temperature $T$ and in $\beta-$equilibrium, 
\begin{equation}
    \label{eq:first-law}
    h(\rho) \equiv  \quant{\pderiv{e}{\rho}}_{T, \beta} = \deriv{e}{\rho} = \frac{p(
    \rho) + e(\rho)}{\rho}\,, 
\end{equation}
 where $e$ and $p$ are the energy density and pressure, while $\rho$ is the rest-mass energy density of baryons.

We choose to directly parameterize the enthalpy for three primary reasons. First, the enthalpy is a monotonically-increasing and slowly-varying function of the baryon density, which is numerically beneficial.  Second, the enthalpy can be intuitively interpreted as a measure of the stiffness of the EoS: a larger enthalpy corresponds to higher pressure and energy density.  Third, and importantly for hydrodynamic simulations, the enthalpy in cold, beta-equibrilated matter is related to other thermodynamic quantities by linear operations, which facilitates analytic calculations and avoids interpolation or numerical integration. 

From the first law, we have
\red{
\begin{align}
    \label{eq:hofrhoderiv}
    \deriv{h}{\log \rho} & =\deriv{e}{\rho} + \deriv{p} {\rho} - h\\
    \label{eq:deriv-props}
    &=   \deriv{p}{\rho} = \deriv{p}{e}\deriv{e}{\rho} \\
    \label{eq:deriv-props-cs2}
   &=h c_s^2\,,
\end{align}}
Equation~\eqref{eq:deriv-props} suggests that $dp/d\rho$ is zero if and only if $dh/d\rho$ is zero. Equation~\eqref{eq:deriv-props-cs2} provides the motivation for our parametrization choices.
Consider, for example, a constant speed of sound  $c_s=c_{s,0}$. Then 
\begin{equation}
    c_{s,0}^2 = c_s^2 = \frac{dp}{de} \implies p = p_0 + c_{s, 0}^2 \Delta e\,,
\end{equation}
with $p_0=p(e_0)$ and $\Delta e\equiv e-e_0$. In this special case Eq.~\eqref{eq:deriv-props-cs2} becomes
\begin{align}
\label{eq:general-css}
 h(\log \rho)&\propto \exp\quant{c_{s,0}^2\log\rho}\nonumber \\
 &\approx \rho_0\left[1 + c_{s,0}^2\log\quant{\rho/\rho_0} + \dots\right]\,,
\end{align}
where $\rho_0$ is some fiducial density. Equation~\eqref{eq:general-css} suggests that if  $c_s^2$ is slowly varying,\footnote{In general, causality and stability bound $0\leq c_s^2 \leq 1$.} the enthalpy can be approximated as exponential in $\log \rho$.  Moreover, a smaller $c_s^2$ accelerates the convergence of the series of Eq.~\eqref{eq:general-css}, though this also depends on the choice of density scale $\rho_0$.  We therefore choose the Taylor expansion in Eq.~\eqref{eq:general-css}  as the starting point of the enthalpy parametrization.
 
 We further select $\log \rho/\rho_0,$ as the independent variable of the parametrization. This choice enables us to better resolve the low-density EoS. Equation~\eqref{eq:general-css} further suggests that  $h(\log \rho) \propto \exp(c_{s,0}^2 \log\rho)$ is analytically and computationally simpler than  $h(\rho)\propto  \rho^{c_{s,0}^2}$ as the Taylor expansion of $\rho^{c_{s,0}^2}$ converges more slowly than the expansion of $\exp(c_{s,0}^2\log \rho)$ for non-integer $c_{s,0}$.

 Lastly, a desirable property of the specific enthalpy is that it is continuous across first-order phase transitions. This can be seen from Eq.~\eqref{eq:deriv-props} where maintaining a constant pressure across the transition guarantees that the enthalpy will be constant as well.  This indicates that across certain weak transitions the enthalpy can be expanded in a basis of continuous functions, unlike, for example, a local polytropic exponent.

\subsection {Decomposition}
\label{sec:decomposition}

Motivated by Eq.~\eqref{eq:general-css}, we introduce a parametrization of $h(\log \rho)$.  Given an EoS in some density region $\rho_{\min} \leq \rho \leq \rho_{\max}$ we select a density scaling parameter $\rho_0 \leq \rho_{\min}$ such that $z \equiv \log(\rho/\rho_0)$ is positive in the relevant density range. Importantly, $\rho_0$ is not necessarily equal to $\rho_{\min}$ thus introducing an additional parameter.   We then write
\begin{equation}
    \label{eq:polynomial-expansion}
    h(z) \approx  h_p(z) \equiv \sum_{i=0}^{i_{\max}} \gamma_i z^i\,, 
\end{equation}
where $h(z)$ is the target enthalpy and $ h_p(z)$ is its approximation.
This polynomial decomposition is motivated by the previous observation that $h(z)$ is approximately exponential in $z$ for nearly constant speeds of sound, corresponding to $\gamma_i \sim c_{s,0}^{2i}/i!$. The rapid convergence of the $\gamma_i$ sequence indicates that the $i>i_{\max}$ terms will be small provided that the speed of sound is slowly varying.

Given that $h(z)$ is positive and increasing and $z_i>0$, catastrophic floating point cancellation in numerical calculations can be avoided by restricting to $\gamma_i \geq 0$.  This guarantees that each term $\gamma_i z^i$ is a small and positive correction to previous terms.  Furthermore, the polynomial expansion of Eq.~\eqref{eq:polynomial-expansion} can be efficiently and stably evaluated with Horner's method~\cite{numericalrecipes}.  Allowing for more general $\gamma_i$ is possible, but this comes at a risk of oscillatory behavior and cancellation of large terms which make convergence predictions difficult.  The implications and rationale behind the choice to set $\gamma_i \geq 0$ are further discussed in App.~\ref{sec:coefficient-numerics}.   

A consequence of setting $\gamma_i \geq 0$ is that Eq.~\eqref{eq:polynomial-expansion} is unable to model certain EoSs, for example the case where $dh/dz =h c_s^2$
is not strictly increasing,  even with $i_{\max} \rightarrow \infty$. Such a non-monotonic speed of sound could be encountered for complicated hadronic models or more generically if non-hadronic degrees of freedom are introduced~\cite{McLerran:2018hbz, Tews:2018kmu, Kapusta:2021ney, Han:2018mtj}.  
We therefore augment Eq.~\eqref{eq:polynomial-expansion} by decomposing $h_t(z) \approx h(z)-h_p(z)$  as a Fourier series
\begin{equation}
 h_t(z) \equiv \sum_{j=1}^{j_{\max}} a_j \sin(jkz) + b_j \cos(jkz)\,, 
 \label{eq:enthalpyFourier}
\end{equation}
where $k$ sets the ``wavelength scale" of the fit.  In a Fourier series $k$ is typically fixed to 
\begin{equation}
k = k_{F} \equiv \frac{2\pi}{z_{\max} - z_{\min}}  =\frac{2\pi} {\log(\rho_{\max}/\rho_{\min})}\,,
\end{equation}
but here we vary it and find that $k \gtrapprox k_F$ leads to good fits. The effect of perturbing $k$ around $k_F$ is small, as we explore in App.~\ref{sec:fitting}. 
The trigonometric expansion of Eq.~\eqref{eq:enthalpyFourier} can also serve as a low-pass filter to remove high-frequency oscillations from the tabulated EoS data that may not be physical or computationally resolvable.
In summary, the enthalpy parametrization is 
\begin{equation}
    \label{eq:expansion}
   h_*(z) \equiv h_t(z) + h_p(z) \approx h(z)\,.
\end{equation}
\begin{figure*}
    \centering
    \includegraphics[width=.99\textwidth]{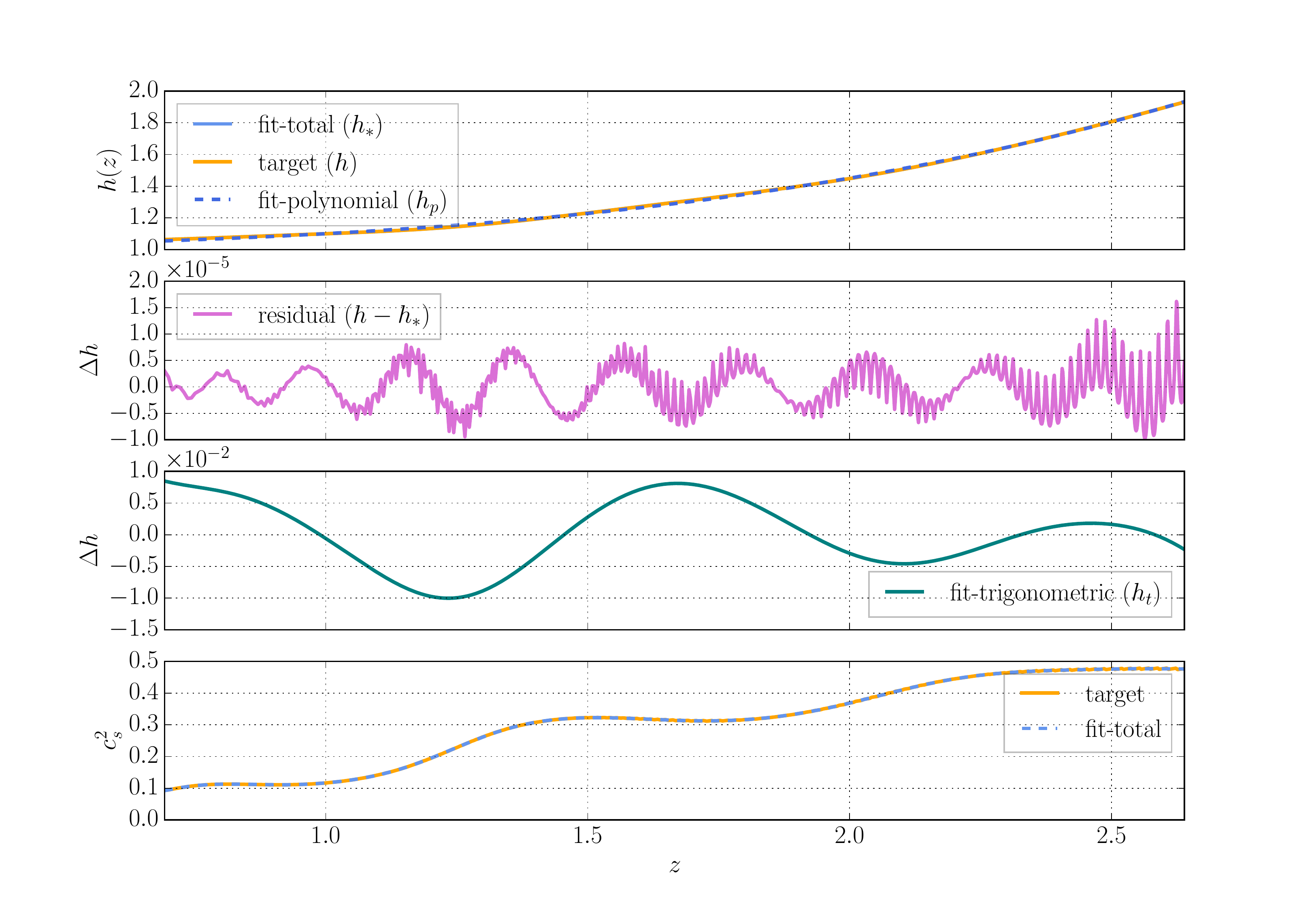}
    \caption{Results of a fit to an EoS drawn from a Gaussian process with the enthalpy parametrization. We plot various thermodynamic quantities as a function of $z$. The fit parameters are $\rho_{\min} = \rho_{\nuc}$, $\rho_{\max} = 7\rhonuc $,  $\rho_0=0.5\rho_{\nuc}$, $k = \pi/(\log(7))$, and $i_{\max} = j_{\max} = 10$.
    \emph{Top Panel:} The tabulated EoS $h$ (solid, orange) and the total fit $h_*$ (solid, light blue).  We also plot the polynomial fit to the EoS $h_p$ (dashed, indigo). Both the total and the polynomial fit are indistinguishable from the tabulated EoS by eye. 
    \emph{Second Panel:} The residuals of the total fit $h- h_*$.  \red{In this metric, the fit demonstrates excellent agreement relative to $h-1 = p/\rho + \eps \gtrsim 1\e{-2}$}
    \emph{Third Panel:}
    The trigonometric fit $h_r=h - h_p$. 
    \emph{Fourth Panel:}
     $(1/h)dh/dz = c_s^2$, for both the tabulated EoS and the total fit.  \red{Heuristically, the speed of sound has a comparable number of plateaus to the number of obvious peaks in $h_t$.}  
    }
    \label{fig:fit-residuals}
\end{figure*}

In Fig.~\ref{fig:fit-residuals} we demonstrate the enthalpy parametrization fit of Eq.~\eqref{eq:expansion} and its polynomial, Eq.~\eqref{eq:polynomial-expansion}, and trigonometric, Eq.~\eqref{eq:enthalpyFourier}, components for a phenomenological EoS drawn from a Gaussian process prior~\cite{Landry:2018jyg, Essick:2019ldf}. 
 The polynomial fit alone is accurate to about $\mathcal O(1\%)$, while the total fit is good to about one part in $10^5$.  For reference, we also plot $c_s^2 = (1/h) dh/dz$, as a measure of the complexity of the EoS.  Even though $c_s^2$ is not globally nearly constant, it is slowly varying and nearly monotonic.  

Given the generic form of the enthalpy parametrization, there is no guarantee that a particular fit will satisfy stability $c_s^2\geq0$ and causality $c_s^2 \leq 1$.  If $h_t(z)=0$, the fit is guaranteed to be stable,  and a sufficient but not necessary condition for causality is $ \gamma_i \leq \gamma_{i-1}/i$, which becomes necessary and sufficient in the case of a constant sound speed.  If $h_t(z)$ is nonzero, \red{then $h_*(z)$ can oscillate, changing on scales of order the most quickly varying Fourier mode}.   Therefore, both conditions must be checked on a grid of spacing
\begin{equation}
   \red{\delta z \lesssim \frac{1}{ j_{\max}k}\,,}
\end{equation}
where, as above, $j_{\max}$ is the index of the fastest varying ``Fourier" mode.  

While an unstable fit to the EoS cannot be tolerated in a numerical simulation, an acausal fit may be used if it is very nearly causal (i.e.  if $\sqrt{c_s^2 - 1}$ is small compared to the velocity resolution of the simulation).  
In practice, however, fits typically are neither acausal nor unstable; if they are it is often a sign that the fit to the EoS is poor and more parameters should be used.

%
\subsection{Computing thermodynamic quantities}
\label{sec:computing-thermo-qs}

Given the expansion of Eq.~\eqref{eq:expansion}, we can analytically compute the thermodynamic quantities needed for GRMHD evolution as formulated in \spectre~\cite{spectrecode, Deppe:2021bhi}, or similar codes~\cite{Mosta:2013gwu}. 
For example, the energy density is
\begin{align}
    \label{eq:energy-density}
    &\deriv{e}{z} = \rho \deriv{e}{\rho} = \rho_ 0 \exp\quant{z}h(z) \Rightarrow\nonumber \\
    &e(z) = \rho_0 \int_{z_0}^z \exp \quant{z'} h(z')dz' + e(z_0)\,.
\end{align}
Since $h(z)$ is expressed in terms of sines, cosines, and polynomials, the integral of Eq.~\eqref{eq:energy-density} can be computed using the following identities 
\begin{align}
    \label{eq:trig-integral}
    &\int \exp\quant{z} \sin(nkz) dz\notag \\
    &=\exp\quant{z} \frac{\sin(nkz) - nk \cos(nkz)}{1+n^2 k^2} + C\,,\\
    \label{eq:polynomial-integral}
    &\int  \exp\quant{z} \frac{z^n}{n!} dz \notag \\
    &= \exp\quant{z}\frac{z^n}{n!} - \int \frac{z^{n-1}}{(n-1)!} \exp \quant{z}dz = \dots \,,
\end{align}
where the ellipses indicate that integration by parts can be repeated until the integral becomes trivial.  
Equation~\eqref{eq:polynomial-integral} is also a gamma function, but it is typically incomplete.
Nonetheless, all integrals can be evaluated analytically and $e(z)$ has an expansion of the form
\begin{align}
    \label{eq:energy-expansion}
    e(z) &=\exp(z) \times \nonumber \\
    &\left(
\sum_i \gamma_i' z^i + \sum_{j}a_j' \sin(kjz) + b_j'\cos(kjz)\right) \nonumber \\
&+ e_*\,, 
\end{align}
where the constant $e_*$ is determined by setting $e(z_{\min}) = e_{\min}$ and the coefficients satisfy
\begin{align}
\label{eq:energy-polynomial-coeffs}
\gamma'_i  &= \frac{1}{i!} \sum_{ i_{\max}\geq  \ell \geq i}   (-1)^{i_{\max}-\ell} \ell!\gamma_{\ell}\,,\\
a'_j &= \frac{a_j}{1+j^2 k^2} + \frac{b_j jk}{1+ j^2 k^2}\,,\\
b'_j &= \frac{b_j}{1+j^2 k^2} - \frac{a_j jk}{1+ j^2 k^2}\,.
\end{align}

The pressure $p(z)$ can also be evaluated analytically with a similar expansion given that
\begin{equation}
    p(z) = \rho_0 h(z)\exp\quant{z} -e(z) = h \rho-e\,.
\end{equation}
This equation showcases the benefits of setting $\gamma_i \geq 0$ in Eq.~\eqref{eq:polynomial-expansion} to avoid cancellations in the enthalpy expansion.  The pressure is computed as the difference of two relatively large quantities, each typically 1--3 orders of magnitude larger than the pressure itself in the relevant density interval.  If the expansion of $h(z)$ additionally had large coefficients (i.e. much larger than the enthalpy) terms of $e(z)$ will be computed by sums of alternating large numbers, which is numerically undesirable.   However, because $\gamma_\ell \sim 1/\ell!$ for EoSs with slowly varying speed of sound, the terms in Eq.~\eqref{eq:energy-polynomial-coeffs} are of comparable size, and about the same size as corresponding terms of $h(z)$.  Thus the terms of $p(z)$ are computed to comparable precision as the terms of  $e(z)$ and $h(z)$.  We find this holds more broadly, even when the speed of sound is not slowly varying, as $\gamma_{\ell}$ is typically decreasing even if it is not decreasing exponentially as in the constant-$c_s^2$ case.

Lastly, we can also analytically compute
\begin{equation}
    \deriv{p}{\rho} = \deriv{h}{z}\,, 
\end{equation}
through
\begin{align}
    \deriv{h}{z} &= \sum_i i \gamma_i z^{i-1} \nonumber \\
    &+ \sum_j jk \left[a_j \cos(jkz) - b_j \sin(jkz) \right]\,.
    \label{eq:dhdz}
\end{align}
As can be seen from Eqs.~\eqref{eq:energy-expansion} and~\eqref{eq:dhdz}, parameters that enter linearly in the original expansion of $h(z)$ also appear linearly in all relevant thermodynamic quantities.

\subsection{Low-Density Stitching}

The \neweos parametrization is best suited for high-density regions where pressure and energy density are comparable. Low-density regions with $p  \ll h\rho \sim e$ might be better fit by direct parametrizations of the pressure. We therefore combine the \neweos parametrization with a simpler low-density parametrization below $ \rho_{\min}$. Incidentally, this density region coincides with the region of validity of nuclear theory calculations~\cite{Tews:2020, Drischler:2016djf, Essick:2020flb} and terrestrial experiments~\cite{Adhikari:2021phr, CREX:2022kgg, Roca-Maza:2015eza, Essick:2021ezp, Essick:2021kjb}. The low-density EoS is therefore better constrained and thus there is reduced need for flexibility in the EoS parametrization. Moreover, the low-density EoS has a reduced impact on NS observables, especially if the simulation resolution is low, such that $(dp/dr) \Delta r > \delta p$, where $\Delta r$ is the grid spacing and $\delta p$ is the difference induced by EoS mismodeling.

A number of options exist for the low-density EoS, including direct parametrizations of nuclear models~\cite{Tews:2020} or chiral effective field theory($\chi$-EFT) results~\cite{Tews:2018iwm,Pang:2021jta}.
Here we select the existing spectral parametrization implementation~\cite{Foucart:2019yzo}, as it is more flexible than single-polytropes, but smoother than piecewise-polytropes and tabulated EoSs.  In certain cases, we explore extending the spectral parametrization up to relatively high densities $\sim 2\rhonuc$ if it can fit the target EoS well-enough in this density regime. Due to the low number of parameters in the spectral parametrization, all degrees of freedom are determined by requiring differentiability of the pressure and continuity of the internal energy at the stitching points. We verify this stitching maintains $\mathcal C^1$ smoothness, see App.~\ref{sec:fitting}. 
\subsection{\red{Free parameters and fitting}}

\begin{figure}
    \centering
    \includegraphics[width=.49\textwidth]{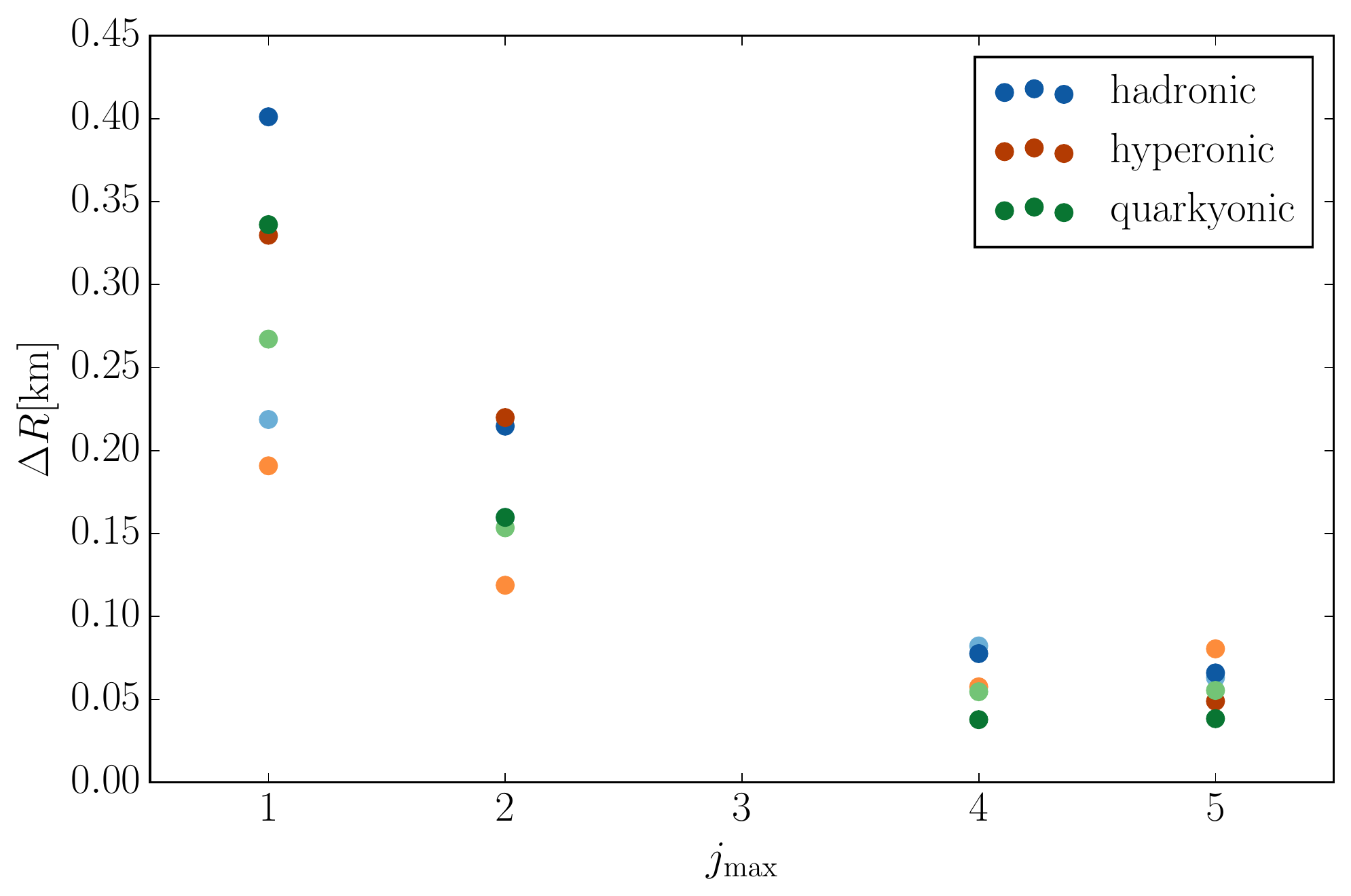}
    \caption{\red{Radius error in fitting Gaussian Process-generate EoSs conditioned on  $\chi$-EFT~\cite{Essick:2021ezv, Essick:2021kjb} with the \neweos{}parametriation. We plot two hadronic-conditioned draws, two quark-conditioned draws, and two hyperonic draws.  This  indicates the draws are from processes conditioned on EoS models of the given type, so that e.g. the hadronic process is consistent with known hadronic EoSs.  Nonetheless the processes use ``agnsotic" kernels which lead to very compatible distributions on EoSs for each of the three cases~\cite{Landry:2018prl, Essick:2019ldf}. A problem with stitching stability affected multiple of the fits at $j_{\max}=3$, so we exclude these.  }}
    \label{fig:chieft-fits}
\end{figure}

\red{The number of free parameters needed  to achieve good fits of arbitrary EoSs will impact the simulation cost.  Indeed, the cost of evaluating any EoS-dependent quantity is proportional to the number of coefficients used for the \neweos{}parametrization.  Therefore it is prudent to use only as many terms as necessary to achieve an accurate fit; accuracy in the context of numerical simulations is measured relative to other simulation errors.  There is no definitive metric for EoS mismodeling error, as the relevant error will depend on the application.  For example, in applications to BNS inspirals, the relevant errors are in GW phase, matter hydrodynamic variables, and magnetic field variables.  When considering the fitness of an EoS parametrization for use in simulations, all these factors should be taken into consideration.

Nonetheless, it is pragmatically necessary to define surrogate goodness-of-fit statistics in order to both fit the \neweos{} parametrization to data and determine approximately if such a fit is good.  We describe the fitting procedure of the parametrization to a tabulated model that we employ in App.~\ref{sec:fitting}.  Briefly, we fit the specific enthalpy $h(z)$ on a linear grid in $z$ but with variable precision, requiring higher precision at lower densities to achieve equal cost across density scales.   However, fitting is not the only way to extract coefficients for use in the \neweos{}parametrization; for example, coefficients to approximate a polytropic EoS are derived in App.~\ref{sec:polytrope appendix} using a Taylor expansion of the specific enthalpy.  Nonetheless, for realistic nuclear models, fitting the specific enthalpy is usually necessary.

One convenient benchmark is to examine the error in radius of a typical neutron star induced by using a \neweos{} parametrization fit as compared to a tabulated model.  In order to demonstrate the general requirements for fitting, we fit a collection of realistic nuclear theoretic EoSs, compute the error in the radius of a $1.4M_{\odot}$ NS ($\Delta R_{\typ}$) and display the results in Table~\ref{tab:realistic-eos-fits}.  These fits are all carried out with $i_{\max} = 12$, and $j_{\max} \leq 5$, and have typical NS radius error of less than $70$\,m. EoS modelling error would therefore likely not be limiting in simulations with $\sim 70$ meter resolution; this is a conservative choice of error measure as realistic simulation errors will likely dominate static errors.   Additionally, we fit phenomenolgical EoSs drawn from a Gaussian process-mixture model priors~\cite{Landry:2018prl, Essick:2019ldf}.  We examine two cases, first are draws from a model-agnostic prior,  which are only loosely informed by nuclear theory calculations.  The second class are Gaussian process draws conditioned on $\chi$-EFT up to $1.5\rhonuc$~\cite{Essick:2021ezv, Essick:2021kjb}.  Both nuclear-theoretic and phenomenological EoSs show comparable fit quality, indicating that the \neweos{}parametrization is able to reproduce a wide range of EoS models.  

We list $j_{\max}$ in Table~\ref{tab:realistic-eos-fits} as we expect that the number of trigonometric terms is the leading-order driver of cost to evaluate the parametrization. We quantify this further in Sec.~\ref{sec:sly-grmhd}.  Contrarily we expect little dependence of evaluation cost of $i_{\max}$ because evaluation of polynomials using Horner's method is extremely efficient.   For realistic EoSs, fine-tuning of low-density stitching and nonlinear parameters can reduce the number of trigonometric correction terms that are required to achieve a good fit.  Even when no fine-tuning is required, typically good fits are achieved with $j_{\max} \sim 4$.  We quantify this in Fig.~\ref{fig:chieft-fits} by showing the error in the radius of a typical star for 6 different $\chi$-EFT informed Gaussian process draws, when no fine tuning of nonlinear or low-density parameters is performed.  The fits are better when more trigonometric correction terms are included, all falling below $100$\,m error by $j_{\max} =4$.  These errors are often due to the EoS at low-densities, and so typically fine-tuning of certain parameters, such as the low density polytropic index, or the energy density of EoS at the stitching density, must be carried out to achieve $\sim 10$-meter-error fits.  In practice, though, this may not be necessary as quantities such as the tidal deformability are determined by the bulk of the matter, interior to the crust, therefore crust modeling errors may be less significant then predicted by using the radius as a metric. These considerations will be especially important for BNS simulations where GW emission is predominately determined by tidal deformability, and other sources of error may overshadow EoS modeling error.

\begin{table}
 \centering
 \caption{\red{A list of EoS fits with the \neweos{}parametrization to nuclear theoretic and phenomenological EoS.  Theoretic EoSs are listed according to the conventions of~\cite{Read:2008iy}.  Phenemenological EoSs are drawn from Gaussian process priors.  The EoSs gp1 and gp2 are drawn from a model agnostic Gaussian process prior~\cite{Landry:2018prl, Essick:2019ldf}. EoSs  gp$\chi$eft1, gp$\chi$eft3, and gp$\chi$eft5  are drawn from Gaussian process priors conditioned on $\chi$-EFT predictions at low densities.  These three EoSs represent draws from hadronic, hyperonic, and quarkyonic conditioned GPs, respectively.}}
  \begin{tabular}{c|ccccc}
EoS& $R_{\typ}\ [\mathrm{km}]$& $\Delta R_{\typ}\ [\mathrm{km}]$& $\Delta M_{\max}\ [M_{\odot}]$& $j_{\max}$& Ref.\\
\hline
alf2& 12.968& -0.028& -0.003& 3& \cite{Alford:2004pf} \\
bsk19& 10.763& -0.006& -0.001& 5& \cite{Potekhin:2013qqa} \\
ap4& 10.595& -0.02& 0.001& 3& \cite{Akmal:1998cf}  \\
H4& 12.931& 0.01& 0.001& 3& \cite{Lackey:2005tk} \\
bbb2& 11.442& -0.05& -0.008& 5& \cite{Baldo:1997ag} \\
eng& 12.306& -0.071& -0.009& 2&  \cite{Engvik:1995gn}\\
mpa1& 11.696& -0.062& -0.003& 5&  \cite{Muther:1987xaa}\\
ms1& 14.223& -0.015& -0.007& 5&\cite{Mueller:1996pm}  \\
qmc700& 11.942& -0.008& -0.002& 5& \cite{Rikovska-Stone:2006gml} \\
sly& 11.873& -0.053& -0.003& 5&\cite{Douchin:2001sv}  \\
wff2& 10.373& -0.049& -0.002& 5& \cite{Wiringa:1988tp}\\
gp1& 12.302& -0.02& -0.007& 4& \cite{Landry:2018prl} \\
gp2& 12.345& -0.024& 0.001& 4&  \cite{Landry:2018prl} \\
gp$\chi$eft1& 10.496& -0.052& 0.001& 5& \cite{Essick:2021ezv} \\
gp$\chi$eft3& 10.509& -0.049& 0.001& 5&  \cite{Essick:2021ezv}\\
gp$\chi$eft5& 10.789& -0.057& -0.002& 5&  \cite{Essick:2021ezv}\\
  \end{tabular}
 \label{tab:realistic-eos-fits}
\end{table}}
\subsection{Use cases}

\red{The primary function of the \neweos{}parametrization is to represent EoS models for use in numerical simulations containing dense matter.  Given the wide range of models of nuclear matter, the \neweos{}parametrization is intentionally very flexible.  Existing parametrizations of the nuclear EoS typically have a handful of parameters, and extending them might be nontrivial.  In contrast, well-interpolated tables have many ``parameters", or tabulation points, some of which we would prefer not to resolve in simulations (such as artificially rapid changes in some pressure derivative).  The \neweos{} balances these requirements in such a way that the maximal level of flexibility can be found without introducing extraneous parameters.  This allows us to resolve EoSs from nuclear theory, Sec.~\ref{sec:dbhf-test}, as well as EoSs which extend or modify nuclear models, Sec.~\ref{sec:dbhf-2507-grmhd}.  Such flexibility is crucial for determining the observational implications of new degrees of freedom at arbitrary density scales.}

Furthermore, the requirements laid out in Sec.~\ref{sec:flexible-eos} are tailored for a specific application of EoS parametrizations, namely numerical simulations involving NSs. These requirements are  domain specific and need not necessarily lead to efficient parametrizations for different applications, for example EoS inference using astrophysical data. Besides the general faithfulness and computational efficiency considerations, EoS parametrizations employed in inference need to satisfy an additional requirement: they must provide a reliable path from the observed data to the EoS constraints. Specifically, the data must be the primary driver of inference while the impact of the EoS parametrization itself must be either minimal or driven by first principles and nuclear theory. Parametrizations that impose a functional form for the EoS in terms of a finite number of parameters may fail this requirement~\cite{Greif:2018njt, Carney:2018sdv}. Specifically, the spectral, piecewise-polytropic, and speed-of-sound parametrizations impose additional phenomenological correlations between different densities that are not guided by nuclear theory but instead by the arbitrary functional form of the parametrization itself~\cite{ Legred:2022pyp}. Though we have not repeated the analysis of~\cite{Legred:2022pyp}, we expect that the enthalpy parametrization has the same pitfall as it possesses many nearly-irrelevant degrees of freedom that are not constrained by current observations and will generically impart correlations between density scales. We therefore caution against using it for inference purposes. 

\section{Parametrization Verification and simulations}
\label{sec:verification}

\red{
In this section, we look in depth at fitting nuclear and phenomenological models with the \neweos\ parametrization and perfom numerical simulations.
First, we use SLy1.35~\cite{Foucart:2019yzo}, a spectral fit to the SLy EoS~\cite{Douchin:2001sv, Read:2008iy} with a low-density polytropic exponent of 1.35962. This represents a nuclear EoS which has been effectively simplified by being fit with a spectral EoS.   Therefore, this test allows us to analyze the performance of the \neweos\ parametrization on a problem where lower dimensional parametrizations are applicable, in terms of both accuracy and computational performance.
We next consider a tabulated DBHF~\cite{Gross-Boelting:1998xsk} EoS, \red{derived from relativistic, \emph{ab initio} calculations of protons and neutrons \emph{dressed} via interactions with one-boson exchange potentials.\footnote{The EoS we use has employed the Bonn A potential defined in Ref.~\cite{Gross-Boelting:1998xsk}.} It is relatively stiff, with a typical NS radius of $\sim 13.5 \mathrm{km}$}.  This allows us to assess the accuracy with which we can fit realistic nuclear models.   We then modify the DBHF EoS using a constant-speed-of-sound parametrization~\cite{Alford:2015dpa} to construct a model with a strong phase transition, DBHF\_2507.  With this we assess the ability of the \neweos\ parametrization to augment realistic low-density models with phenomenological extensions inspired by nuclear theory.  

Using the three models presented above, we study the evolution of isolated NSs by numerical simulation. As in Ref.~\cite{Deppe:2021bhi}, we work with \spectre\, within the Cowling approximation and examine NS modes that are sourced by density perturbations due to numerical noise. We neglect spacetime dynamics and magnetic fields, which will likely be most relevant in crust physics where magnetic and matter pressure are comparable.
We run each simulation for 40,000  CFL-limited time steps~\cite{CFL}.
For the DG-FD hybrid solver of \spectre\, we use a sixth-order ($\text{P}_5$) discontinuous Galerkin scheme where each element uses $6^3$ Gauss-Lobatto points on the mesh. If an element switches its mesh from discontinuous Galerkin to finite difference, we use $11^3$ uniformly spaced grid points for finite difference cells.
The finite difference solver needs to compute the solution (in our case $\rho$, $p$, and $Wv^i$, where $W$ is the Lorentz factor and $v^i$ the spatial velocity) at cell interfaces (halfway between grid points). We compute these using two different reconstruction schemes: the widely employed monotonized central~\cite{VANLEER1977276} and a positivity-preserving adaptive order scheme which was recently implemented in \spectre~\cite{PPAO-inprep}. In the $n$-th order adaptive scheme, we first try reconstructing the finite-difference interface values with a degree $n-1$ polynomial without any limiting procedure. If the reconstructed values are (i) not positive or (ii) trigger a certain oscillation-detecting criterion, we repeat the reconstruction with progressively lower-order methods. In this work we use the fifth-order adaptive scheme which first tries reconstruction with a quartic polynomial and switches to monotonized central if the reconstructed values fail to satisfy the conditions described above. Finally, if the monotonized central reconstruction did not produce positive values at the interface, first-order reconstruction is used. 

}

\begin{table*}
 \centering
 \caption{
    Analysis settings for the \spectre\, simulations with fits to the SLy1.35, DBHF, DBHF\_2507 EoSs, and a polytropic EoS. \emph{Labels} are chosen to serve as unique identifiers for the runs. The integer suffix represents the approximate grid spacing of the run in meters. \emph{Elts} describes the number of computational elements used in the (three-dimensional) domain. \emph{FD $\Delta x$} represents the finite-difference grid spacing, in meters, of the finite-difference cells in each element when using finite-difference instead of discontinuous-Galerkin methods; this is the primary measure of resolution of the run and allows for easy comparison to other codes.  $\rho_c$ represents the initial central (rest-mass) density of the NS being simulated.  \red{\emph{Cost}, in cpu-minutes per CFL-limited timestep, is the approximate cost of computing one time step in this simulation.  While runtime depends on an array of factors and may not always be indicative of EoS evaluation speed, differences of $\gtrsim 20\%$ represent on otherwise identical runs likely represent EoS-induced slowdown}.  \emph{Figs} represents which figures contain plots pertaining to this run.  \emph{Info} represents the section in which more information about the EoS can be found.  \red{\emph{Radius} represents the TOV radius of the NS being simulated. \emph{Elts/D} represents the approximate number of computational elements across the diameter of the star. }
 }
 {\renewcommand{\arraystretch}{1.2}
 \begin{tabular}{@{\extracolsep{0.25cm}} c cccccccc}
    \hline
    \hline

Label & Elts & FD $\Delta x$ (m)\footnote{We express resolution in finite difference grid spacing for easy comparison to finite difference codes.} & $\rho_{c} (1/M_{\odot}^2)$ & Cost(cpum/st) &  Figs & Info & Radius (km) & Elts/D\\
\hline 

\spectralslylevthreelabel & \spectralslylevthreeelements & \spectralslylevthreefdgridspacing & \spectralslylevthreerhoc & \spectralslylevthreecost &  \spectralslylevthreefigs & \spectralslylevthreeinfo &\spectralslylevthreeradius
&\spectralslylevthreecperd\\
\enthalpyslylevthreelabel & \enthalpyslylevthreeelements & \enthalpyslylevthreefdgridspacing & \enthalpyslylevthreerhoc & \enthalpyslylevthreecost &  \enthalpyslylevthreefigs & \enthalpyslylevthreeinfo & \enthalpyslylevthreeradius&
\enthalpyslylevthreecperd\\

\spectraldbhflevthreelabel & \spectraldbhflevthreeelements & \spectraldbhflevthreefdgridspacing & \spectraldbhflevthreerhoc & \spectraldbhflevthreecost &  \spectraldbhflevthreefigs & \spectraldbhflevthreeinfo & \spectraldbhflevthreeradius& \spectraldbhflevthreecperd\\
\enthalpydbhflevthreelabel & \enthalpydbhflevthreeelements & \enthalpydbhflevthreefdgridspacing & \enthalpydbhflevthreerhoc & \enthalpydbhflevthreecost &  \enthalpydbhflevthreefigs & \enthalpydbhflevthreeinfo & \enthalpydbhflevthreeradius&
\enthalpydbhflevthreecperd\\
\enthalpyptlevthreelabel & \enthalpyptlevthreeelements & \enthalpyptlevthreefdgridspacing & \enthalpyptlevthreerhoc & \enthalpyptlevthreecost &  \enthalpyptlevthreefigs & \enthalpyptlevthreeinfo & \enthalpyptlevthreeradius&
\enthalpyptlevthreecperd\\
\enthalpyptadordlevfourlabel & \enthalpyptadordlevfourelements & \enthalpyptadordlevfourfdgridspacing & \enthalpyptadordlevfourrhoc & \enthalpyptadordlevfourcost &  \enthalpyptadordlevfourfigs & \enthalpyptadordlevfourinfo & \enthalpyptadordlevfourradius&
\enthalpyptadordlevfourcperd\\
\enthalpyptmonotlevfourlabel & \enthalpyptmonotlevfourelements & \enthalpyptmonotlevfourfdgridspacing & \enthalpyptmonotlevfourrhoc & \enthalpyptmonotlevfourcost & \enthalpyptmonotlevfourfigs & \enthalpyptmonotlevfourinfo & \enthalpyptmonotlevfourradius&
\enthalpyptmonotlevfourcperd\\
\enthalpypolytropelevthreelabel & \enthalpypolytropelevthreeelements & \enthalpypolytropelevthreefdgridspacing & \enthalpypolytropelevthreerhoc & \enthalpypolytropelevthreecost &  \enthalpypolytropelevthreefigs & \enthalpypolytropelevthreeinfo & \enthalpypolytropelevthreeradius&
\enthalpypolytropelevthreecperd\\
\polytropicpolytropelevthreelabel & \polytropicpolytropelevthreeelements & \polytropicpolytropelevthreefdgridspacing & \polytropicpolytropelevthreerhoc & \polytropicpolytropelevthreecost &  \polytropicpolytropelevthreefigs & \polytropicpolytropelevthreeinfo & \polytropicpolytropelevthreeradius& \polytropicpolytropelevthreecperd\\
\enthalpysmoothptlevthreelabel & \enthalpysmoothptlevthreeelements & \enthalpysmoothptlevthreefdgridspacing & \enthalpysmoothptlevthreerhoc & \enthalpysmoothptlevthreecost& \enthalpysmoothptlevthreefigs & \enthalpysmoothptlevthreeinfo & \enthalpysmoothptlevthreeradius& \enthalpysmoothptlevthreecperd\\
\hline
 \end{tabular}
 
 }
 \label{tab:runs-performed}
\end{table*}

\subsection{SLy1.35}
\label{sec:sly-test}

\subsubsection{SLy1.35: fit results}
\label{sec:sly-static}

\begin{figure}
    \centering
\includegraphics[width=.49\textwidth]{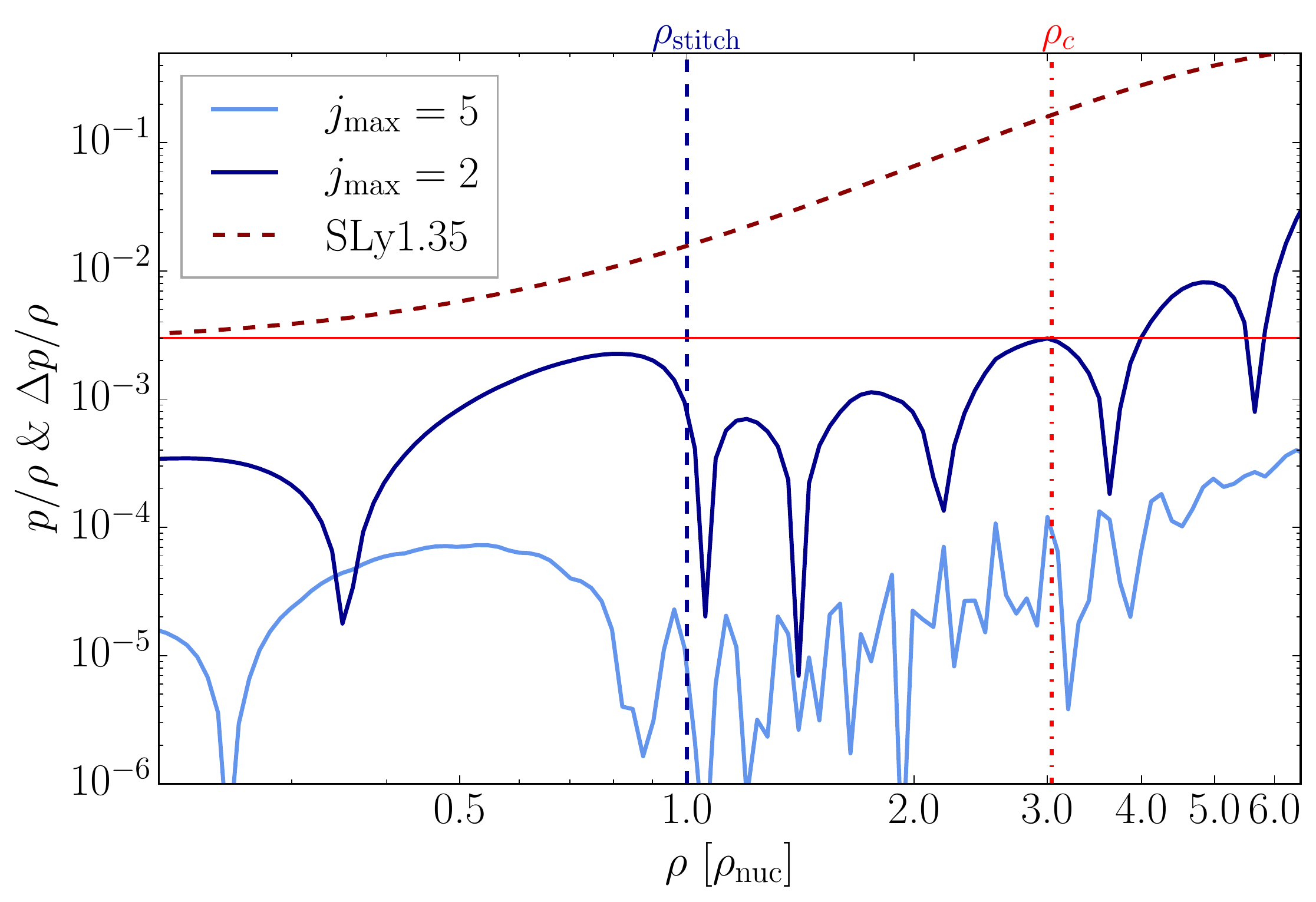}
    \caption{Fitting SLy1.35 (a spectral model of SLy) with the \neweos parametrization, expressed through the difference in pressure divided by the density.  The SLy1.35 EoS value for $p(\rho)/\rho$ is marked by a maroon dashed line for comparison to the residuals.  The $j_{\max} = 5$ and $j_{\max}=2$  fit residuals are marked in light blue and indigo.  
    The vertical blue dashed line marks  the stitching density between the enthalpy and the spectral parametrizations, while the vertical red dot-dashed line marks the central density of the NS we simulate in Sec.~\ref{sec:sly-static}. The solid red horizontal line marks an error level of $3\e{-3}$ for comparison with fits in Sec.~\ref{sec:dbhf-static}; errors below $3\e{-3}$ at $(1,3)\rhonuc$ serve as a heuristic for a good fit.}
    \label{fig:sly1p35comparison-micro}
\end{figure}

We fit SLy1.35 with the enthalpy parametrization and show the error in pressure divided by density as a function of density in Fig.~\ref{fig:sly1p35comparison-micro}.   We vary the number of trigonometric terms in Eq.~\eqref{eq:enthalpyFourier} and show results with $j_{\max}=2$ and $j_{\max}=5$. The $j_{\max}=5$ fit shows exceptional agreement; the error measure, $\Delta p/\rho$, is  near or below $ 1\e{-4}$ over essentially the entire domain.  The $j_{\max}=5$ fit shows increased error, though $\Delta (p/\rho)$ remains near or below $3\e{-3}$ above $\rhonuc$.  We stitch to a spectral parametrization below $\rhonuc$, marked in the Fig.~\ref{fig:sly1p35comparison-micro} as a vertical dashed blue line.  Even though the low-density behavior of the EoS is a spectral EoS fitting a spectral EoS, it is not guaranteed the low-density fit is good, because we prioritize smooth stitching to the \neweos solution above accurate low-density EoS modeling, see App.~\ref{sec:fitting}.  In line with this, we see a significantly better low-density fit for $j_{\max}=5$.

\begin{figure}
    \centering
   \includegraphics[width=.49\textwidth]{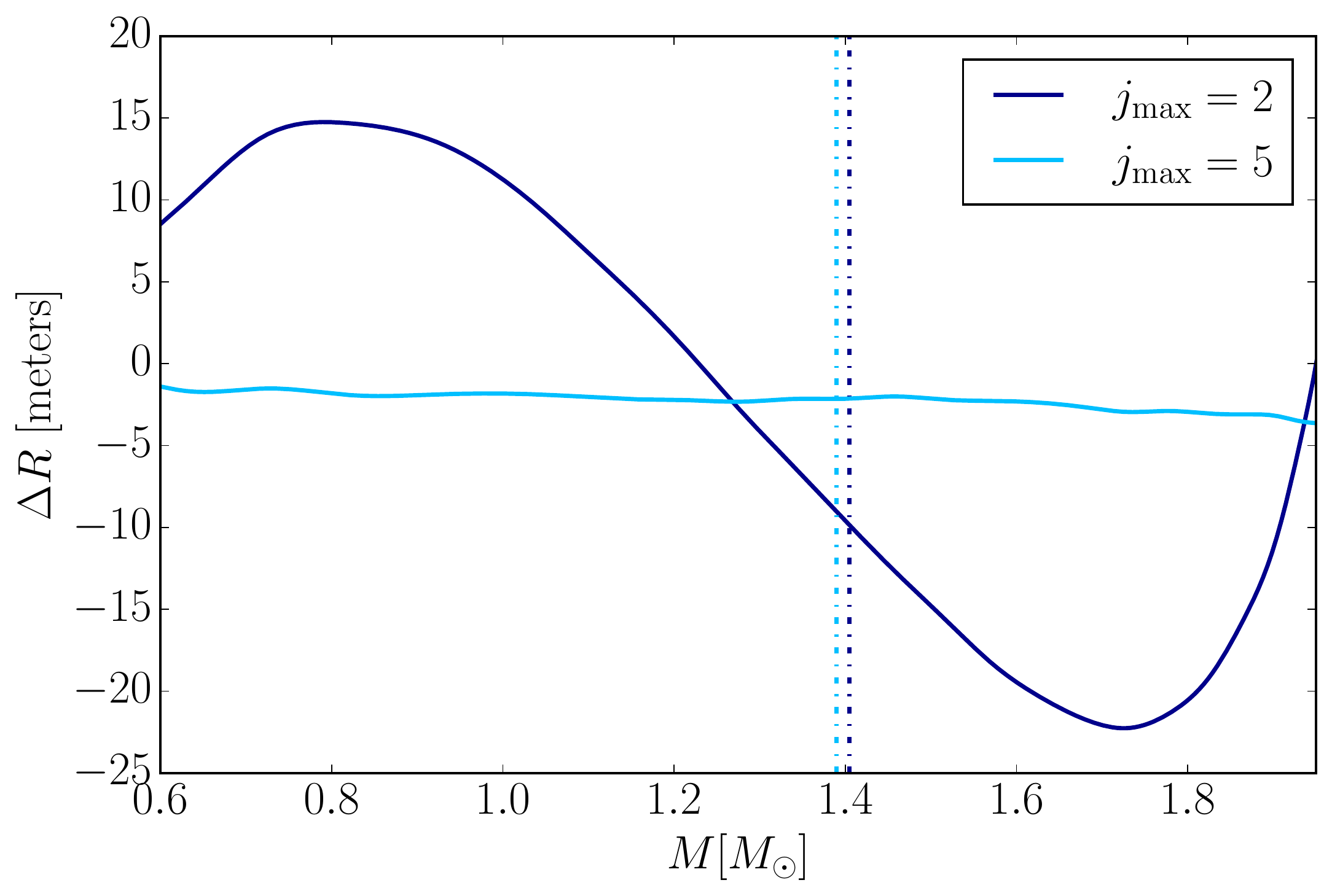}
    \caption{Radius error $\Delta R$ as a function of mass for the SLy1.35 EoS and the $j_{\max}=10$ and $j_{\max}=5$ \neweos parametrization fits.  We mark the mass of the stars with central density $\rho_c\sim 3.04\rhonuc$, (simulated in Sec.~\ref{sec:sly-grmhd}), with dashed-dot lines. Consistently with the microscopic comparison of Fig.~\ref{fig:sly1p35comparison-micro}, the enthalpy fit can reproduce macroscopic quantities with excellent agreement.  The error decreases with more trigonometric terms, but always remains small compared to  $200$\,m grid resolution.}
    \label{fig:sly1p35comparison-macro}
\end{figure}
%

\subsubsection{SLy1.35: Relativistic simulations}
\label{sec:sly-grmhd}

\begin{figure}
    \centering
     \includegraphics[width=.49\textwidth]{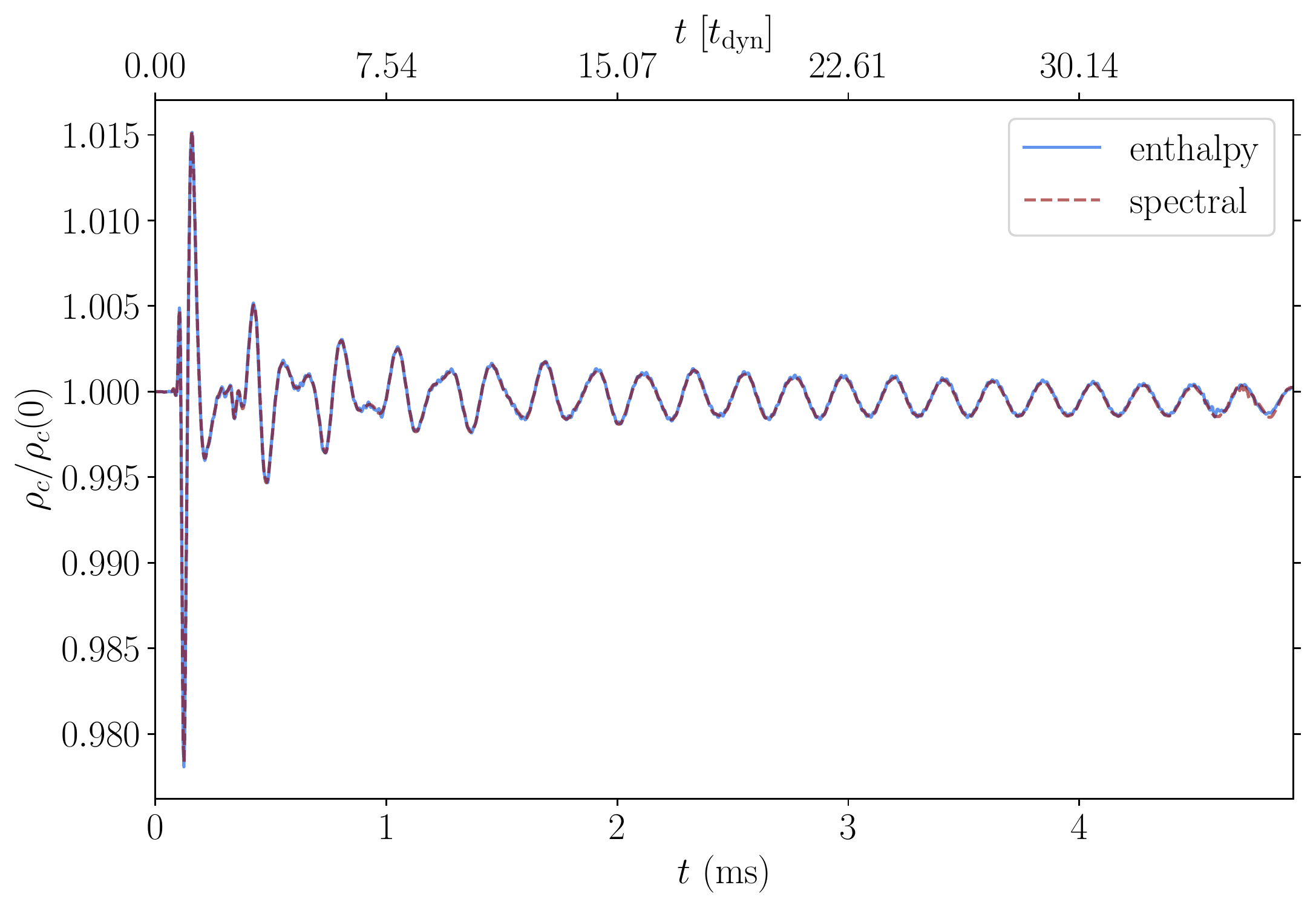}
    \includegraphics[width=.49\textwidth]{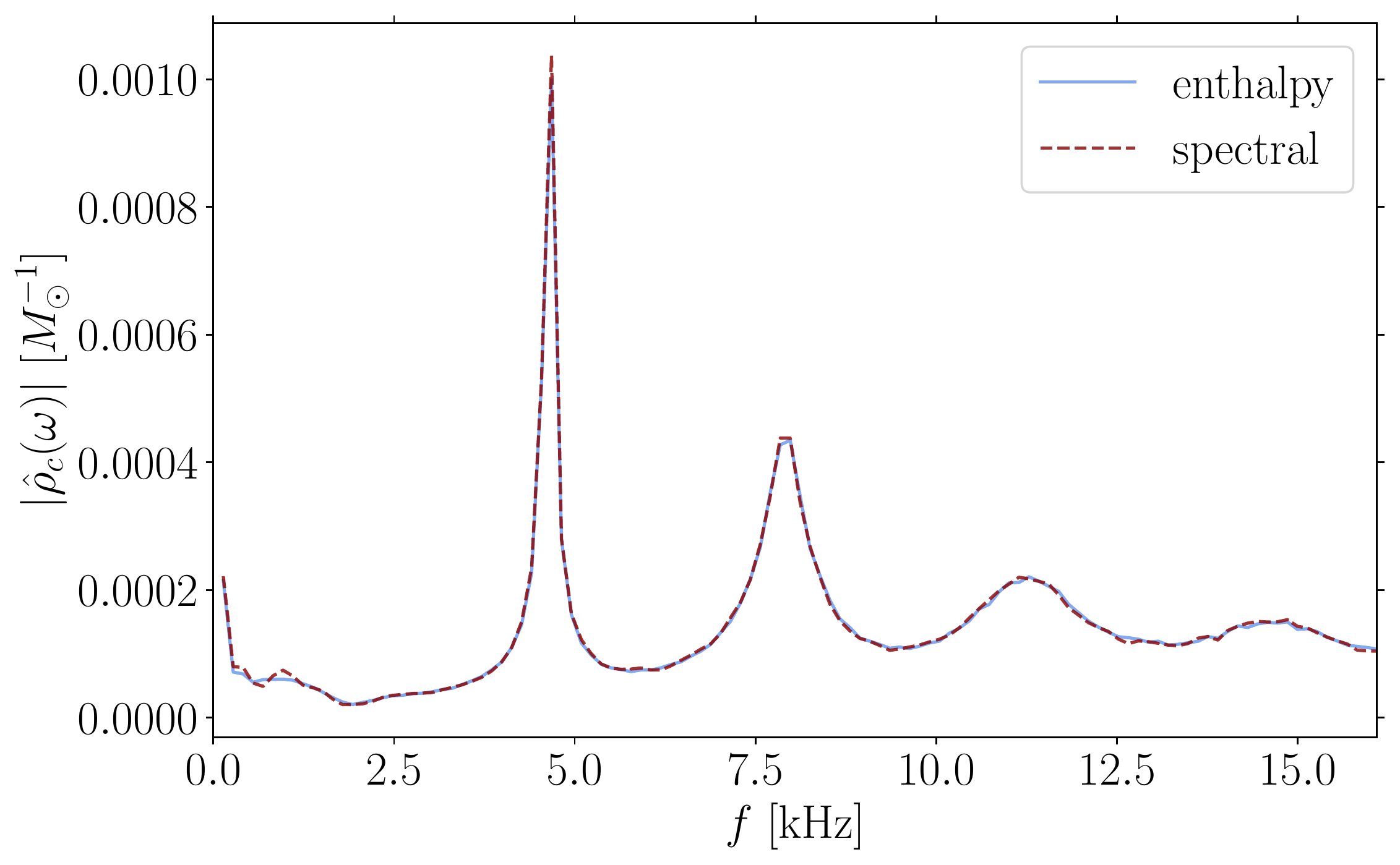}
    \caption{NS central density as a function of time (top panel) and its spectrum (bottom panel) for \spectre\, simulations with SLy1.35 (red dashed) and its $j_{\max}=10$ enthalpy fit (blue solid).  These runs are labeled \spectralslylevthreelabel\ and \enthalpyslylevthreelabel\ in Table~\ref{tab:runs-performed}.  
    In both plots the curves are nearly indistinguishable.  We plot times in both milliseconds (ms), and dynamical times ($t_{\mathrm{dyn}}\equiv 1/\sqrt{\rhoc}$)}
        \label{fig:sly-compare-oscillation}
\end{figure}

We carry out simulations directly with SLy1.35 using the defining spectral expansion~\cite{Foucart:2019yzo} as well as the $j_{\max}=5$ enthalpy parametrization fit; details are given in Table~\ref{tab:runs-performed}.  We evolve a NS with an initial central density of $\sim3.04 \rhonuc$ which has a Tolman-Oppenheimer-Volkoff (TOV)~\cite{Oppenheimer:1939ne} mass of about $1.4 M_{\odot}$ and a radius of about $11.5~\mathrm{km}$, see Figs.~\ref{fig:sly1p35comparison-micro} and~\ref{fig:sly1p35comparison-macro}.  The simulation resolution corresponds approximately to a $220$\,m finite difference grid spacing.
We plot the central density as a function of time and its spectrum
\begin{equation}
    \hat \rhoc (\omega) = \int_0^T \rhoc(t) e^{-i\omega t}dt\,,
\end{equation}
in  Fig.~\ref{fig:sly-compare-oscillation} and find essentially identical evolution between the spectral and enthalpy fits, in line with expectations from the static tests of Sec.~\ref{sec:sly-static}.  This demonstrates that the \neweos parametrization is able to faithfully reproduce results from lower-dimensional parametrizations.

\begin{table}
    \begin{tabular}{|l||*{2}{c|}}\hline
    \backslashbox{Parametrization}{Evaluation Cost (ns)}
    &\makebox[3em]{$p(\rho)$}&\makebox[3em]{$\eps(\rho)$} \\\hline\hline
    enthalpy, $j_{\max}=5$& \benchenthalpyslymorepressurefromdensity & \benchenthalpyslymoreinternalenergyfromdensity\\\hline
    enthalpy, $j_{\max}=2$& \benchenthalpyslypressurefromdensity&\benchenthalpyslyinternalenergyfromdensity\\\hline
    spectral &\benchspectralpressurefromdensity&\benchspectralinternalenergyfromdensity\\\hline
    \end{tabular}
    \caption{Evaluation cost in nanoseconds for the spectral and two different enthalpy fits to SLy1.35 for the pressure and internal energy evaluated at $\rho= 5\e{-4} M_{\odot}^2$. The spectral parametrization has a shorter (longer) pressure (internal energy) evaluation time. The enthalpy evaluation cost further increases with the number of trigonometric terms employed.}
    \label{tab:sly-profile}
\end{table}

 With regards to computational cost, the enthalpy parametrization results in an overall 15\% increase in total runtime compared to the spectral parametrization on similar hardware.  To isolate the EoS evaluation cost, we benchmark the $p(\rho)$ and $\epsilon(\rho)$ evaluation in Table~\ref{tab:sly-profile}.  The two parametrizations have comparable evaluation times though exact numbers are sensitive to the number of trigonometric terms in the enthalpy case. 
While $p(\rho)$ evaluation is in general faster with the spectral parametrization, the opposite is true for $\eps(\rho)$. This is because the spectral parametrization needs to perform a quadrature to calculate $\eps(\rho)$, see Sec.~\ref{sec:flexible-eos}.  In the \neweos parametrization, trigonometric terms cause slowdowns, although even with $2\times j_{\max}=10$ terms the $p(\rho)$ cost does not exceed a factor of $4$. Further studies with single-polytropic nuclear EoSs suggest that this disadvantage effectively disappears if $j_{\max}=0$, see App.~\ref{sec:polytrope appendix}.

\subsection{DBHF}
\label{sec:dbhf-test}

\subsubsection{DBHF: Fitting a tabulated nuclear model}
\label{sec:dbhf-static}

\begin{figure}
    \centering
    \includegraphics[width=.49\textwidth]{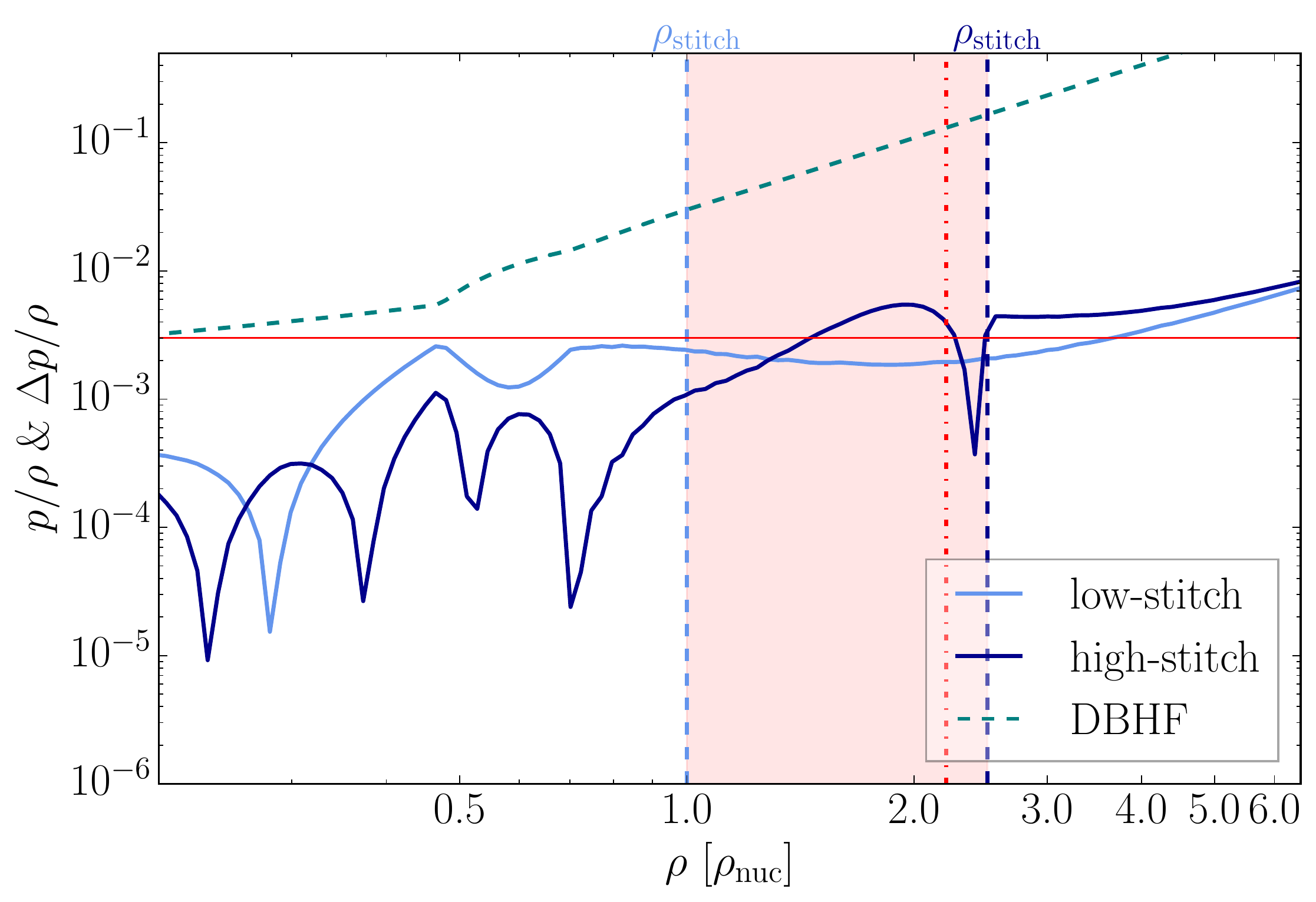}

    \caption{Same as Fig.~\ref{fig:sly1p35comparison-micro} but for the DBHF nuclear EoS model and two enthalpy fits that are stitched to a spectral parametrization at $\rhonuc$ (low-stitch, light blue) and $2.5\rhonuc$ (high-stitch, indigo).  We also plot the tabulated DBHF model $p(\rho)/\rho$ in dashed-teal for reference. The vertical dashed lines denote the stitching densities.  We also mark the value $\Delta p/ \rho = 3\e{-3}$ as a solid red horizontal line for reference.  We mark the central density of the star we simulate in Sec.~\ref{sec:dbhf-grmhd} with a vertical red, dot-dash line.}
    \label{fig:low_stitch_dbhf}
\end{figure}

\begin{figure}
    \centering
    \includegraphics[width=.49\textwidth]{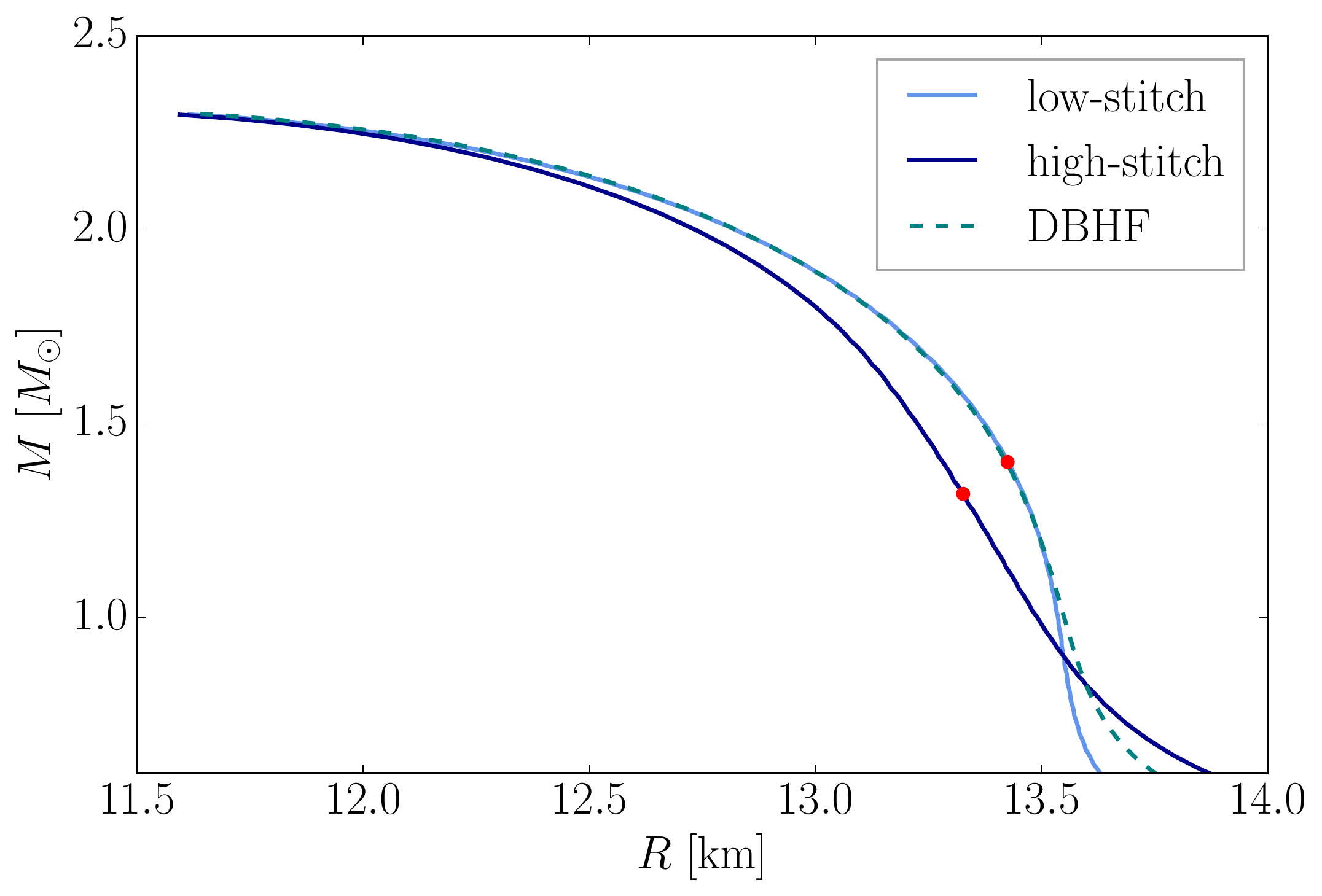}
    \caption{The NS mass-radius relation but for the DBHF nuclear model and two enthalpy fits that are stitched to a spectral parametrization at $\rhonuc$ (low-stitch, light blue) and $2.5\rhonuc$ (high-stitch, indigo). We find visibly improved fits to the $M$--$R$ relation when the \neweos parametrization extends down to lower densities.  Red dots mark the NSs we evolve in Sec.~\ref{sec:dbhf-grmhd}.}
    \label{fig:dbhf-macro}
\end{figure}

We fit the tabulated DBHF EoS with the \neweos parametrization and further explore the effect of low-density stitching to the spectral parametrization by probing two different stitch densities: $\rhonuc $ and $2.5\rhonuc$; these fits are referred to as ``low-stitch'' and ``high-stitch'' in what follows.
In the high-stitch case we use $j_{\max} = 10$ trigonometric terms, while we find that $j_{\max} = 5$ is enough for the low-stitch one. See App.~\ref{sec:fitting} for more details.  We examine the microscopic and macroscopic performance of both fits in Figs.~\ref{fig:low_stitch_dbhf} and~\ref{fig:dbhf-macro}. On the microscopic side, the low-stitch fit achieves higher accuracy above $ 1.1\rhonuc$, but worse accuracy below.

The macroscopic side presents a clearer picture.
When we use the enthalpy parametrization to describe the EoS down to a density of $\rhonuc$, we obtain excellent agreement with the tabulated EoS, with radius differences ${\cal{O}}(1)$\,m for astrophysically relevant NS masses. However, when we stitch to the spectral parametrization at $2.5\rhonuc$ the radius error increases to $\mathcal O(100)$\,m at $1.4M_{\odot}$.
The improved agreement between the $2.5 \rhonuc$ and the $\rhonuc$ stitching fits can be attributed to the high accuracy of the \neweos parametrization in the range $\rhonuc$ to $2.5\rhonuc$, as seen in Fig.~\ref{fig:low_stitch_dbhf}. 
The difference in the two errors is particularly pronounced near $2\rhonuc$, consistent with the observed strong correlation between the pressure at twice saturation density and radius of a $1.4 M_{\odot}$ star~\cite{Lattimer:2000nx}.  
This further establishes the importance of the enthalpy parametrization as a flexible EoS parametrization at nuclear saturation and above, in this case it appears errors in $p/ \rho$ must be at most $3\e{-3}$ to achieve high-precision reproduction of astrophysical observables. However, this is not an indication that the spectral parametrization \emph{cannot} fit DBHF well, it just cannot fit DBHF well while maintaining $\mathcal C^1$ pressure smoothness at the high-density transition to the \neweos parametrization and the low density transition to the crust; see App.~\ref{sec:fitting}. 
%

\subsubsection{DBHF: Relativistic simulations}
\label{sec:dbhf-grmhd}

We next turn  to \spectre\, simulations using the DBHF fits \red{from the previous section}.
Since  the low- and high-stitch fits predict $\mathcal O(100)$\,m differences for $R_{1.4}$, we target a resolution at that level in order to resolve their effect.  We select a NS central density of $2.21 \rhonuc$ that lies between the two stitching densities of $\rhonuc$ and $2.5\rhonuc$, see Figs.~\ref{fig:low_stitch_dbhf} and~\ref{fig:dbhf-macro}. Run details and settings are given in Table~\ref{tab:runs-performed}.
As a consequence, the NS resulting from the high-stitch EoS is fully described by the spectral part of the EoS. In the low-stitch EoS, the NS is described with the enthalpy parametrization out to $r/R \approx 7/8$, about two-thirds of the coordinate volume of the star.  

While the high-stitch EoS does represent a spectral fit to the DBHF EoS, the spectral parameters are selected by the requirement that the spectral parametrization reproduces the correct low-density behavior of DBHF and is smoothly stitched to the enthalpy parametrization.  It is important to note that a better spectral fit to any particular astrophysical quantity, such as $R_{1.4}$ may be possible, but the fit accuracy is typically lower than in the enthalpy parametrization case.  Even if the few degrees of freedom in the spectral model are fit to minimize errors in astrophysical observables, fixing the low-density behavior of the EoS often results in $R(M)$ deviations of 50\,m or more~\cite{Foucart:2019yzo}. In what follows, we leverage the mismatch of Fig.~\ref{fig:dbhf-macro} to examine how well we can resolve EoSs with $\sim 100$\,m radius differences in simulations with similar resolution. Since the high-stitch NS is fully described by the spectral parametrization,  this test also serves as a comparison of runtimes between the spectral and \neweos parametrizations.    We do not utilize the tabulated version of DBHF because \spectre\ currently cannot perform GRMHD simulations using tables.

\begin{figure}
    \centering
    \includegraphics[width=.49\textwidth]{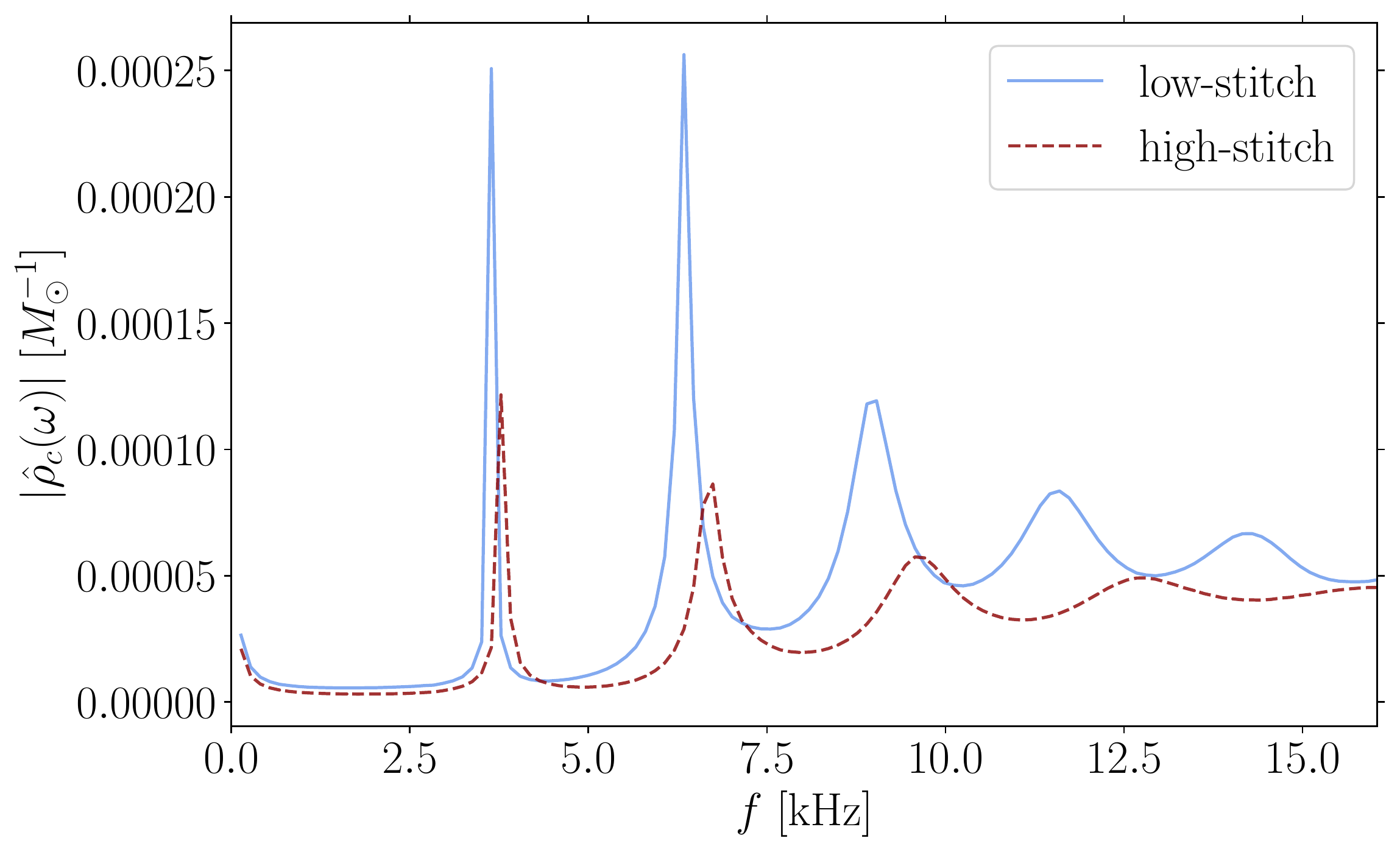}
    \caption{NS central density spectrum for \spectre\, simulations with enthalpy fits to the DBHF nuclear EoS that are stitched to the spectral parametrization at $\rhonuc$ (low-stitch, blue solid) and $2.5\rhonuc$ (high-stitch, red dashed). These runs are labeled \enthalpydbhflevthreelabel\ and \spectraldbhflevthreelabel\ respectively in Table~\ref{tab:runs-performed}. The simulated star has a central density of $ \approx 2.2 \rhonuc$; it is marked in Figs.~\ref{fig:low_stitch_dbhf} and~\ref{fig:dbhf-macro}. In the red case, the NS is completely described by the spectral EoS as its central density is below $2.5\rhonuc$.}
    \label{fig:dbhf-grmhd-comparison}
\end{figure}

We carry out simulations as detailed in Table~\ref{tab:runs-performed} and plot the spectrum of the central density of each star in Fig.~\ref{fig:dbhf-grmhd-comparison}. 
The two spectra disagree both in the location of the NS modes and their strength, a consequence of the EoS mismodeling shown in Figs.~\ref{fig:low_stitch_dbhf} and~\ref{fig:dbhf-macro}. In particular, the fundamental radial modes disagree by nearly $3.5\%$, a difference of $\sim 130\,\mathrm{Hz}$ in this case.  
Table~\ref{tab:runs-performed} further shows the simulation runtime which is comparable in the $130$\,m resolution case; each run took about a day on $\sim 70$ processing elements.  Based on the benchmarking results of Sec.~\ref{sec:sly-static}, total EoS evaluation time should be comparable for the two runs, as the spectral parametrization evalautes the pressure about twice as quickly as the enthalpy parametrization, but evaluates the internal energy about 3 times slower. Since the number of pressure and internal energy evaluations throughout the entire simulation is not known \emph{a priori}, we cannot preemptively conclude which should run faster, though it is likely the difference would be small. This is reflected in the runtime; differences of $10 \%$ are found, with the enthalpy parametrization running slightly faster.  Nonetheless, this could be due to an array of confounding factors such as task allocation efficiency, and hardware differences. We therefore conclude that the enthalpy parametrization, despite having more flexibility, is not slower than  lower dimensional parametrizations at these resolutions for practical problems.

\subsection{DBHF\_2507: Phase transitions}
\label{sec:dbhf-2507-test}

We now turn our attention to EoS with strong phase transitions and study both smooth and non-smooth (i.e., piecewise) EoSs. We base our studies on DBHF\_2507 which is constructed by combining DBHF with the constant-speed-of-sound phenomenological parametrization for strong phase transitions~\cite{Alford:2015dpa}. We select a transition density of $\rho_t = 2.5\rhonuc$ and latent heat ratio $\Delta e/ e=0.7$~\cite{Han:2018mtj,Chatziioannou:2019yko}.
The pressure remains constant during the phase transition, while above that it has a constant speed of sound with $c_s^2 = 1$. The induced phase transition causes a second stable branch to appear in the $M$--$R$ relation above masses $\sim 1.6 M_{\odot}$.

\red{\subsubsection{DBHF\_2507: piecewise parametrization}}
\label{sec:dbhf-2507-static}

In its original form described above, DBHF\_2507 is piecewise smooth, and it can be represented effectively by a piecewise version of the \neweos parametrization.  Below the phase transition we use either the low- or high-stitch fits from Sec.~\ref{sec:dbhf-test}, and transition to a new \neweos segment after the transition. In the high-stitch case, the hadronic part of the EoS is completely described with the spectral parametrization. During the transition, DBHF\_2507 possesses a formally constant pressure as a function of density; however, constructing TOV solutions using the method of Lindblom~\cite{Lindblom:1992} \textemdash{} the TOV method implemented in \spectre { }\textemdash{} requires $dh/dz = dp/d\rho$ to be strictly positive. Therefore, in the transition region we modify the EoS to exhibit $dh/dz =\smallchange$, where $\smallchange$ is some quantity large enough to guarantee that $h(z)$ is numerically invertible, but still small enough to have a small impact on the TOV solution relative to the target resolution.\footnote{At a central density of $5.07 \rhonuc$, the difference in radius induced by using $\delta =1\e{-4}$ instead of $ \delta =1\e{-3}$ is less than $5$~m.}  After the end of the phase transition, $h(z)$ is given by a constant speed of sound form, see Eq.~\eqref{eq:general-css}; this is similar to the procedure demonstrated in~\cite{Gieg:2019yzq}. 
 
 The advantage of the \neweos parametrization in this problem is that it is able to model constant-speed-of-sound matter (see App.~\ref{sec:polytrope appendix} for the polytropic case) efficiently and with no fine-tuning.  Compare this to polytropic (or spectral) models, which can only model constant-speed-of-sound matter well when 
\begin{equation}
    \Gamma \equiv \frac{\rho}{p}\deriv{p}{\rho} = \frac{\rho}{p} h c_s^2= \frac{p+e}{p} c_s^2
\end{equation}
is slowly varying.  This is typically not true until some density greater than the phase transition, where $p = p_0 + c_s^2 \Delta e \approx c_s^2 \Delta e$, especially if $c_s^2$ is small compared to $1$, such as models where $c_s^2 =1/3$ in the core~\cite{Kurkela:2009gj, Tews:2018kmu}.  In contrast, the enthalpy parametrization can model constant-speed-of-sound matter to arbitrary precision, and benchmarking results demonstrate that in such cases it can even outperform polytropic EoSs by up to 25\%.  
  
\begin{figure}
    \centering
    \includegraphics[width=.49\textwidth]{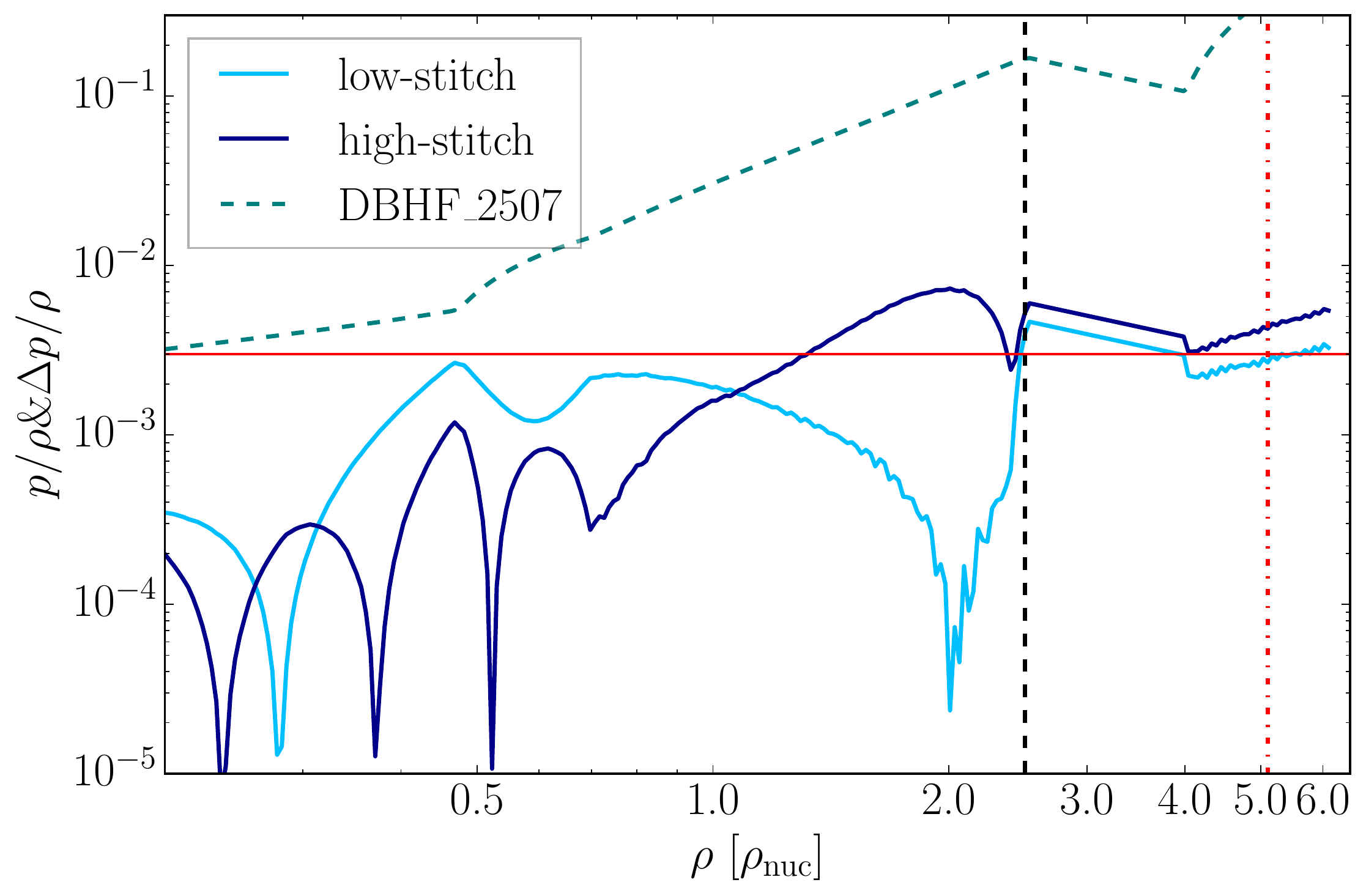}
    \includegraphics[width=.49\textwidth]{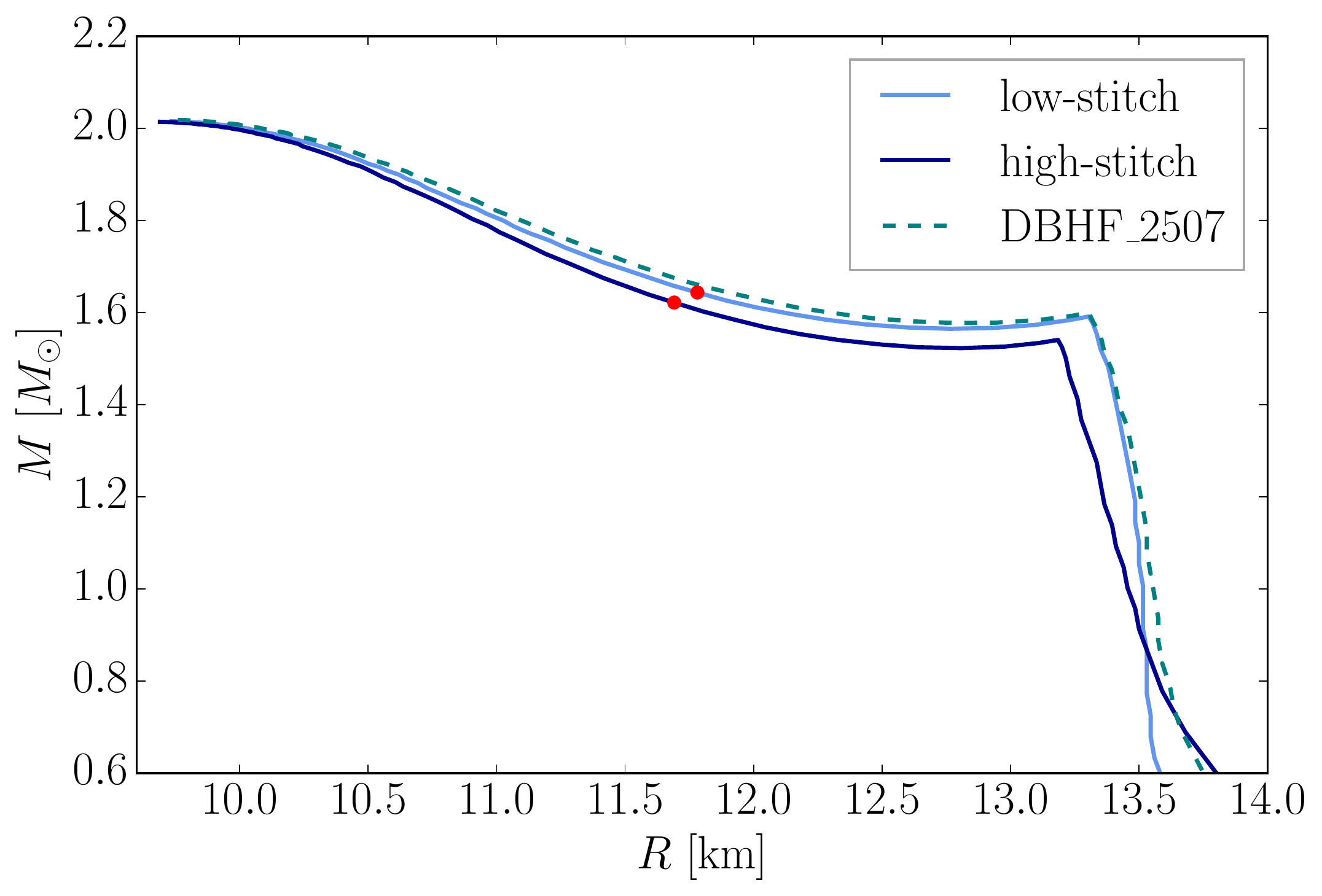}
    
    \caption{Same as Figs.~\ref{fig:low_stitch_dbhf} \& \ref{fig:dbhf-macro} but for the DBHF\_2507 EoS. The procedure by which the low- and high-stitch EoSs of Sec.~\ref{sec:dbhf-static} are extended through the phase transition is described in Sec.~\ref{sec:dbhf-2507-test}. Consistent with Fig.~\ref{fig:dbhf-macro}, the low-stitch case can more accurately reproduce the parameters of the phase transition, though caution must be exercised when comparing to tabulated models as differences in interpolation in this case can be substantial.  See the text of Sec.~\ref{sec:dbhf-2507-grmhd}. In the top panel, the black vertical dashed line marks the onset of the phase transition.  The red dot-dashed line in the top panel  and the red dots in the bottom panel mark the NSs we use in subsequent simulations, analogous to Figs.~\ref{fig:low_stitch_dbhf} \& \ref{fig:dbhf-macro} respectively.}
 \label{fig:dbhf-2507-macro}
\end{figure}

We plot the $M$--$R$ curve in Fig.~\ref{fig:dbhf-2507-macro}, using the low-stitch and high-stitch fits for the hadronic part of the EoS as discussed in Sec.~\ref{sec:dbhf-static}.  The low-stitch EoS shows better agreement with DBHF\_2507, consistent with previous results; see Fig.~\ref{fig:dbhf-macro}.  The transition mass and radius for the low-stitched model are functionally identical to the tabulated values with errors of $\lesssim 0.01 \,M_{\odot}$ and $\lesssim 1$\,m.   In the high-stitch case the errors increase to $0.05 \,M_{\odot}$ and $\sim 100$\,m.  Nonetheless, errors decrease with increasing central density and the maximum mass $\Mmax$ is consistent to $\sim 0.01\,M_{\odot}$ for both fits. This indicates that the \neweos\ parametrization can produce effective EoS fits at high densities even when extending a (relatively) poor low-density fit.\footnote{Such comparisons to tabulated models might be difficult to interpret, as a $1\%$ interpolation inconsistency in $\rho(p)$ can lead to differences of $\mathcal O (100\  \mathrm{m})$ on the second stable branch.  This problem is more pronounced here as the DBHF\_2507  construction requires computing $\rho(e) = 2.5\rhonuc$ via table-based root finding, a procedure that depends on the interpolation strategy and, in turn, affects the transition mass and radius. For hadronic EoSs this issue is suppressed, as differences in interpolation are smoothed over by the integration of the TOV equations. The \neweos parametrization, having an analytic expression for $e(\rho)$, does not face this issue.}

\red{\subsubsection{DBHF\_2507: Relativistic simulations}}
\label{sec:dbhf-2507-grmhd}

We perform \spectre\, simulations with both the high- and low-stitch EoSs and NSs with central density of $\rho_c= 4.67\rhonuc$, above the transition density from nuclear to quark matter; see the red dots in Fig.~\ref{fig:dbhf-2507-macro}. Preliminary results with low spatial resolutions demonstrated that for such $\rho_c > \rho_{\mathrm{t}}$, the NS undergoes strong density oscillations that are quickly damped.  Given that the fundamental mode is long-lived~\cite{Kokkotas:2000up}, this short damping timescale is probably related to numerical dissipation.
Therefore we perform and compare simulations at various grid resolutions to ensure convergence, increasing the number of computational elements while the number of grid points inside each element is fixed.
The main results presented below correspond to a $\sim 70$~m resolution. We also restrict to the low-stitch fit since the differences between the low- and high-stitch fits are likely resolvable for $<130$~m resolution, see Fig.~\ref{fig:dbhf-2507-macro}.

\begin{figure}
    \centering
    \includegraphics[width=.49\textwidth]{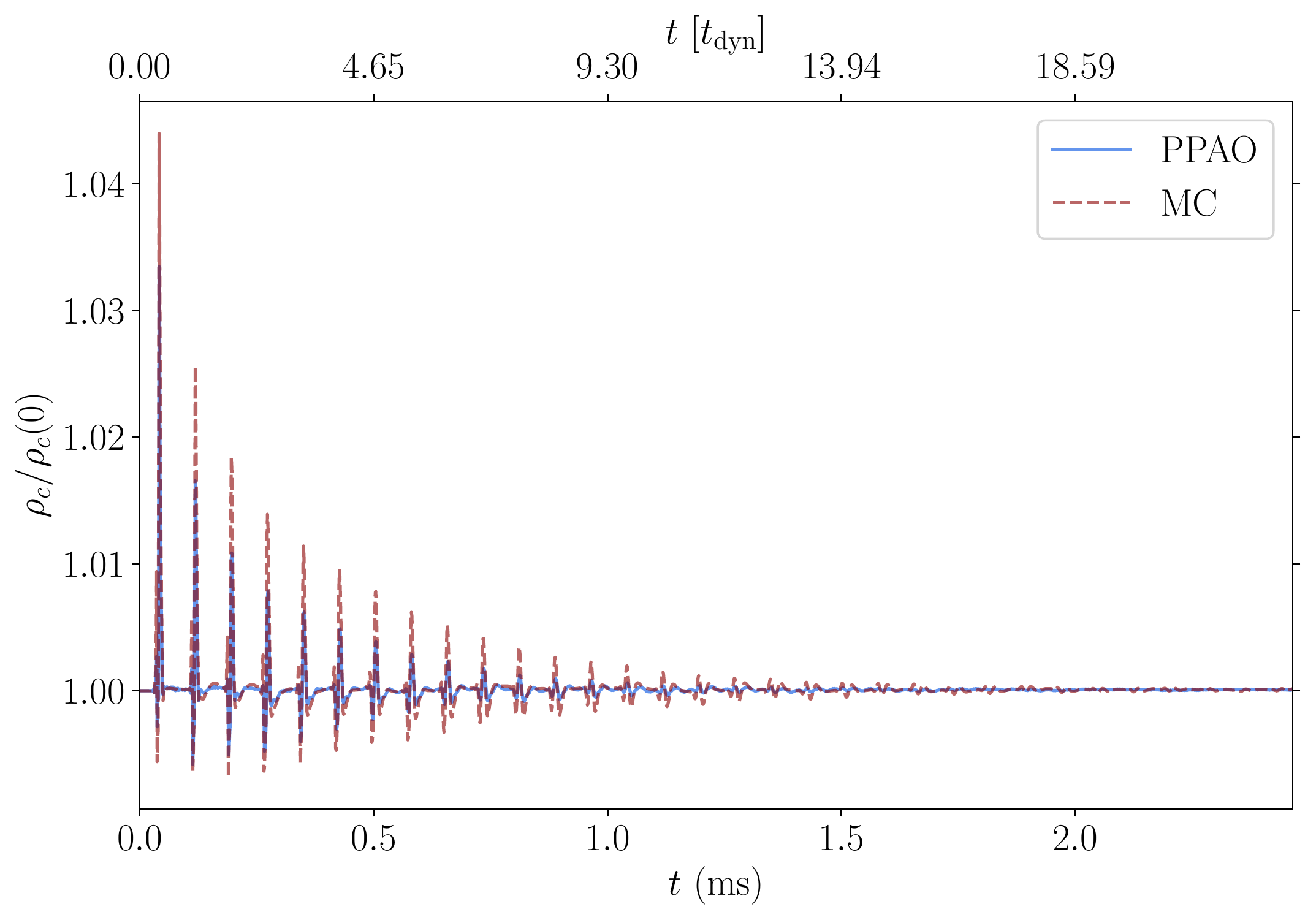}
    \includegraphics[width=.49\textwidth]{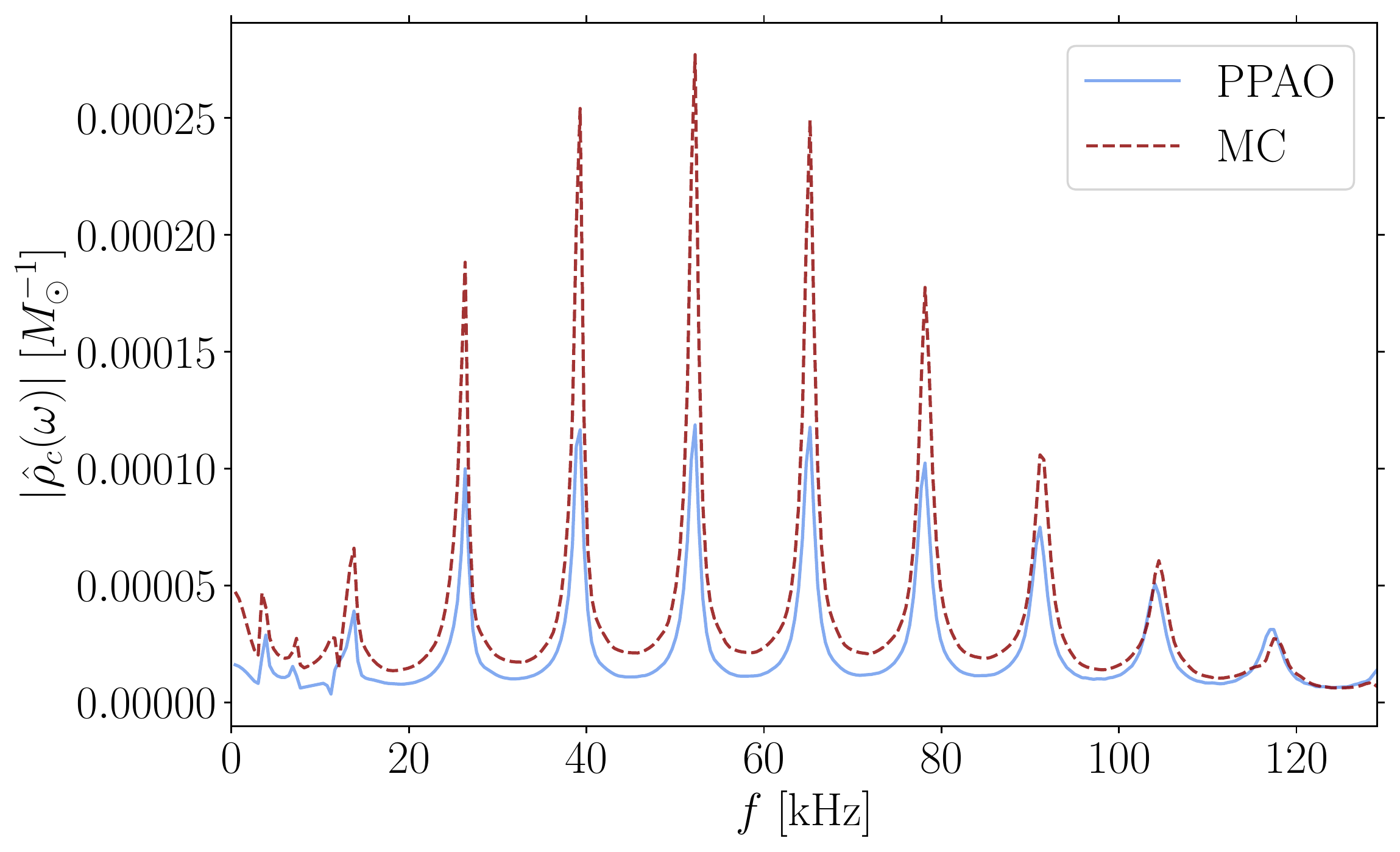}
    \caption{Normalized NS central density as a function of time (top panel) and its spectrum (bottom panel) for \spectre {} simulations with enthalpy fits to the DBHF\_2507 nuclear EoS that are stitched to the spectral parametrization at $\rhonuc$ for two different choices of finite-difference reconstruction schemes. The adapative order reconstructor is marked in blue and the monotonized central reconstructor is marked in red.  Run details are listed as \enthalpyptadordlevfourlabel\ and \enthalpyptmonotlevfourlabel\ respectively in Table~\ref{tab:runs-performed}. The simulated star has a central density of $\sim 4.67\rhonuc$; it is marked in Fig.~\ref{fig:dbhf-2507-macro}.  We find excellent agreement on mode frequencies but slight differences in power distribution.}
    \label{fig:dbhf-2507-sim}
\end{figure}

\begin{figure*}[t]
\includegraphics[width=\textwidth]{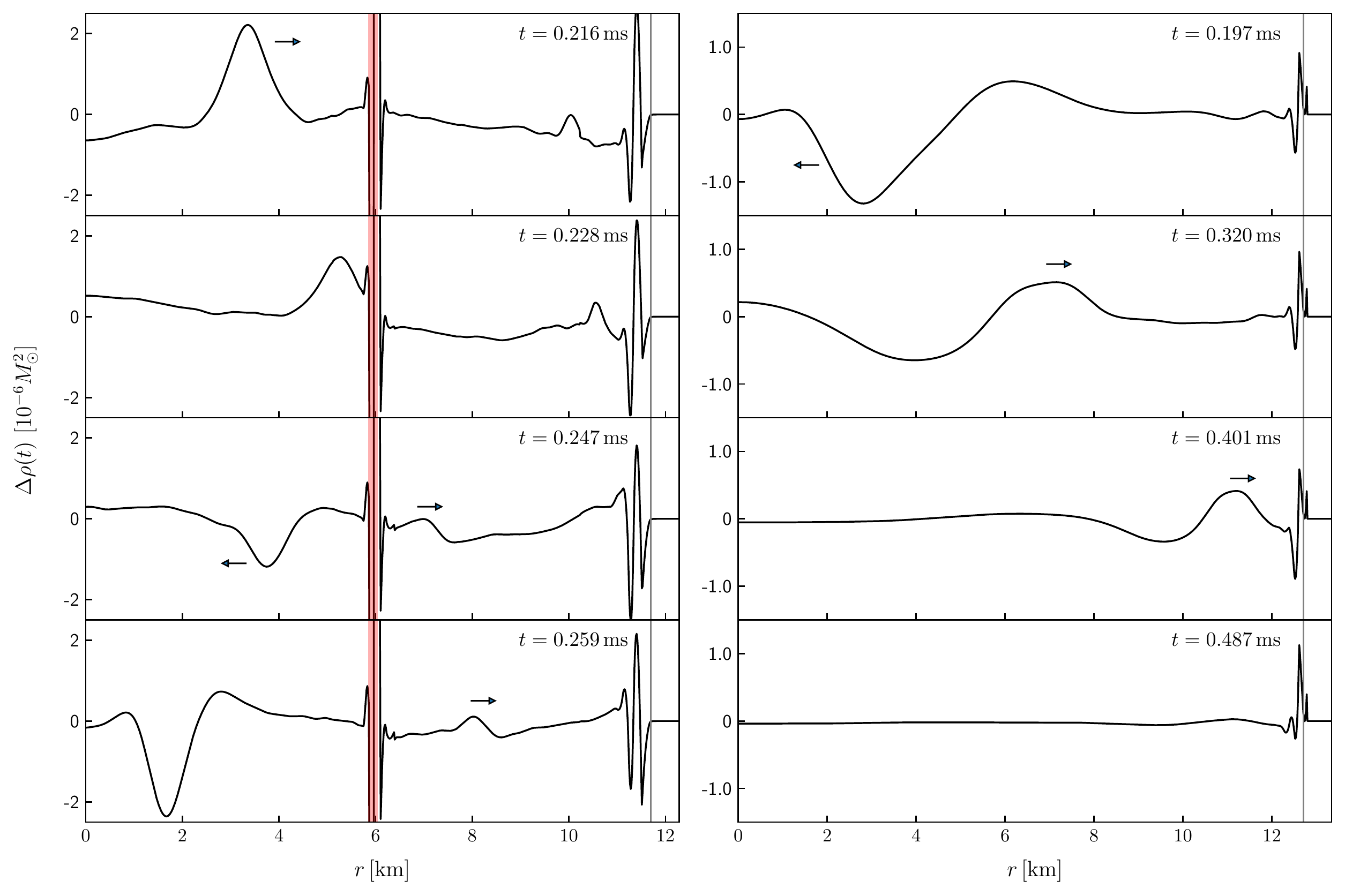}
    \caption{Rest-mass density profile relative to the initial profile $\Delta \rho \equiv \rho(r,t) - \rho(r,0)$ as a function of radius for a hybrid star described with DBHF\_2507 (left panel; details in Sec.~\ref{sec:dbhf-2507-grmhd}) and a simple polytrope with $\Gamma=2$ (right panel; details in App.~\ref{sec:polytrope appendix}) for different times (top to bottom). We denote the NS surface with a vertical solid gray line in each panel and the quark-hadronic boundary with a vertical red line in the left panel. We show snapshots of the density at four different times in order to examine the dynamical behavior of the density oscillations. For the hybrid star (left) density perturbations are partially transmitted and reflected at the quark-hadronic boundary, while for the polytrope (right) the wave smoothly propagates back and forth within the NS interior. Small black arrows highlight the wave packet and its traveling direction. The hybrid star snapshots are from the run \enthalpyptadordlevfourlabel\ and the polytrope snapshots are from a simulation with identical domain, finite-difference reconstruction scheme, and central density, but a polytropic EoS \eqref{eq:polytrope-specific} in place of DBHF\_2507. See \polytropicpolytropelevthreelabel{} for details of a lower-resolution polytropic simulation.}
    \label{fig:shock-snapshots}
\end{figure*}

We plot the central density and the spectrum in the top and bottom panels of Fig.~\ref{fig:dbhf-2507-sim} for both reconstruction schemes. We find good agreement in the frequency and damping time of the density modes, though the monotonized central scheme predicts more than double the power of the adaptive order method below  $\sim 80$~kHz. Interestingly, we find that the presence of a quark core in the NS changes the spectrum qualitatively, c.f., Fig.~\ref{fig:dbhf-grmhd-comparison}.
The spectrum is now dominated by modes in the ${\cal{O}}(10)$\,kHz range, an order of magnitude higher than the hadronic NS case of Fig.~\ref{fig:dbhf-grmhd-comparison}.  The spacing of the modes, about $16$~kHz, is of order $c/2R_{\mathrm{core}}$, where $R_{\mathrm{core}}\sim6$~km is the radius of the quark core. We attribute this to density perturbations that are confined to the quark core and are only weakly coupled to the bulk behavior of the star across the transition.  In order to confirm this, we plot the density profile of the star \red{extracted from the run \enthalpyptadordlevfourlabel}  as a function of radius in Fig.~\ref{fig:shock-snapshots} at different times.  Most of the oscillation power sourced in the quark core is reflected back into the core at the quark-hadronic boundary, with only a small fraction getting transmitted into the hadronic region. The reflected pulse gets inverted (fixed-end reflection); this is consistent with theoretical expectations since the sound speed changes from $c_s^2 = 1$ in the quark core to $c_s^2 \sim 0.3$ in the hadronic region at the boundary.

The initial perturbation needed to drive these modes is provided by numerical noise near the transition, with $\mathcal{O}(50-100 \mathrm{kHz})$ being the scale of the sound crossing frequency of a computational element of our domain.  High frequency modes, in particular $p$-modes being primarily confined to the core of the star is in line with expectations for radial modes~\cite{Sen:2022kva}.

Simulation runtimes are provided in Table~\ref{tab:runs-performed}. 
We find that the adaptive-order simulation (\enthalpyptadordlevfourlabel) has a $\sim20\%$ longer runtime than the monotonized central (\enthalpyptmonotlevfourlabel) simulation, which is expected.
Turning to the EoS and comparing \enthalpyptlevthreelabel\  and \enthalpydbhflevthreelabel\, at identical resolutions, number of CPUs, and reconstruction schemes, we find a $\mathcal O(40\%)$ increase in runtime for DBHF\_2507 as compared to DBHF. 
We do not attribute this runtime slowdown to increased EoS evaluation time, as the enthalpy EoS employed beyond the phase transition uses no trigonometric correction terms, and thus is nearly computationally identical to the polytrope profiled in Table~\ref{tab:polytrope-profile}. This indicates that individual EoS evaluations (above the transition) are actually somewhat cheaper than in either of the fits to DBHF,  discussed in Sec.~\ref{sec:dbhf-test} (below the transition they are identical). Instead, we attribute the slowdown to the non-smoothness of this EoS; $p(\rho)$ is not analytically $\mathcal C^1$ across the phase transition, and $\rho(p)$ is an incredibly sensitive function near the transition.  Since our default primitive recovery scheme requires root-finding to determine $\rho(p)$~\cite{Kastaun:2020uxr}, this can result in significant slowdowns. 

\begin{figure}
    \centering
    \includegraphics[width=.49\textwidth]{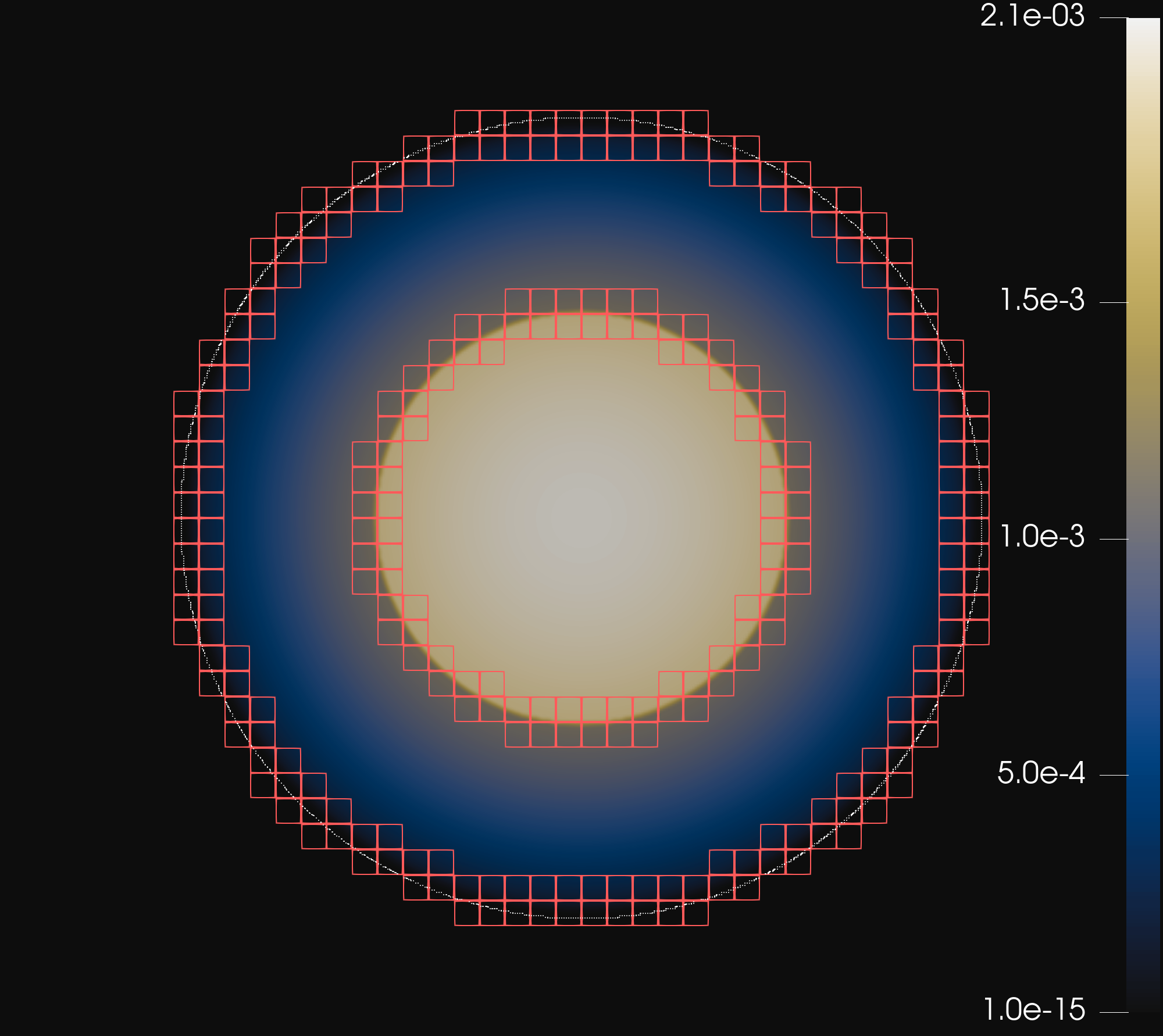}
    \caption{NS rest mass density (colorbar, units of $M_{\odot}^{-2}$) on the $y-z$ plane at $t=100M_\odot$ for the run \enthalpyptadordlevfourlabel. Red marks subdomain elements where finite-difference is used. The finite-difference cells are confined near the NS surface (outer circle, white solid line) and the phase transition layer (inner circle) where discontinuities are expected. The majority of the star is still evolved with the more computationally efficient discontinuous-Galerkin method.}
    \label{fig:paraview-pt}
\end{figure}

Comparing the simulations \enthalpyptlevthreelabel\, and \enthalpyptmonotlevfourlabel, we also find that refining the grid resolution from 134~m to 67~m results in only a 4-fold increase in runtime, compared to the expected $(134/67)^3 = 8$.\footnote{\red{Since we run simulations with the fixed number of time steps,} refining the grid does not lead to any major slowdown from time stepping.  This would not be the case if we ran to a fixed final time.}
We attribute this lower-than-expected increase to the DG-FD hybrid scheme~\cite{Deppe2022method}. At higher resolutions each cell is smaller, therefore finite-difference cells are more tightly concentrated in regions with discontinuities. This
results in a lower fraction of the NS reverting to the slower finite-difference scheme from the faster discontinuous Galerkin scheme.  We display a slice of the NS for \enthalpyptadordlevfourlabel\, in Fig.~\ref{fig:paraview-pt}, and mark the cells which have reverted to finite-difference in red.  The majority of the NS interior is indeed using the discontinuous Galerkin method, with finite difference being used only at the phase-transition and surface interfaces.

\subsubsection{Smooth transitions}
\label{sec:dbhf-smoothed}

\begin{figure}
    \centering
    \includegraphics[width=.49\textwidth]{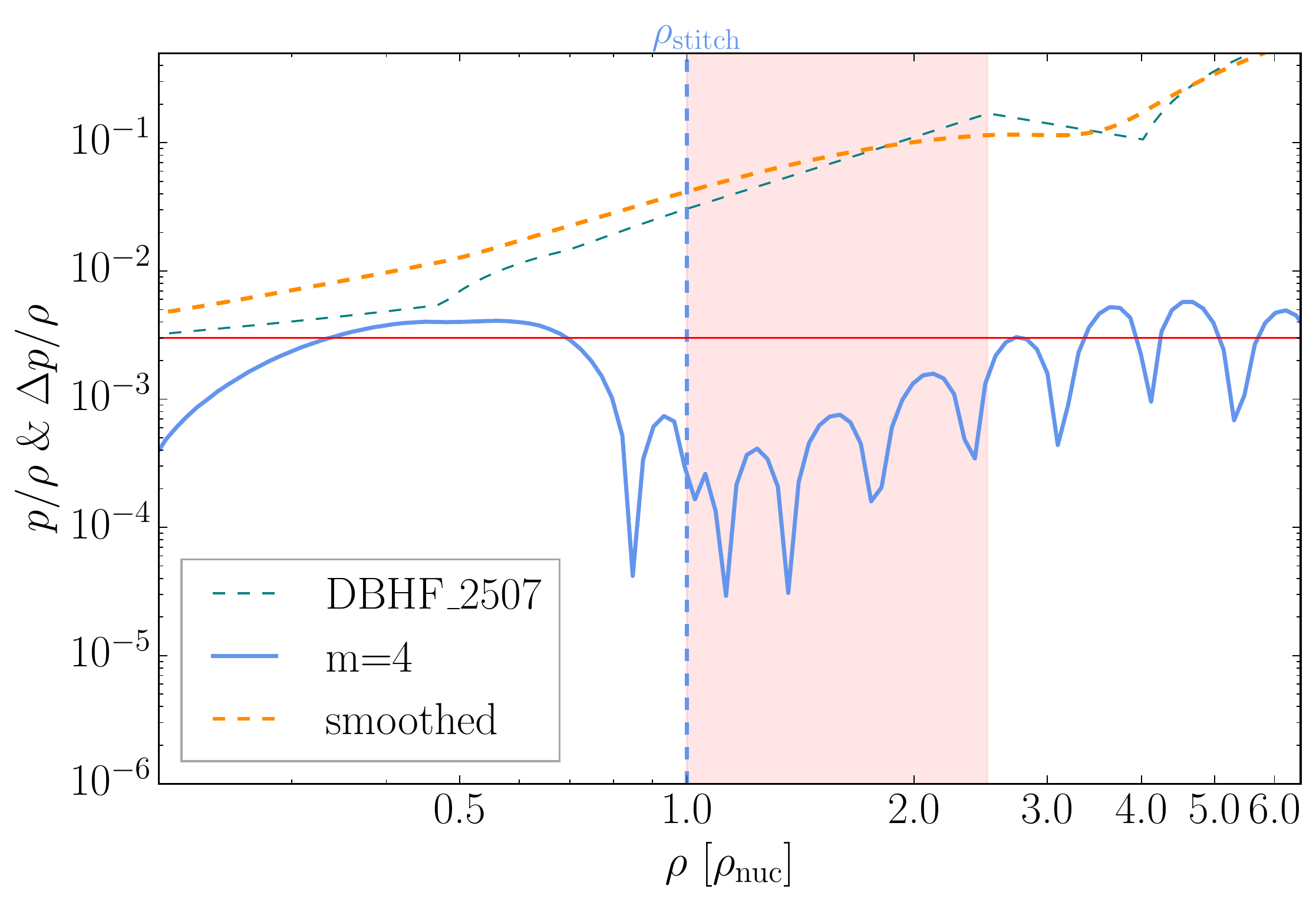}
    \caption{Same as Fig.~\ref{fig:low_stitch_dbhf} but for the smoothed version of DBHF\_2507 constructed in Sec.~\ref{sec:dbhf-smoothed}. The enthalpy parametrization achieves sub-$1\%$ errors in the most relevant region, $1-2\rhonuc$.}
    \label{fig:dbhf-2507-smoothed}
\end{figure}

Due to its flexibility, the \neweos parametrization can also be used to model smoother transitions in the EoS, such as those that may arise from a crossover transition. To demonstrate this, we begin with DBHF\_2507 and average the speed of sound at nearby points over the entire EoS.  This is distinct from the small perturbation added to the EoS in in Sec.~\ref{sec:dbhf-2507-grmhd}, as in this case we expect the resulting EoS to be well described by a smooth interpolant.   To demonstrate this, we fit this new smoothed EoS with $j_{\max}=4$ trigonometric terms and display the fit in Fig.~\ref{fig:dbhf-2507-smoothed}.  We find sub-$1\%$ agreement in the $1-2\rhonuc$ density region which most directly affects macroscopic observables. Relative errors are typically higher below the enthalpy-spectral transition point (set to $\rhonuc$ here).  This is because smoothing the EoS is done by locally averaging the speed of sound, so that after averaging the speed of sound is locally close to being constant.  For polytropic and nearly polytropic EoSs the speed of sound is not nearly constant at low densities, so a spectral parametrization cannot effectively fit the smooth $c_s^2$ EoS.

We expect simulating NSs with EoSs displaying smooth but rapidly varying speeds of sound to be slower. This is for two primary reasons.  First,  fitting EoSs to some fixed degree of precision for more complicated EoSs typically requires adding more parameters, increasing evaluation time.  Second, EoSs with rapid changes in $dp/d\rho$ tend to slow down primitive recovery, as the function $\rho(p)$ must be evaluated by root-finding.  Even though the EoS in this case is analytically smooth, root-finding algorithms require more evaluations if the function is quickly varying. 

\begin{figure}
    \centering
    \includegraphics[width=.49\textwidth]{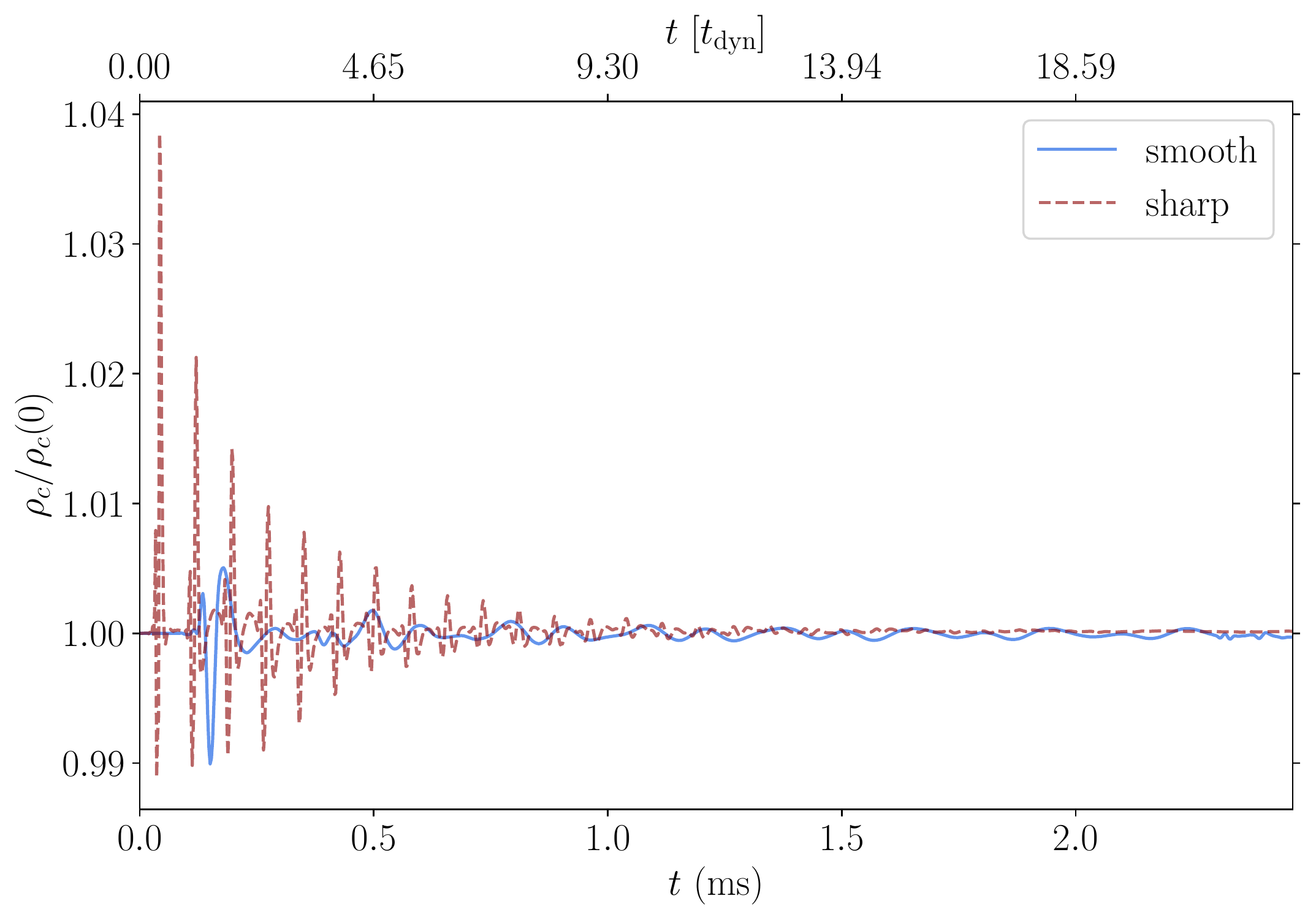}
    \caption{\red{NS central density as a function of time for the run \enthalpysmoothptlevthreelabel\ (blue, solid) and \enthalpyptlevthreelabel\ (red, dashed).  The smooth fit poorly reflects the mode structure of the true DBHF\_2507 EoS, even though the behavior of the microscopic EoS is qualitatively similar.}}
    \label{fig:dbhf_2507_smooth_vs_sharp}
\end{figure}

\red{We perform a run with identical central density, $\rho \sim 4.67\rhonuc$ for 10,000 CFL limited time steps, in order to bound performance decreases. We display the results in Table~\ref{tab:runs-performed} as \enthalpysmoothptlevthreelabel\ }.  We find that the EoS presented in Fig.~\ref{fig:dbhf-2507-smoothed} requires a comparable time per evolution step to \enthalpydbhflevthreelabel\, \red{indicating the EoS is sufficiently smooth to not induce a  large slowdown at this resolution.  We further plot the oscillations of the central density of both the smooth-transition DBHF\_2507 model (smooth) and \enthalpyptlevthreelabel\ (sharp) in Fig.~\ref{fig:dbhf_2507_smooth_vs_sharp}.  We find that the smoothed fit does not lead to the characteristic decoupling of core modes, meaning that such a model would be a poor representation of the true DBHF\_2507 EoS, even if it is able to reproduce other characteristics of DBHF\_2507, such as a small radius near $M_{\max}$.}

In general, we expect smooth EoSs to be most effectively represented by globally smooth parametrizations, while EoSs with discontinuities will be better modeled by piecewise parametrizations.  In addition, nonsmoothness can lead to a loss of accuracy in simulations~\cite{Foucart:2019yzo}, so an additional trade-off may exist between accuracy and performance in the choice to use a smooth versus a piecewise representation.   The \neweos parameterization is flexible enough to be effective in both the piecewise EoS and the smooth EoS cases.

\section{Discussion}
\label{sec:discussion}

We introduced a new enthalpy-based parametrization for the cold nuclear EoS that can capture a wide range nuclear models and their phenomenological extensions using polynomials and trigonometric terms.  The \emph{\neweos parametrization} emphasizes \emph{flexibility}, as it is able to effectively model both smooth and non-smooth nuclear models, and \emph{computational performance} as its evaluation cost scales with the number of parameters used. For example, it displays comparable performance to single-polytrope parametrizations for the case of polytropic EoSs, while the computational cost scales with the number of fit parameters for more complex (such as non-smooth) models.  This trade-off between computational performance and flexibility, allows us to tune EoS fits to the resolution requirements of the problem at hand.  

Computational performance is achieved by inexpensive evaluation of the various thermodynamic quantities.  
The $p(\rho)$ evaluation cost does not exceed $\mathcal O(4)$ times that of a polytropic EoS for any case we investigated, even when many trigonometric terms are used.   In cases where this slowdown is significant, the enthalpy parametrization may be sped up significantly by using Clenshaw's method~\cite{numericalrecipes}.  We obtain faster evaluation of $\epsilon(\rho)$ than the existing spectral parametrization in all cases, 
as the latter evaluates $\eps$ numerically, while the enthalpy parametrization computes all thermodynamic quantities analytically.  
Overall, the additional computational cost of the enthalpy parametrization on top of other existing parametrizations is always smaller than the cost of other simulation components.

With the caveat that quantifying EoS fitting accuracy is subtle and depends on the parameters one compares, we overall find that the \neweos parametrization is able to successfully fit nuclear models.  In principle and in the context of numerical simulations, EoS parametrizations need only fit the nuclear EoS as well as the simulation resolution.  Nonetheless, even subpercent errors in the pressure near $\rhonuc-2\rhonuc$ can lead to $\sim 100$~m differences in NS radii. In contrast to lower-dimensional or less flexible parametrizations, we show that the \neweos parametrization is able to fit tabulated and phenomenological nuclear models to effectively arbitrary precision by using additional parameters. The optimal number of parameters is then determined by balancing accuracy and computational cost for a given numerical resolution.

The \neweos parametrization's flexibility allows us to efficiently and with little fine tuning represent both smooth and non-smooth nuclear models. The latter may correspond to models with strong phase transitions that we can fit and numerically evolve using \spectre. Our simulations demonstrate that we can stably evolve such stars in the Cowling approximation. However, studying the evolution of hybrid hadronic-quark NSs away from an unstable EoS branch that falls between the hadronic and the quark branches~\cite{Espino:2021adh} hinges on full metric evolution coupled to GRMHD. \spectre's hybrid DG-FD scheme is crucial for the computational performance of these simulations. The DG-FD scheme allows phase transitions to be modeled with  lower-order finite-difference methods while continuing to use higher-order discontinuous-Galerkin methods throughout the individual hadronic and quark regions.  This leads to better computational scaling than might be expected upon mesh refinement, as better resolution of boundaries (such as the quark-hadronic matter boundary) within the star reduces the amount of the domain which uses the slower finite-difference approach.

The enthalpy parametrization is a step toward ensuring that numerical simulations can efficiently represent a wide range of nuclear phenomenology. Accurate simulations of NSs will continue being crucial for the interpretation of new astrophysical and  experimental data. Even with current EoS constraints, the space of potential BNS phenomenology is large, and many questions remain regarding the impact of magnetic fields, instabilities, temperature effects~\cite{Carbone:2019pkr,Raithel:2019gws} and transport physics. Future steps include extending the applicability of the enthalpy parametrization beyond cold, beta-equilibrated nuclear matter, and incorporating more physical effects in \spectre\ simulations.

The simulations presented here were performed with \spectre{} commit hash 2df19579a84385b3d5ab4663e3da7e33012e0355.
The earliest release of \spectre{} with this commit is version 2023.01.13~\cite{spectre_jan_23_release}.  Input files for the runs performed, including enthalpy fit parameters for each EoS studied, are available on Github~\cite{PaperGithub}.

\acknowledgements

I.L. thanks Tianqi Zhao for helpful conversations in preparing this manuscript.   The authors thank Reed Essick, Ingo Tews, Phil Landry, and Achim Schwink for access to $\chi$-EFT conditioned Gaussian process draws.  Charm++/Converse~\cite{laxmikant_kale_2020_3972617} was developed by the Parallel
Programming Laboratory in the Department of Computer Science at the University
of Illinois at Urbana-Champaign. This project made use of python libraries including \texttt{scipy} and \texttt{numpy}~\cite{scipy, numpy}. Figures were produced using \texttt{matplotlib}~\cite{matplotlib} and \texttt{ParaView}~\cite{paraview}. Computations were performed with the Wheeler cluster at
Caltech, which is supported by a grant from the Sherman Fairchild Foundation and Caltech.
This work was supported in part by the Sherman Fairchild Foundation at Caltech and Cornell, as well as by NSF
Grants No.~PHY-2011961, No.~PHY-2011968, and No.~OAC-1931266 at Caltech and by NSF
Grants No.~PHY-1912081 and No.~OAC-1931280 at Cornell.
IL and KC acknowledge support from the Department of Energy under award number DE-SC0023101.
FF gratefully acknowledges support from the Department of Energy, Office of Science, Office of Nuclear Physics, under contract number DE-AC02-05CH11231, from NASA through grant 
80NSSC22K0719, and from the NSF through grant AST-2107932.
The authors are grateful for computational resources provided by the LIGO Laboratory and supported by National Science Foundation Grants PHY-0757058 and PHY-0823459. 

\appendix

\section{Fitting the enthalpy parametrization}
\label{sec:fitting}

%
\begin{figure}
    \centering
    \includegraphics[width=.49\textwidth]{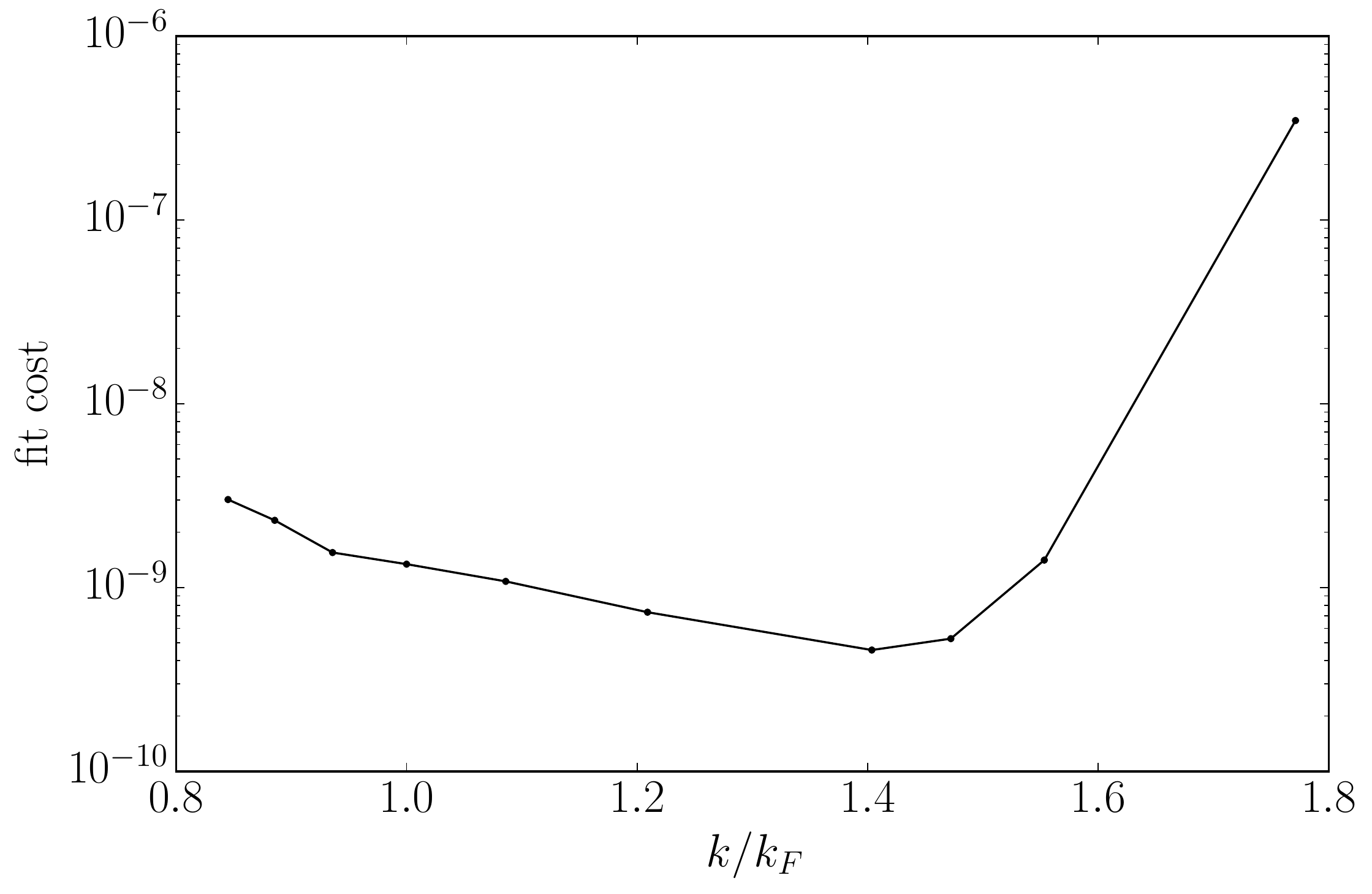}
    \caption{Cost, Eq~\ref{eq:cost}, in arbitrary units,  of the fit to the phenomenological EoS of Fig.~\ref{fig:fit-residuals} as a function of $k$.  The minimum occurs at $k$ slightly larger than $k_F$, in this case near $k= 1.4 k_F$. }
    \label{fig:cost_of_k}
\end{figure}

In this Appendix we provide details about the procedure with which we fit some tabulated EoS data with the enthalpy parametrization which includes the following parameters:
\begin{itemize}
\item The upper $\rho_{\max}$ and lower $\rho_{\min}$ density limits are chosen based on the densities of interest.  For NS simulations, reasonable values are $\rho_{\min} =\rhonuc$, $\rho_{\max} = 7\rhonuc$, but the upper limit depends on the maximum density expected in the simulation.  
\item The scaling parameters $\rho_0$ and the wavenumber of the trigonometric correction terms $k$ are not fit, but rather fixed.  When we extend a model via a constant speed of sound, Sec.~\ref{sec:dbhf-2507-grmhd}, the choice of $\rho_0$ is determined by the modeling problem. When the EoS is fit, $\rho_0$ is chosen $\rho_0 \in (0, \rho_{\min}]$, so that $z_{\max} = \log(\rho_{\max}/\rho_0) \lesssim i_{\max}$, see Sec.~\ref{sec:coefficient-numerics}. We find $\rho_0 = \rho_{\min}/2$ 
 is generally a robust choice.  Analogous to how $\rho_0$ controls the scale of polynomial terms, $k$ controls the scale of trigonometric oscillations.  As described in Sec.~\ref{sec:neweos}, typically $k\approx k_F$ is a good choice, but small perturbations $k\in [k_F/2, 2k_F]$ may improve the fit quality for certain problems, depending on the details of the EoS.   Figure~\ref{fig:cost_of_k} shows that the effect of varying $k$ is small for the particular test problem displayed in Fig.~\ref{fig:fit-residuals}.

\item The parameters $i_{\max}$ and $j_{\max}$ determine the number of polynomials and trigonometric terms respectively; see Eq.~\eqref{eq:polynomial-expansion} and Eq.~\eqref{eq:enthalpyFourier}.  The quality of the fit is a strong function of $i_{\max}$ and $j_{\max}$, but increasing $j_{\max}$ above $\sim10$ comes at a considerable computational cost even at low resolutions. On the other hand the cost of increasing $i_{\max}$ is small, typically of order $2\%$ or less of the total cost of the $p(\rho)$ evaluation per additional polynomial term.

\item The coefficients of the polynomial $\gamma_i$, Eq.~\eqref{eq:polynomial-expansion}, and the trigonometric $a_j, b_j$, Eq.~\eqref{eq:enthalpyFourier}, expansion are fit through a linear least-squares approach.

\item The energy density of the EoS at the stitching point, $e_{\min} = e(z_{\min})$.  This is the integration constant associated with solving $de/dz = \rho h$.  This parameter is constrained by $\rho_{\min} h_{\min} - e_{\min} = p_{\min} \geq 0$.   In principle $e(z_{\min})$ can be computed from EoS tables, but in practice EoS tables may be too coarsely tabulated, or may contain violations of the first law of thermodynamics at levels which significantly affect the computed value of $\epsilon$.  For example, a fractional error of $1\e{-3}$ in $e_{\min}$ will often translate to a fractional error of $\sim 1\e{-1}$ in $\epsilon$, which therefore shifts the value of $p(z)$ by $10\%$, as $h \rho = p + \epsilon$ is fixed by the parametrization.  Therefore in certain cases it is more effective to treat $e_{\min}$ as a free parameter, and further use it to optimize the values of $p(\rho)$.   In practice $e_{\min}$ is set by the low-density EoS parametrization to guarantee thermodynamic consistency.  
\end{itemize}.   
Given a target EoS with enthalpy $h(z_i)$ at discrete densities $z_i$, the linear fit is based on minimizing the cost function
\begin{equation}
    C(a_j, b_j, c_i) = \sum_k \quant{\frac{h_{*}(z_k; a_j, b_j, c_i) - h(z_k)}{\sigma(z)}}^2\,,
    \label{eq:cost}
\end{equation}
where $h_*$ is given in Eq.~\eqref{eq:expansion}.
The factor $\sigma(z)$ is the fit tolerance which can be chosen such that the fit is optimal at different density regions. We choose to target similar relative uncertainty on the non-rest-mass component of the enthalpy density $(h-1)\rho= p + \red{\epsilon \rho}$ across density scales: $\sigma(z) \propto \rho(z) \propto \exp(z)$.  Overall, the tolerance scales as $1/\rho$, so the fit is relatively better (with respect to $h$) at low densities.  The energy density at the stitching point is then selected;  if the tabulated EoS is sufficiently high-resolution, it can be computed by, e.g., the trapezoidal rule.  Otherwise, there is no canonical choice for this value, we choose it to maximize agreement with tabulated $p(\rho)$ at high densities.  

Finally, the EoS fit is completed by stitching to some other EoS parametrizaton at $\rho_{\mathrm{stitch}} = \rho_{\min}$.  In the majority of cases this is the spectral parametrization, though we also explore another enthalpy segment in Sec.~\ref{sec:dbhf-2507-grmhd} and a polytrope in App.~\ref{sec:polytrope appendix}. 
The low-density spectral EoS itself transitions to a lower-density polytrope at some fixed reference density $\rho_r$.  Following Ref.~\cite{Foucart:2019yzo}, we define $x \equiv \log(\rho/\rho_r)$ and write the spectral  pressure as
\begin{equation}
    \label{eq:spectral-expansion}
    p_{s}(x)=\begin{cases} p_0 \exp\quantb{\Gamma_0 x} &x \leq 0\,,\\
    p_0 \exp\quantb{ \sum_{i=0}^{3} \frac{1}{i+1}\Gamma_i x^{i+1}}& x>0\,,\end{cases}
\end{equation}
where $p_0$ controls the overall pressure, and $\Gamma_0, \Gamma_i$ are the spectral coefficients.
The low-density behavior fixes $\Gamma_0$, while requiring a $\mathcal C^1$ transition to the \neweos parametrization, i.e., continuity in pressure, energy density, and pressure derivative fixes $3$ more parameters.  In practice  because $e_{\min} = e_{\stitch}$ is an integration constant in the \neweos\ parametrization, we can freely set it to the value computed for $e_{\stitch}$ from the low-density parametrization, guaranteeing exact consistency.    

The remaining $1$ degree of freedom is selected by either maximizing smoothness across the lower-density transition to the polytrope or maximizing accuracy of the low-density EoS.  
Smoothness is prioritized when the stitching density is below the core density of typical NS.  Then, we set $\Gamma_1 = 0.0$~\cite{Foucart:2019yzo}, guaranteeing that Eq.~\eqref{eq:spectral-expansion} is $\mathcal C^2$ across the transition $\rho_r$.  This typically produces good fits to the overall $M$--$R$ curve for the entire EoS.
If the spectral parametrization is stitched to the enthalpy parametrization at a higher density (near the core density of astrophysical NSs as is the case in the high-stitch fit of Sec.\ref{sec:dbhf-test}) we instead allow $\Gamma_1$ to vary, choosing it to maximize the agreement of the total parametrized EoS with the target.  Both strategies typically result in machine-precision level $\mathcal C^1$-stitching to the \neweos parametrization, with residuals much smaller than mismodeling in the low-density regime. 

\section{Approximating a single polytrope}
\label{sec:polytrope appendix}

As an example of the strategy for fitting a target EoS with the enthalpy parametrization, we consider a single-polytrope. In this case, the enthalpy coefficients can be computed analytically. The general goal is to express the EoS in the form of Eq.~\eqref{eq:general-css}, i.e., compute the enthalpy as a function of log-density.
 
The polytropic exponent is defined as
\begin{equation}
    \label{eq:gamma-def}
    \Gamma(z) \equiv  \deriv{\log p}{\log \rho} =  \frac{\rho}{p} \deriv{h}{z} =\frac{\rho}{p}\deriv{\quant{\frac{1}{\rho}\deriv{e}{z}}}{z}\,.
\end{equation}
For a constant polytropic exponent $\Gamma(z) = \Gamma_0$ and using the identity
\begin{equation}
    p(z) = h(z)\rho(z) - e(z) = \deriv{e}{z} - e(z)\,,
\end{equation}
Eq.~\eqref{eq:gamma-def} becomes
\begin{equation}
\deriv{^2e}{z^2}-(\Gamma_0+1) \deriv{e}{z} +\Gamma_0 e = 0\,.
\end{equation}
The solution to this differential equation is
\begin{equation}
    e(z) = (e_0-\rho_0)\exp(\Gamma_0 z) + \rho_0\exp(z)\,,
\end{equation}
where we have enforced $e(z=0)=e_0$ and $e(z\to -\infty)\to\rho(z)$\footnote{This is equivalent to assuming the specific internal energy $\epsilon$ is $0$ in ordinary, low-density cold matter.  This can be done by defining the baryon ``rest" mass to be the average mass of a baryon in the outer crust of a NS (despite the fact these baryons may be bound in, e.g. iron and therefore differ from the mass of a free neutron/proton by up to $1\%$).} and the enthalpy is
\begin{equation}
    h(z) = \frac{1}{\rho}\deriv{e}{z}= \frac{e_0-\rho_0}{\rho_0}\Gamma_0 \exp{\quantb{(\Gamma_0-1)z}}+ 1\,.
\end{equation}
Comparing with Eq.~\eqref{eq:polynomial-expansion} the polynomial coefficients of the enthalpy expansion are $\gamma_i = h_0 \quant{\Gamma_0 - 1}^i/i!$, with $h_0 = \Gamma_0\quant{e_0-\rho_0}/\rho_0 $, if $i\neq 0$, and $\gamma_0  =h_0 \quant{\Gamma_0 - 1} + 1$. 
In practice, evaluating the polynomial expansion of Eq.~\eqref{eq:polynomial-expansion} requires many floating point operations.  Nonetheless, this computation is not necessarily slower than evaluating a simple polytrope if $\Gamma_0$ is not an integer, because floating-point exponentiation typically at least an order of magnitude slower than multiplication and addition.

We use this \neweos parametrization of the polytrope model to compare against the direct single-polytrope \spectre\ implementation and verify the predicted Cowling-approximation NS modes~\cite{Font:2001ew, Deppe:2021bhi}.  
The low-density EoS in the \neweos parametrization case is stitched to the exact polytropic expression 
\begin{equation}
    \label{eq:polytrope-specific}
    P(\rho) = \frac{100}{M_{\odot}^{-2.0}} \rho ^{2.0}\,.
\end{equation}

\begin{figure}
    \centering
    \includegraphics[width=.49\textwidth]{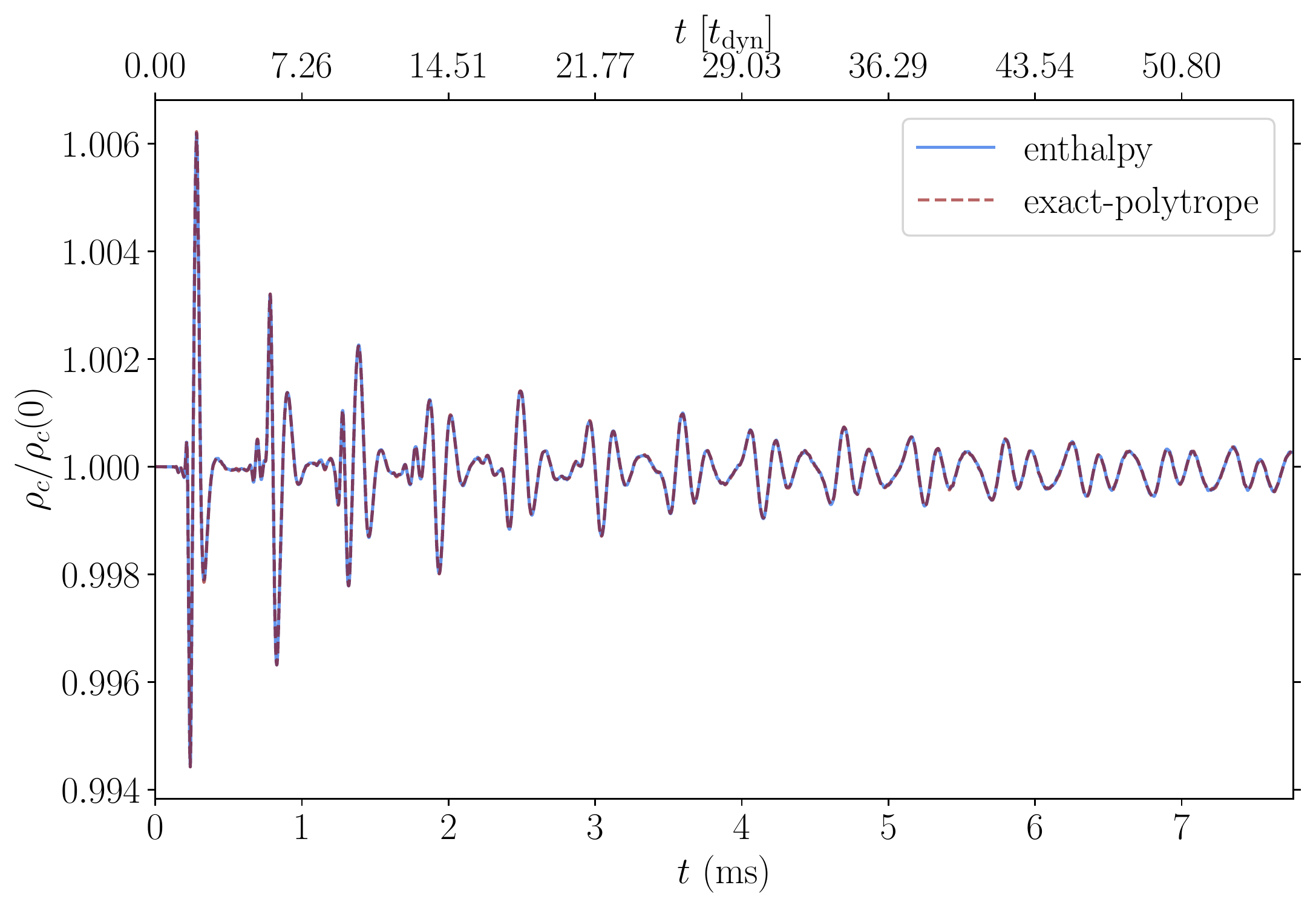}
    \includegraphics[width=.49\textwidth]{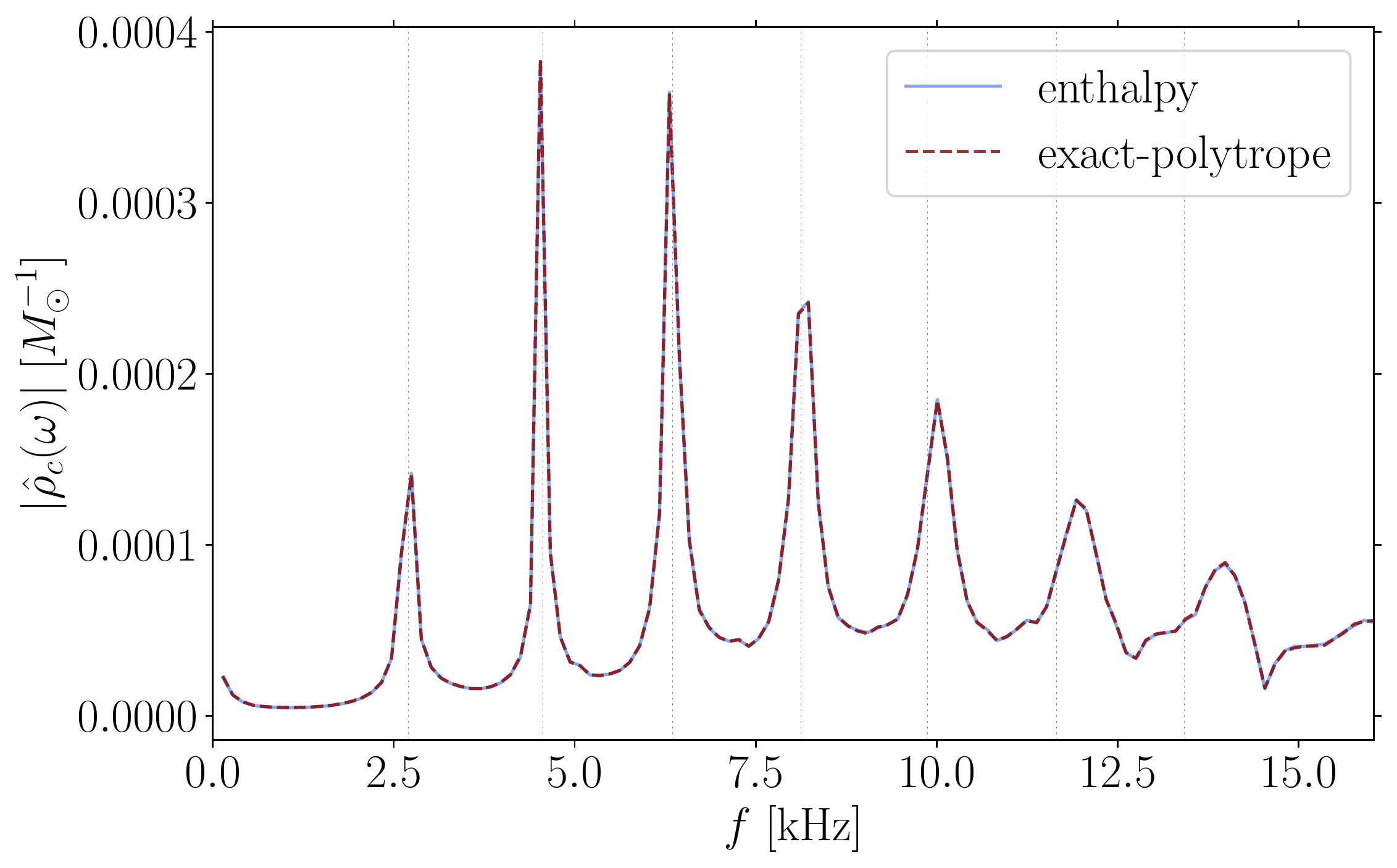}
    \caption{NS central density as a function of time (top panel) and its spectrum (bottom panel) for \spectre\ simulations with an enthalpy fit to a single polytrope with $\Gamma_0=2.0$ (blue) and a direct single-polytropic parametrization (red, dashed,). Known Cowling frequencies~\cite{Font:2001ew} are marked as dashed vertical lines.  The spectra are identical by eye.}
    \label{fig:polytrope-comparison}
\end{figure}

We evolve a NS with central density 
$1.28 \e{-3} M_{\odot}^{-2} \approx 2.84 \rhonuc $, which is the same as the stars evolved in Refs.~\cite{Font:2001ew, Deppe:2021bhi}.
The number of terms necessary in the polynomial expansion depends on the desired accuracy. For a resolution of $130$\,m we find that $i_{\max} = 8$ is more than sufficient. This is consistent with theoretical expectations, as the first neglected term, is of order $1/9! \approx 2\e{-6}$, indicating errors should be of this scale or smaller.   Results are shown in Fig.~\ref{fig:polytrope-comparison} where the enthalpy fit to the polytrope and the direct single-polytropic parametrization return essentially identical results.  We display the run details in Table~\ref{tab:runs-performed}, as \enthalpypolytropelevthreelabel\ and \polytropicpolytropelevthreelabel.

We find effectively no difference in the runtime for the simulations using each of the polytropic and enthalpy parametrizations.  Examining the cost of individual EoS calls in Table~\ref{tab:polytrope-profile}, we find the enthalpy parametrization is somewhat faster in both $p(\rho)$ and $\eps(\rho)$ evaluation, indicating that in this case, EoS evaluation time is not a significant contribution to runtime.  This speedup is also expected to extend to constant-speed-of-sound matter, which has an identical functional form to a polytropic EoS when expanded in the \neweos/ parametrization, the only difference being the values of the coefficients.  

The reason the polytropic EoS evaluation  is not faster despite having a very simple analytic expression is the inefficiency of floating-point exponentiation.   In the case of our test problem the floating point exponent \texttt{2.0} is only known during runtime, and so the compiler cannot optimize EoS calls.  If the exponent is known to be an integer at compile time, the calls can be evaluated using repeated multiplication.  We implement this improvement for this particular test problem, and find the cost of polytropic  $p(\rho)$ evaluations to be $10$~ns on identical hardware, indicating a 5-fold improvement.  Nonetheless, this speedup is not reflected in the total evolution runtime
for resolutions of at least $\sim 120$~m as the EoS evaluation cost is subdominant to other simulation components.
\begin{table}
    \begin{tabular}{|l||*{2}{c|}}\hline
    \backslashbox{Parametrization}{Evaluation Cost (ns)}
    &\makebox[3em]{$p(\rho)$}&\makebox[3em]{$\eps(\rho)$} \\\hline\hline
    enthalpy & \benchenthalpypressurefromdensity & \benchenthalpyinternalenergyfromdensity\\\hline
    polytrope &\benchpolytropepressurefromdensity&\benchpolytropeinternalenergyfromdensity\\\hline
    \end{tabular}
    \caption{Performance (in nanoseconds) of the enthalpy and single-polytropic fits to a single polytrope in evaluating the pressure and internal energy at $\rho= 5.0\e{-4} M_{\odot}^2$. The enthalpy parametrization outperforms then polytrope in both cases.}
    \label{tab:polytrope-profile}
\end{table}
%

\section{Numerical considerations for enthalpy coefficients}
\label{sec:coefficient-numerics}

Here we expand upon numerical considerations for choices of polynomial, coefficients $\gamma_i$.   The choice, $\gamma_i\geq 0$ in Eq.~\eqref{eq:polynomial-expansion} effectively bounds the number of terms in the polynomial expansion which can be practically used.   To see this, consider  $z_{\max} \equiv \log\quant{\rho_{\max}/\rho_0}$.  Coefficients $\gamma_i$ must satisfy $\gamma_iz_{\max}^i \lesssim \mathcal{O}(\gamma_0) \sim 1$, otherwise they would be larger than the total enthalpy in this region, which is typically also of this scale.  With this in mind, we consider $i$ as ``too large" if $i\gg z_{\max}$, as any term which satisfies $\gamma_i z_{\max}^i \lesssim \gamma_0$ has 
\begin{equation}
    \label{eq:ratio-of-term}
    \frac{\gamma_i z^i}{\gamma_i z_{\max}^i} = \quant{\frac{z}{z_{\max}}}^i\,,
\end{equation}
 small except when $z$ is nearly $z_{\max}$.  That is, the degree of freedom is only relevant at the highest densities, and this density region shrinks as $i$ gets larger.
For typical scales, such as $\rho_0 = 0.5\rhonuc$, $\rho_{\min} = \rhonuc$, and $\rho_{\max}= 7\rhonuc$, then $z_{\max}\sim\log(14) \approx 2.6$.  

One can decrease $\rho_0$ to increase the relevant value of $z_{\max}$, but this requires adding more parameters, which may not be desirable, since many of them may be irrelevant, or degenerate.   One way to view this, is that in the Taylor expansion of the exponential function (equivalently the expansion of $h(z)$ in a constant speed of sound case), the term $z^i/i!$ is the largest term on $z\in (i-1, i)$, and is generally decreasing in relevance away from this region relative to other terms.  Therefore the $i$\textsuperscript{th} term of this expansion is most relevant near $z\lesssim i$, and is unimportant far from this region. This implies that flexibility is essentially equidistributed in $\log(\rho)$ for this approximation (the same argument applies to the spectral parametrization), and that higher polynomial terms cannot resolve features at low densities.  Instead, we choose to switch to a new function basis at this point, optimized to capture the largest scale features at lowest order of approximation.

\newpage
\bibliography{references}

\begin{thebibliography}{125}%
\makeatletter
\providecommand \@ifxundefined [1]{%
 \@ifx{#1\undefined}
}%
\providecommand \@ifnum [1]{%
 \ifnum #1\expandafter \@firstoftwo
 \else \expandafter \@secondoftwo
 \fi
}%
\providecommand \@ifx [1]{%
 \ifx #1\expandafter \@firstoftwo
 \else \expandafter \@secondoftwo
 \fi
}%
\providecommand \natexlab [1]{#1}%
\providecommand \enquote  [1]{``#1''}%
\providecommand \bibnamefont  [1]{#1}%
\providecommand \bibfnamefont [1]{#1}%
\providecommand \citenamefont [1]{#1}%
\providecommand \href@noop [0]{\@secondoftwo}%
\providecommand \href [0]{\begingroup \@sanitize@url \@href}%
\providecommand \@href[1]{\@@startlink{#1}\@@href}%
\providecommand \@@href[1]{\endgroup#1\@@endlink}%
\providecommand \@sanitize@url [0]{\catcode `\\12\catcode `\$12\catcode
  `\&12\catcode `\#12\catcode `\^12\catcode `\_12\catcode `\%12\relax}%
\providecommand \@@startlink[1]{}%
\providecommand \@@endlink[0]{}%
\providecommand \url  [0]{\begingroup\@sanitize@url \@url }%
\providecommand \@url [1]{\endgroup\@href {#1}{\urlprefix }}%
\providecommand \urlprefix  [0]{URL }%
\providecommand \Eprint [0]{\href }%
\providecommand \doibase [0]{http://dx.doi.org/}%
\providecommand \selectlanguage [0]{\@gobble}%
\providecommand \bibinfo  [0]{\@secondoftwo}%
\providecommand \bibfield  [0]{\@secondoftwo}%
\providecommand \translation [1]{[#1]}%
\providecommand \BibitemOpen [0]{}%
\providecommand \bibitemStop [0]{}%
\providecommand \bibitemNoStop [0]{.\EOS\space}%
\providecommand \EOS [0]{\spacefactor3000\relax}%
\providecommand \BibitemShut  [1]{\csname bibitem#1\endcsname}%
\let\auto@bib@innerbib\@empty
\bibitem [{\citenamefont {Abbott}\ \emph
  {et~al.}(2017{\natexlab{a}})\citenamefont {Abbott} \emph
  {et~al.}}]{TheLIGOScientific:2017qsa}%
  \BibitemOpen
  \bibfield  {author} {\bibinfo {author} {\bibfnamefont {B.~P.}\ \bibnamefont
  {Abbott}} \emph {et~al.} (\bibinfo {collaboration} {LIGO Scientific
  Collaboration, Virgo Collaboration}),\ }\href {\doibase
  10.1103/PhysRevLett.119.161101} {\bibfield  {journal} {\bibinfo  {journal}
  {Phys. Rev. Lett.}\ }\textbf {\bibinfo {volume} {119}},\ \bibinfo {pages}
  {161101} (\bibinfo {year} {2017}{\natexlab{a}})},\ \Eprint
  {http://arxiv.org/abs/1710.05832} {arXiv:1710.05832 [gr-qc]} \BibitemShut
  {NoStop}%
\bibitem [{\citenamefont {Abbott}\ \emph
  {et~al.}(2017{\natexlab{b}})\citenamefont {Abbott} \emph
  {et~al.}}]{LIGOScientific:2017ync}%
  \BibitemOpen
  \bibfield  {author} {\bibinfo {author} {\bibfnamefont {B.~P.}\ \bibnamefont
  {Abbott}} \emph {et~al.} (\bibinfo {collaboration} {LIGO Scientific, Virgo,
  Fermi GBM, INTEGRAL, IceCube, AstroSat Cadmium Zinc Telluride Imager Team,
  IPN, Insight-Hxmt, ANTARES, Swift, AGILE Team, 1M2H Team, Dark Energy Camera
  GW-EM, DES, DLT40, GRAWITA, Fermi-LAT, ATCA, ASKAP, Las Cumbres Observatory
  Group, OzGrav, DWF (Deeper Wider Faster Program), AST3, CAASTRO, VINROUGE,
  MASTER, J-GEM, GROWTH, JAGWAR, CaltechNRAO, TTU-NRAO, NuSTAR, Pan-STARRS,
  MAXI Team, TZAC Consortium, KU, Nordic Optical Telescope, ePESSTO, GROND,
  Texas Tech University, SALT Group, TOROS, BOOTES, MWA, CALET, IKI-GW
  Follow-up, H.E.S.S., LOFAR, LWA, HAWC, Pierre Auger, ALMA, Euro VLBI Team, Pi
  of Sky, Chandra Team at McGill University, DFN, ATLAS Telescopes, High Time
  Resolution Universe Survey, RIMAS, RATIR, SKA South Africa/MeerKAT}),\ }\href
  {\doibase 10.3847/2041-8213/aa91c9} {\bibfield  {journal} {\bibinfo
  {journal} {Astrophys. J. Lett.}\ }\textbf {\bibinfo {volume} {848}},\
  \bibinfo {pages} {L12} (\bibinfo {year} {2017}{\natexlab{b}})},\ \Eprint
  {http://arxiv.org/abs/1710.05833} {arXiv:1710.05833 [astro-ph.HE]}
  \BibitemShut {NoStop}%
\bibitem [{\citenamefont {Dietrich}\ \emph {et~al.}(2021)\citenamefont
  {Dietrich}, \citenamefont {Hinderer},\ and\ \citenamefont
  {Samajdar}}]{Dietrich:2020eud}%
  \BibitemOpen
  \bibfield  {author} {\bibinfo {author} {\bibfnamefont {T.}~\bibnamefont
  {Dietrich}}, \bibinfo {author} {\bibfnamefont {T.}~\bibnamefont {Hinderer}},
  \ and\ \bibinfo {author} {\bibfnamefont {A.}~\bibnamefont {Samajdar}},\
  }\href {\doibase 10.1007/s10714-020-02751-6} {\bibfield  {journal} {\bibinfo
  {journal} {Gen. Rel. Grav.}\ }\textbf {\bibinfo {volume} {53}},\ \bibinfo
  {pages} {27} (\bibinfo {year} {2021})},\ \Eprint
  {http://arxiv.org/abs/2004.02527} {arXiv:2004.02527 [gr-qc]} \BibitemShut
  {NoStop}%
\bibitem [{\citenamefont {Chatziioannou}(2020)}]{Chatziioannou:2020pqz}%
  \BibitemOpen
  \bibfield  {author} {\bibinfo {author} {\bibfnamefont {K.}~\bibnamefont
  {Chatziioannou}},\ }\href {\doibase 10.1007/s10714-020-02754-3} {\bibfield
  {journal} {\bibinfo  {journal} {Gen. Rel. Grav.}\ }\textbf {\bibinfo {volume}
  {52}},\ \bibinfo {pages} {109} (\bibinfo {year} {2020})},\ \Eprint
  {http://arxiv.org/abs/2006.03168} {arXiv:2006.03168 [gr-qc]} \BibitemShut
  {NoStop}%
\bibitem [{\citenamefont {Piekarewicz}(2022)}]{Piekarewicz:2022ycz}%
  \BibitemOpen
  \bibfield  {author} {\bibinfo {author} {\bibfnamefont {J.}~\bibnamefont
  {Piekarewicz}},\ }\href@noop {} {\  (\bibinfo {year} {2022})},\ \Eprint
  {http://arxiv.org/abs/2209.14877} {arXiv:2209.14877 [nucl-th]} \BibitemShut
  {NoStop}%
\bibitem [{\citenamefont {Abbott}\ \emph {et~al.}(2020)\citenamefont {Abbott}
  \emph {et~al.}}]{Abbott:2020uma}%
  \BibitemOpen
  \bibfield  {author} {\bibinfo {author} {\bibfnamefont {B.~P.}\ \bibnamefont
  {Abbott}} \emph {et~al.} (\bibinfo {collaboration} {LIGO Scientific,
  Virgo}),\ }\href {\doibase 10.3847/2041-8213/ab75f5} {\bibfield  {journal}
  {\bibinfo  {journal} {Astrophys. J. Lett.}\ }\textbf {\bibinfo {volume}
  {892}},\ \bibinfo {pages} {L3} (\bibinfo {year} {2020})},\ \Eprint
  {http://arxiv.org/abs/2001.01761} {arXiv:2001.01761 [astro-ph.HE]}
  \BibitemShut {NoStop}%
\bibitem [{\citenamefont {Cromartie}\ \emph {et~al.}(2019)\citenamefont
  {Cromartie} \emph {et~al.}}]{Cromartie:2019kug}%
  \BibitemOpen
  \bibfield  {author} {\bibinfo {author} {\bibfnamefont {H.~T.}\ \bibnamefont
  {Cromartie}} \emph {et~al.},\ }\href {\doibase 10.1038/s41550-019-0880-2}
  {\bibfield  {journal} {\bibinfo  {journal} {Nature Astron.}\ }\textbf
  {\bibinfo {volume} {4}},\ \bibinfo {pages} {72} (\bibinfo {year} {2019})},\
  \Eprint {http://arxiv.org/abs/1904.06759} {arXiv:1904.06759} \BibitemShut
  {NoStop}%
\bibitem [{\citenamefont {Fonseca}\ \emph {et~al.}(2021)\citenamefont {Fonseca}
  \emph {et~al.}}]{Fonseca:2021wxt}%
  \BibitemOpen
  \bibfield  {author} {\bibinfo {author} {\bibfnamefont {E.}~\bibnamefont
  {Fonseca}} \emph {et~al.},\ }\href {\doibase 10.3847/2041-8213/ac03b8}
  {\bibfield  {journal} {\bibinfo  {journal} {Astrophys. J. Lett.}\ }\textbf
  {\bibinfo {volume} {915}},\ \bibinfo {pages} {L12} (\bibinfo {year}
  {2021})},\ \Eprint {http://arxiv.org/abs/2104.00880} {arXiv:2104.00880
  [astro-ph.HE]} \BibitemShut {NoStop}%
\bibitem [{\citenamefont {Miller}\ \emph {et~al.}(2019)\citenamefont {Miller},
  \citenamefont {Chirenti},\ and\ \citenamefont {Lamb}}]{Miller:2019nzo}%
  \BibitemOpen
  \bibfield  {author} {\bibinfo {author} {\bibfnamefont {M.~C.}\ \bibnamefont
  {Miller}}, \bibinfo {author} {\bibfnamefont {C.}~\bibnamefont {Chirenti}}, \
  and\ \bibinfo {author} {\bibfnamefont {F.~K.}\ \bibnamefont {Lamb}},\
  }\href@noop {} {\  (\bibinfo {year} {2019})},\ \Eprint
  {http://arxiv.org/abs/1904.08907} {arXiv:1904.08907 [astro-ph.HE]}
  \BibitemShut {NoStop}%
\bibitem [{\citenamefont {Miller}\ \emph {et~al.}(2021)\citenamefont {Miller}
  \emph {et~al.}}]{Miller:2021qha}%
  \BibitemOpen
  \bibfield  {author} {\bibinfo {author} {\bibfnamefont {M.~C.}\ \bibnamefont
  {Miller}} \emph {et~al.},\ }\href {\doibase 10.3847/2041-8213/ac089b}
  {\bibfield  {journal} {\bibinfo  {journal} {Astrophys. J. Lett.}\ }\textbf
  {\bibinfo {volume} {918}},\ \bibinfo {pages} {L28} (\bibinfo {year}
  {2021})},\ \Eprint {http://arxiv.org/abs/2105.06979} {arXiv:2105.06979
  [astro-ph.HE]} \BibitemShut {NoStop}%
\bibitem [{\citenamefont {Riley}\ \emph {et~al.}(2019)\citenamefont {Riley}
  \emph {et~al.}}]{Riley:2019yda}%
  \BibitemOpen
  \bibfield  {author} {\bibinfo {author} {\bibfnamefont {T.~E.}\ \bibnamefont
  {Riley}} \emph {et~al.},\ }\href {\doibase 10.3847/2041-8213/ab481c}
  {\bibfield  {journal} {\bibinfo  {journal} {Astrophys. J. Lett.}\ }\textbf
  {\bibinfo {volume} {887}},\ \bibinfo {pages} {L21} (\bibinfo {year}
  {2019})},\ \Eprint {http://arxiv.org/abs/1912.05702} {arXiv:1912.05702
  [astro-ph.HE]} \BibitemShut {NoStop}%
\bibitem [{\citenamefont {Riley}\ \emph {et~al.}(2021)\citenamefont {Riley}
  \emph {et~al.}}]{Riley:2021pdl}%
  \BibitemOpen
  \bibfield  {author} {\bibinfo {author} {\bibfnamefont {T.~E.}\ \bibnamefont
  {Riley}} \emph {et~al.},\ }\href {\doibase 10.3847/2041-8213/ac0a81}
  {\bibfield  {journal} {\bibinfo  {journal} {Astrophys. J. Lett.}\ }\textbf
  {\bibinfo {volume} {918}},\ \bibinfo {pages} {L27} (\bibinfo {year}
  {2021})},\ \Eprint {http://arxiv.org/abs/2105.06980} {arXiv:2105.06980
  [astro-ph.HE]} \BibitemShut {NoStop}%
\bibitem [{\citenamefont {Antoniadis}\ \emph {et~al.}(2013)\citenamefont
  {Antoniadis}, \citenamefont {Freire}, \citenamefont {Wex}, \citenamefont
  {Tauris}, \citenamefont {Lynch} \emph {et~al.}}]{Antoniadis:2013pzd}%
  \BibitemOpen
  \bibfield  {author} {\bibinfo {author} {\bibfnamefont {J.}~\bibnamefont
  {Antoniadis}}, \bibinfo {author} {\bibfnamefont {P.~C.}\ \bibnamefont
  {Freire}}, \bibinfo {author} {\bibfnamefont {N.}~\bibnamefont {Wex}},
  \bibinfo {author} {\bibfnamefont {T.~M.}\ \bibnamefont {Tauris}}, \bibinfo
  {author} {\bibfnamefont {R.~S.}\ \bibnamefont {Lynch}},  \emph {et~al.},\
  }\href {\doibase 10.1126/science.1233232} {\bibfield  {journal} {\bibinfo
  {journal} {Science}\ }\textbf {\bibinfo {volume} {340}},\ \bibinfo {pages}
  {1233232} (\bibinfo {year} {2013})},\ \Eprint
  {http://arxiv.org/abs/1304.6875} {arXiv:1304.6875 [astro-ph.HE]} \BibitemShut
  {NoStop}%
\bibitem [{\citenamefont {Adhikari}\ \emph {et~al.}(2021)\citenamefont
  {Adhikari} \emph {et~al.}}]{Adhikari:2021phr}%
  \BibitemOpen
  \bibfield  {author} {\bibinfo {author} {\bibfnamefont {D.}~\bibnamefont
  {Adhikari}} \emph {et~al.} (\bibinfo {collaboration} {PREX}),\ }\href
  {\doibase 10.1103/PhysRevLett.126.172502} {\bibfield  {journal} {\bibinfo
  {journal} {Phys. Rev. Lett.}\ }\textbf {\bibinfo {volume} {126}},\ \bibinfo
  {pages} {172502} (\bibinfo {year} {2021})},\ \Eprint
  {http://arxiv.org/abs/2102.10767} {arXiv:2102.10767 [nucl-ex]} \BibitemShut
  {NoStop}%
\bibitem [{\citenamefont {Adhikari}\ \emph {et~al.}(2022)\citenamefont
  {Adhikari} \emph {et~al.}}]{CREX:2022kgg}%
  \BibitemOpen
  \bibfield  {author} {\bibinfo {author} {\bibfnamefont {D.}~\bibnamefont
  {Adhikari}} \emph {et~al.} (\bibinfo {collaboration} {CREX}),\ }\href@noop {}
  {\  (\bibinfo {year} {2022})},\ \Eprint {http://arxiv.org/abs/2205.11593}
  {arXiv:2205.11593 [nucl-ex]} \BibitemShut {NoStop}%
\bibitem [{\citenamefont {Abbott}\ \emph {et~al.}(2018)\citenamefont {Abbott}
  \emph {et~al.}}]{Abbott:2018exr}%
  \BibitemOpen
  \bibfield  {author} {\bibinfo {author} {\bibfnamefont {B.~P.}\ \bibnamefont
  {Abbott}} \emph {et~al.} (\bibinfo {collaboration} {LIGO Scientific
  Collaboration, Virgo Collaboration}),\ }\href {\doibase
  10.1103/PhysRevLett.121.161101} {\bibfield  {journal} {\bibinfo  {journal}
  {Phys. Rev. Lett.}\ }\textbf {\bibinfo {volume} {121}},\ \bibinfo {pages}
  {161101} (\bibinfo {year} {2018})},\ \Eprint
  {http://arxiv.org/abs/1805.11581} {arXiv:1805.11581 [gr-qc]} \BibitemShut
  {NoStop}%
\bibitem [{\citenamefont {Essick}(2021)}]{Essick:2021ezv}%
  \BibitemOpen
  \bibfield  {author} {\bibinfo {author} {\bibfnamefont {R.}~\bibnamefont
  {Essick}},\ }\href@noop {} {\  (\bibinfo {year} {2021})},\ \Eprint
  {http://arxiv.org/abs/2111.04244} {arXiv:2111.04244 [astro-ph.HE]}
  \BibitemShut {NoStop}%
\bibitem [{\citenamefont {Pang}\ \emph {et~al.}(2021)\citenamefont {Pang},
  \citenamefont {Tews}, \citenamefont {Coughlin}, \citenamefont {Bulla},
  \citenamefont {Van Den~Broeck},\ and\ \citenamefont
  {Dietrich}}]{Pang:2021jta}%
  \BibitemOpen
  \bibfield  {author} {\bibinfo {author} {\bibfnamefont {P.~T.~H.}\
  \bibnamefont {Pang}}, \bibinfo {author} {\bibfnamefont {I.}~\bibnamefont
  {Tews}}, \bibinfo {author} {\bibfnamefont {M.~W.}\ \bibnamefont {Coughlin}},
  \bibinfo {author} {\bibfnamefont {M.}~\bibnamefont {Bulla}}, \bibinfo
  {author} {\bibfnamefont {C.}~\bibnamefont {Van Den~Broeck}}, \ and\ \bibinfo
  {author} {\bibfnamefont {T.}~\bibnamefont {Dietrich}},\ }\href {\doibase
  10.3847/1538-4357/ac19ab} {\bibfield  {journal} {\bibinfo  {journal}
  {Astrophys. J.}\ }\textbf {\bibinfo {volume} {922}},\ \bibinfo {pages} {14}
  (\bibinfo {year} {2021})},\ \Eprint {http://arxiv.org/abs/2105.08688}
  {arXiv:2105.08688 [astro-ph.HE]} \BibitemShut {NoStop}%
\bibitem [{\citenamefont {Raaijmakers}\ \emph {et~al.}(2020)\citenamefont
  {Raaijmakers} \emph {et~al.}}]{Raaijmakers:2019dks}%
  \BibitemOpen
  \bibfield  {author} {\bibinfo {author} {\bibfnamefont {G.}~\bibnamefont
  {Raaijmakers}} \emph {et~al.},\ }\href {\doibase 10.3847/2041-8213/ab822f}
  {\bibfield  {journal} {\bibinfo  {journal} {Astrophys. J. Lett.}\ }\textbf
  {\bibinfo {volume} {893}},\ \bibinfo {pages} {L21} (\bibinfo {year}
  {2020})},\ \Eprint {http://arxiv.org/abs/1912.11031} {arXiv:1912.11031
  [astro-ph.HE]} \BibitemShut {NoStop}%
\bibitem [{\citenamefont {Raaijmakers}\ \emph {et~al.}(2021)\citenamefont
  {Raaijmakers}, \citenamefont {Greif}, \citenamefont {Hebeler}, \citenamefont
  {Hinderer}, \citenamefont {Nissanke}, \citenamefont {Schwenk}, \citenamefont
  {Riley}, \citenamefont {Watts}, \citenamefont {Lattimer},\ and\ \citenamefont
  {Ho}}]{Raaijmakers:2021uju}%
  \BibitemOpen
  \bibfield  {author} {\bibinfo {author} {\bibfnamefont {G.}~\bibnamefont
  {Raaijmakers}}, \bibinfo {author} {\bibfnamefont {S.~K.}\ \bibnamefont
  {Greif}}, \bibinfo {author} {\bibfnamefont {K.}~\bibnamefont {Hebeler}},
  \bibinfo {author} {\bibfnamefont {T.}~\bibnamefont {Hinderer}}, \bibinfo
  {author} {\bibfnamefont {S.}~\bibnamefont {Nissanke}}, \bibinfo {author}
  {\bibfnamefont {A.}~\bibnamefont {Schwenk}}, \bibinfo {author} {\bibfnamefont
  {T.~E.}\ \bibnamefont {Riley}}, \bibinfo {author} {\bibfnamefont {A.~L.}\
  \bibnamefont {Watts}}, \bibinfo {author} {\bibfnamefont {J.~M.}\ \bibnamefont
  {Lattimer}}, \ and\ \bibinfo {author} {\bibfnamefont {W.~C.~G.}\ \bibnamefont
  {Ho}},\ }\href@noop {} {\  (\bibinfo {year} {2021})},\ \Eprint
  {http://arxiv.org/abs/2105.06981} {arXiv:2105.06981 [astro-ph.HE]}
  \BibitemShut {NoStop}%
\bibitem [{\citenamefont {Landry}\ \emph {et~al.}(2020)\citenamefont {Landry},
  \citenamefont {Essick},\ and\ \citenamefont
  {Chatziioannou}}]{Landry:2020vaw}%
  \BibitemOpen
  \bibfield  {author} {\bibinfo {author} {\bibfnamefont {P.}~\bibnamefont
  {Landry}}, \bibinfo {author} {\bibfnamefont {R.}~\bibnamefont {Essick}}, \
  and\ \bibinfo {author} {\bibfnamefont {K.}~\bibnamefont {Chatziioannou}},\
  }\href {\doibase 10.1103/PhysRevD.101.123007} {\bibfield  {journal} {\bibinfo
   {journal} {Phys. Rev. D}\ }\textbf {\bibinfo {volume} {101}},\ \bibinfo
  {pages} {123007} (\bibinfo {year} {2020})},\ \Eprint
  {http://arxiv.org/abs/2003.04880} {arXiv:2003.04880 [astro-ph.HE]}
  \BibitemShut {NoStop}%
\bibitem [{\citenamefont {Legred}\ \emph {et~al.}(2021)\citenamefont {Legred},
  \citenamefont {Chatziioannou}, \citenamefont {Essick}, \citenamefont {Han},\
  and\ \citenamefont {Landry}}]{Legred:2021}%
  \BibitemOpen
  \bibfield  {author} {\bibinfo {author} {\bibfnamefont {I.}~\bibnamefont
  {Legred}}, \bibinfo {author} {\bibfnamefont {K.}~\bibnamefont
  {Chatziioannou}}, \bibinfo {author} {\bibfnamefont {R.}~\bibnamefont
  {Essick}}, \bibinfo {author} {\bibfnamefont {S.}~\bibnamefont {Han}}, \ and\
  \bibinfo {author} {\bibfnamefont {P.}~\bibnamefont {Landry}},\ }\href
  {\doibase 10.1103/PhysRevD.104.063003} {\bibfield  {journal} {\bibinfo
  {journal} {Phys. Rev. D}\ }\textbf {\bibinfo {volume} {104}},\ \bibinfo
  {pages} {063003} (\bibinfo {year} {2021})}\BibitemShut {NoStop}%
\bibitem [{\citenamefont {{Drischler}}\ \emph {et~al.}(2016)\citenamefont
  {{Drischler}}, \citenamefont {{Carbone}}, \citenamefont {{Hebeler}},\ and\
  \citenamefont {{Schwenk}}}]{DrischlerCarbone2016}%
  \BibitemOpen
  \bibfield  {author} {\bibinfo {author} {\bibfnamefont {C.}~\bibnamefont
  {{Drischler}}}, \bibinfo {author} {\bibfnamefont {A.}~\bibnamefont
  {{Carbone}}}, \bibinfo {author} {\bibfnamefont {K.}~\bibnamefont
  {{Hebeler}}}, \ and\ \bibinfo {author} {\bibfnamefont {A.}~\bibnamefont
  {{Schwenk}}},\ }\href {\doibase 10.1103/PhysRevC.94.054307} {\bibfield
  {journal} {\bibinfo  {journal} {\prc}\ }\textbf {\bibinfo {volume} {94}},\
  \bibinfo {eid} {054307} (\bibinfo {year} {2016})},\ \Eprint
  {http://arxiv.org/abs/1608.05615} {arXiv:1608.05615 [nucl-th]} \BibitemShut
  {NoStop}%
\bibitem [{\citenamefont {Ant\'on}\ \emph {et~al.}(2006)\citenamefont
  {Ant\'on}, \citenamefont {Zanotti}, \citenamefont {Miralles}, \citenamefont
  {Mart\'i}, \citenamefont {Ib\'a\~nez}, \citenamefont {Font},\ and\
  \citenamefont {Pons}}]{2006ApJ...637..296A}%
  \BibitemOpen
  \bibfield  {author} {\bibinfo {author} {\bibfnamefont {L.}~\bibnamefont
  {Ant\'on}}, \bibinfo {author} {\bibfnamefont {O.}~\bibnamefont {Zanotti}},
  \bibinfo {author} {\bibfnamefont {J.~A.}\ \bibnamefont {Miralles}}, \bibinfo
  {author} {\bibfnamefont {J.~M.}\ \bibnamefont {Mart\'i}}, \bibinfo {author}
  {\bibfnamefont {J.~M.}\ \bibnamefont {Ib\'a\~nez}}, \bibinfo {author}
  {\bibfnamefont {J.~A.}\ \bibnamefont {Font}}, \ and\ \bibinfo {author}
  {\bibfnamefont {J.~A.}\ \bibnamefont {Pons}},\ }\href {\doibase
  10.1086/498238} {\bibfield  {journal} {\bibinfo  {journal} {The Astrophysical
  Journal}\ }\textbf {\bibinfo {volume} {637}},\ \bibinfo {pages} {296}
  (\bibinfo {year} {2006})}\BibitemShut {NoStop}%
\bibitem [{\citenamefont {Font}(2008)}]{Font:2008fka}%
  \BibitemOpen
  \bibfield  {author} {\bibinfo {author} {\bibfnamefont {J.~A.}\ \bibnamefont
  {Font}},\ }\href {\doibase 10.12942/lrr-2008-7} {\bibfield  {journal}
  {\bibinfo  {journal} {Living Rev. Rel.}\ }\textbf {\bibinfo {volume} {11}},\
  \bibinfo {pages} {7} (\bibinfo {year} {2008})}\BibitemShut {NoStop}%
\bibitem [{\citenamefont {Baumgarte}\ and\ \citenamefont
  {Shapiro}(2010)}]{baumgarte_shapiro_2010}%
  \BibitemOpen
  \bibfield  {author} {\bibinfo {author} {\bibfnamefont {T.~W.}\ \bibnamefont
  {Baumgarte}}\ and\ \bibinfo {author} {\bibfnamefont {S.~L.}\ \bibnamefont
  {Shapiro}},\ }\href {\doibase 10.1017/CBO9781139193344} {\emph {\bibinfo
  {title} {Numerical Relativity: Solving Einstein's Equations on the
  Computer}}}\ (\bibinfo  {publisher} {Cambridge University Press},\ \bibinfo
  {year} {2010})\BibitemShut {NoStop}%
\bibitem [{\citenamefont {Baiotti}\ and\ \citenamefont
  {Rezzolla}(2017)}]{Baiotti:2016qnr}%
  \BibitemOpen
  \bibfield  {author} {\bibinfo {author} {\bibfnamefont {L.}~\bibnamefont
  {Baiotti}}\ and\ \bibinfo {author} {\bibfnamefont {L.}~\bibnamefont
  {Rezzolla}},\ }\href {\doibase 10.1088/1361-6633/aa67bb} {\bibfield
  {journal} {\bibinfo  {journal} {Rept. Prog. Phys.}\ }\textbf {\bibinfo
  {volume} {80}},\ \bibinfo {pages} {096901} (\bibinfo {year} {2017})},\
  \Eprint {http://arxiv.org/abs/1607.03540} {arXiv:1607.03540 [gr-qc]}
  \BibitemShut {NoStop}%
\bibitem [{\citenamefont {Radice}\ \emph {et~al.}(2020)\citenamefont {Radice},
  \citenamefont {Bernuzzi},\ and\ \citenamefont {Perego}}]{Radice:2020ddv}%
  \BibitemOpen
  \bibfield  {author} {\bibinfo {author} {\bibfnamefont {D.}~\bibnamefont
  {Radice}}, \bibinfo {author} {\bibfnamefont {S.}~\bibnamefont {Bernuzzi}}, \
  and\ \bibinfo {author} {\bibfnamefont {A.}~\bibnamefont {Perego}},\ }\href
  {\doibase 10.1146/annurev-nucl-013120-114541} {\bibfield  {journal} {\bibinfo
   {journal} {Ann. Rev. Nucl. Part. Sci.}\ }\textbf {\bibinfo {volume} {70}},\
  \bibinfo {pages} {95} (\bibinfo {year} {2020})},\ \Eprint
  {http://arxiv.org/abs/2002.03863} {arXiv:2002.03863 [astro-ph.HE]}
  \BibitemShut {NoStop}%
\bibitem [{\citenamefont {Foucart}(2020)}]{Foucart:2020ats}%
  \BibitemOpen
  \bibfield  {author} {\bibinfo {author} {\bibfnamefont {F.}~\bibnamefont
  {Foucart}},\ }\href {\doibase 10.3389/fspas.2020.00046} {\bibfield  {journal}
  {\bibinfo  {journal} {Front. Astron. Space Sci.}\ }\textbf {\bibinfo {volume}
  {7}},\ \bibinfo {pages} {46} (\bibinfo {year} {2020})},\ \Eprint
  {http://arxiv.org/abs/2006.10570} {arXiv:2006.10570 [astro-ph.HE]}
  \BibitemShut {NoStop}%
\bibitem [{\citenamefont {Kyutoku}\ \emph {et~al.}(2021)\citenamefont
  {Kyutoku}, \citenamefont {Shibata},\ and\ \citenamefont
  {Taniguchi}}]{Kyutoku:2021icp}%
  \BibitemOpen
  \bibfield  {author} {\bibinfo {author} {\bibfnamefont {K.}~\bibnamefont
  {Kyutoku}}, \bibinfo {author} {\bibfnamefont {M.}~\bibnamefont {Shibata}}, \
  and\ \bibinfo {author} {\bibfnamefont {K.}~\bibnamefont {Taniguchi}},\ }\href
  {\doibase 10.1007/s41114-021-00033-4} {\bibfield  {journal} {\bibinfo
  {journal} {Living Rev. Rel.}\ }\textbf {\bibinfo {volume} {24}},\ \bibinfo
  {pages} {5} (\bibinfo {year} {2021})},\ \Eprint
  {http://arxiv.org/abs/2110.06218} {arXiv:2110.06218 [astro-ph.HE]}
  \BibitemShut {NoStop}%
\bibitem [{\citenamefont {{Rezzolla}}\ and\ \citenamefont
  {{Zanotti}}(2013)}]{2013rehy.book.....R}%
  \BibitemOpen
  \bibfield  {author} {\bibinfo {author} {\bibfnamefont {L.}~\bibnamefont
  {{Rezzolla}}}\ and\ \bibinfo {author} {\bibfnamefont {O.}~\bibnamefont
  {{Zanotti}}},\ }\href@noop {} {\emph {\bibinfo {title} {{Relativistic
  Hydrodynamics}}}}\ (\bibinfo  {publisher} {Oxford University Press},\
  \bibinfo {year} {2013})\BibitemShut {NoStop}%
\bibitem [{\citenamefont {Margalit}\ and\ \citenamefont
  {Metzger}(2017)}]{Margalit:2017dij}%
  \BibitemOpen
  \bibfield  {author} {\bibinfo {author} {\bibfnamefont {B.}~\bibnamefont
  {Margalit}}\ and\ \bibinfo {author} {\bibfnamefont {B.~D.}\ \bibnamefont
  {Metzger}},\ }\href {\doibase 10.3847/2041-8213/aa991c} {\bibfield  {journal}
  {\bibinfo  {journal} {Astrophys. J.}\ }\textbf {\bibinfo {volume} {850}},\
  \bibinfo {pages} {L19} (\bibinfo {year} {2017})},\ \Eprint
  {http://arxiv.org/abs/1710.05938} {arXiv:1710.05938 [astro-ph.HE]}
  \BibitemShut {NoStop}%
\bibitem [{\citenamefont {Radice}\ and\ \citenamefont
  {Dai}(2019)}]{Radice:2018ozg}%
  \BibitemOpen
  \bibfield  {author} {\bibinfo {author} {\bibfnamefont {D.}~\bibnamefont
  {Radice}}\ and\ \bibinfo {author} {\bibfnamefont {L.}~\bibnamefont {Dai}},\
  }\href {\doibase 10.1140/epja/i2019-12716-4} {\bibfield  {journal} {\bibinfo
  {journal} {Eur. Phys. J. A}\ }\textbf {\bibinfo {volume} {55}},\ \bibinfo
  {pages} {50} (\bibinfo {year} {2019})},\ \Eprint
  {http://arxiv.org/abs/1810.12917} {arXiv:1810.12917 [astro-ph.HE]}
  \BibitemShut {NoStop}%
\bibitem [{\citenamefont {Shibata}\ \emph {et~al.}(2019)\citenamefont
  {Shibata}, \citenamefont {Zhou}, \citenamefont {Kiuchi},\ and\ \citenamefont
  {Fujibayashi}}]{Shibata:2019ctb}%
  \BibitemOpen
  \bibfield  {author} {\bibinfo {author} {\bibfnamefont {M.}~\bibnamefont
  {Shibata}}, \bibinfo {author} {\bibfnamefont {E.}~\bibnamefont {Zhou}},
  \bibinfo {author} {\bibfnamefont {K.}~\bibnamefont {Kiuchi}}, \ and\ \bibinfo
  {author} {\bibfnamefont {S.}~\bibnamefont {Fujibayashi}},\ }\href {\doibase
  10.1103/PhysRevD.100.023015} {\bibfield  {journal} {\bibinfo  {journal}
  {Phys. Rev. D}\ }\textbf {\bibinfo {volume} {100}},\ \bibinfo {pages}
  {023015} (\bibinfo {year} {2019})},\ \Eprint
  {http://arxiv.org/abs/1905.03656} {arXiv:1905.03656 [astro-ph.HE]}
  \BibitemShut {NoStop}%
\bibitem [{\citenamefont {K\"oppel}\ \emph {et~al.}(2019)\citenamefont
  {K\"oppel}, \citenamefont {Bovard},\ and\ \citenamefont
  {Rezzolla}}]{Koppel:2019pys}%
  \BibitemOpen
  \bibfield  {author} {\bibinfo {author} {\bibfnamefont {S.}~\bibnamefont
  {K\"oppel}}, \bibinfo {author} {\bibfnamefont {L.}~\bibnamefont {Bovard}}, \
  and\ \bibinfo {author} {\bibfnamefont {L.}~\bibnamefont {Rezzolla}},\ }\href
  {\doibase 10.3847/2041-8213/ab0210} {\bibfield  {journal} {\bibinfo
  {journal} {Astrophys. J. Lett.}\ }\textbf {\bibinfo {volume} {872}},\
  \bibinfo {pages} {L16} (\bibinfo {year} {2019})},\ \Eprint
  {http://arxiv.org/abs/1901.09977} {arXiv:1901.09977 [gr-qc]} \BibitemShut
  {NoStop}%
\bibitem [{\citenamefont {Annala}\ \emph {et~al.}(2022)\citenamefont {Annala},
  \citenamefont {Gorda}, \citenamefont {Katerini}, \citenamefont {Kurkela},
  \citenamefont {N\"attil\"a}, \citenamefont {Paschalidis},\ and\ \citenamefont
  {Vuorinen}}]{Annala:2021gom}%
  \BibitemOpen
  \bibfield  {author} {\bibinfo {author} {\bibfnamefont {E.}~\bibnamefont
  {Annala}}, \bibinfo {author} {\bibfnamefont {T.}~\bibnamefont {Gorda}},
  \bibinfo {author} {\bibfnamefont {E.}~\bibnamefont {Katerini}}, \bibinfo
  {author} {\bibfnamefont {A.}~\bibnamefont {Kurkela}}, \bibinfo {author}
  {\bibfnamefont {J.}~\bibnamefont {N\"attil\"a}}, \bibinfo {author}
  {\bibfnamefont {V.}~\bibnamefont {Paschalidis}}, \ and\ \bibinfo {author}
  {\bibfnamefont {A.}~\bibnamefont {Vuorinen}},\ }\href {\doibase
  10.1103/PhysRevX.12.011058} {\bibfield  {journal} {\bibinfo  {journal} {Phys.
  Rev. X}\ }\textbf {\bibinfo {volume} {12}},\ \bibinfo {pages} {011058}
  (\bibinfo {year} {2022})},\ \Eprint {http://arxiv.org/abs/2105.05132}
  {arXiv:2105.05132 [astro-ph.HE]} \BibitemShut {NoStop}%
\bibitem [{\citenamefont {Camilletti}\ \emph {et~al.}(2022)\citenamefont
  {Camilletti}, \citenamefont {Chiesa}, \citenamefont {Ricigliano},
  \citenamefont {Perego}, \citenamefont {Lippold}, \citenamefont {Padamata},
  \citenamefont {Bernuzzi}, \citenamefont {Radice}, \citenamefont {Logoteta},\
  and\ \citenamefont {Guercilena}}]{Camilletti:2022jms}%
  \BibitemOpen
  \bibfield  {author} {\bibinfo {author} {\bibfnamefont {A.}~\bibnamefont
  {Camilletti}}, \bibinfo {author} {\bibfnamefont {L.}~\bibnamefont {Chiesa}},
  \bibinfo {author} {\bibfnamefont {G.}~\bibnamefont {Ricigliano}}, \bibinfo
  {author} {\bibfnamefont {A.}~\bibnamefont {Perego}}, \bibinfo {author}
  {\bibfnamefont {L.~C.}\ \bibnamefont {Lippold}}, \bibinfo {author}
  {\bibfnamefont {S.}~\bibnamefont {Padamata}}, \bibinfo {author}
  {\bibfnamefont {S.}~\bibnamefont {Bernuzzi}}, \bibinfo {author}
  {\bibfnamefont {D.}~\bibnamefont {Radice}}, \bibinfo {author} {\bibfnamefont
  {D.}~\bibnamefont {Logoteta}}, \ and\ \bibinfo {author} {\bibfnamefont
  {F.~M.}\ \bibnamefont {Guercilena}},\ }\href {\doibase
  10.1093/mnras/stac2333} {\  (\bibinfo {year} {2022}),\
  10.1093/mnras/stac2333},\ \Eprint {http://arxiv.org/abs/2204.05336}
  {arXiv:2204.05336 [astro-ph.HE]} \BibitemShut {NoStop}%
\bibitem [{\citenamefont {Typel}\ \emph {et~al.}(2022)\citenamefont {Typel}
  \emph {et~al.}}]{Typel:2022lcx}%
  \BibitemOpen
  \bibfield  {author} {\bibinfo {author} {\bibfnamefont {S.}~\bibnamefont
  {Typel}} \emph {et~al.},\ }\href@noop {} {\  (\bibinfo {year} {2022})},\
  \Eprint {http://arxiv.org/abs/2203.03209} {arXiv:2203.03209 [astro-ph.HE]}
  \BibitemShut {NoStop}%
\bibitem [{\citenamefont {Siegel}\ \emph {et~al.}(2018)\citenamefont {Siegel},
  \citenamefont {M\"osta}, \citenamefont {Desai},\ and\ \citenamefont
  {Wu}}]{Siegel:2017sav}%
  \BibitemOpen
  \bibfield  {author} {\bibinfo {author} {\bibfnamefont {D.~M.}\ \bibnamefont
  {Siegel}}, \bibinfo {author} {\bibfnamefont {P.}~\bibnamefont {M\"osta}},
  \bibinfo {author} {\bibfnamefont {D.}~\bibnamefont {Desai}}, \ and\ \bibinfo
  {author} {\bibfnamefont {S.}~\bibnamefont {Wu}},\ }\href {\doibase
  10.3847/1538-4357/aabcc5} {\bibfield  {journal} {\bibinfo  {journal}
  {Astrophys. J.}\ }\textbf {\bibinfo {volume} {859}},\ \bibinfo {pages} {71}
  (\bibinfo {year} {2018})},\ \Eprint {http://arxiv.org/abs/1712.07538}
  {arXiv:1712.07538 [astro-ph.HE]} \BibitemShut {NoStop}%
\bibitem [{\citenamefont {Read}\ \emph
  {et~al.}(2009{\natexlab{a}})\citenamefont {Read}, \citenamefont {Lackey},
  \citenamefont {Owen},\ and\ \citenamefont {Friedman}}]{Read:2008iy}%
  \BibitemOpen
  \bibfield  {author} {\bibinfo {author} {\bibfnamefont {J.~S.}\ \bibnamefont
  {Read}}, \bibinfo {author} {\bibfnamefont {B.~D.}\ \bibnamefont {Lackey}},
  \bibinfo {author} {\bibfnamefont {B.~J.}\ \bibnamefont {Owen}}, \ and\
  \bibinfo {author} {\bibfnamefont {J.~L.}\ \bibnamefont {Friedman}},\ }\href
  {\doibase 10.1103/PhysRevD.79.124032} {\bibfield  {journal} {\bibinfo
  {journal} {Phys. Rev. D}\ }\textbf {\bibinfo {volume} {79}},\ \bibinfo
  {pages} {124032} (\bibinfo {year} {2009}{\natexlab{a}})},\ \Eprint
  {http://arxiv.org/abs/0812.2163} {arXiv:0812.2163 [astro-ph]} \BibitemShut
  {NoStop}%
\bibitem [{\citenamefont {Foucart}\ \emph {et~al.}(2019)\citenamefont
  {Foucart}, \citenamefont {Duez}, \citenamefont {Gudinas}, \citenamefont
  {Hebert}, \citenamefont {Kidder}, \citenamefont {Pfeiffer},\ and\
  \citenamefont {Scheel}}]{Foucart:2019yzo}%
  \BibitemOpen
  \bibfield  {author} {\bibinfo {author} {\bibfnamefont {F.}~\bibnamefont
  {Foucart}}, \bibinfo {author} {\bibfnamefont {M.~D.}\ \bibnamefont {Duez}},
  \bibinfo {author} {\bibfnamefont {A.}~\bibnamefont {Gudinas}}, \bibinfo
  {author} {\bibfnamefont {F.}~\bibnamefont {Hebert}}, \bibinfo {author}
  {\bibfnamefont {L.~E.}\ \bibnamefont {Kidder}}, \bibinfo {author}
  {\bibfnamefont {H.~P.}\ \bibnamefont {Pfeiffer}}, \ and\ \bibinfo {author}
  {\bibfnamefont {M.~A.}\ \bibnamefont {Scheel}},\ }\href {\doibase
  10.1103/PhysRevD.100.104048} {\bibfield  {journal} {\bibinfo  {journal}
  {Phys. Rev. D}\ }\textbf {\bibinfo {volume} {100}},\ \bibinfo {pages}
  {104048} (\bibinfo {year} {2019})},\ \Eprint
  {http://arxiv.org/abs/1908.05277} {arXiv:1908.05277 [gr-qc]} \BibitemShut
  {NoStop}%
\bibitem [{\citenamefont {Raithel}\ and\ \citenamefont
  {Paschalidis}(2022)}]{Raithel:2022san}%
  \BibitemOpen
  \bibfield  {author} {\bibinfo {author} {\bibfnamefont {C.~A.}\ \bibnamefont
  {Raithel}}\ and\ \bibinfo {author} {\bibfnamefont {V.}~\bibnamefont
  {Paschalidis}},\ }\href {\doibase 10.1103/PhysRevD.106.023015} {\bibfield
  {journal} {\bibinfo  {journal} {Phys. Rev. D}\ }\textbf {\bibinfo {volume}
  {106}},\ \bibinfo {pages} {023015} (\bibinfo {year} {2022})},\ \Eprint
  {http://arxiv.org/abs/2204.00698} {arXiv:2204.00698 [gr-qc]} \BibitemShut
  {NoStop}%
\bibitem [{\citenamefont {Lindblom}(2010)}]{Lindblom:2010bb}%
  \BibitemOpen
  \bibfield  {author} {\bibinfo {author} {\bibfnamefont {L.}~\bibnamefont
  {Lindblom}},\ }\href {\doibase 10.1103/PhysRevD.82.103011} {\bibfield
  {journal} {\bibinfo  {journal} {Phys. Rev. D}\ }\textbf {\bibinfo {volume}
  {82}},\ \bibinfo {pages} {103011} (\bibinfo {year} {2010})},\ \Eprint
  {http://arxiv.org/abs/1009.0738} {arXiv:1009.0738 [astro-ph.HE]} \BibitemShut
  {NoStop}%
\bibitem [{\citenamefont {Greif}\ \emph {et~al.}(2019)\citenamefont {Greif},
  \citenamefont {Raaijmakers}, \citenamefont {Hebeler}, \citenamefont
  {Schwenk},\ and\ \citenamefont {Watts}}]{Greif:2018njt}%
  \BibitemOpen
  \bibfield  {author} {\bibinfo {author} {\bibfnamefont {S.~K.}\ \bibnamefont
  {Greif}}, \bibinfo {author} {\bibfnamefont {G.}~\bibnamefont {Raaijmakers}},
  \bibinfo {author} {\bibfnamefont {K.}~\bibnamefont {Hebeler}}, \bibinfo
  {author} {\bibfnamefont {A.}~\bibnamefont {Schwenk}}, \ and\ \bibinfo
  {author} {\bibfnamefont {A.~L.}\ \bibnamefont {Watts}},\ }\href {\doibase
  10.1093/mnras/stz654} {\bibfield  {journal} {\bibinfo  {journal} {Mon. Not.
  Roy. Astron. Soc.}\ }\textbf {\bibinfo {volume} {485}},\ \bibinfo {pages}
  {5363} (\bibinfo {year} {2019})},\ \Eprint {http://arxiv.org/abs/1812.08188}
  {arXiv:1812.08188 [astro-ph.HE]} \BibitemShut {NoStop}%
\bibitem [{\citenamefont {{Lindblom}}(1992)}]{Lindblom:1992}%
  \BibitemOpen
  \bibfield  {author} {\bibinfo {author} {\bibfnamefont {L.}~\bibnamefont
  {{Lindblom}}},\ }\href {\doibase 10.1086/171882} {\bibfield  {journal}
  {\bibinfo  {journal} {\apj}\ }\textbf {\bibinfo {volume} {398}},\ \bibinfo
  {pages} {569} (\bibinfo {year} {1992})}\BibitemShut {NoStop}%
\bibitem [{\citenamefont {Deppe}\ \emph
  {et~al.}(2022{\natexlab{a}})\citenamefont {Deppe}, \citenamefont {Throwe},
  \citenamefont {Kidder}, \citenamefont {Vu}, \citenamefont {H\'ebert},
  \citenamefont {Moxon}, \citenamefont {Armaza}, \citenamefont {Bonilla},
  \citenamefont {Kim}, \citenamefont {Kumar}, \citenamefont {Lovelace},
  \citenamefont {Macedo}, \citenamefont {Nelli}, \citenamefont {O'Shea},
  \citenamefont {Pfeiffer}, \citenamefont {Scheel}, \citenamefont {Teukolsky},
  \citenamefont {Wittek} \emph {et~al.}}]{spectrecode}%
  \BibitemOpen
  \bibfield  {author} {\bibinfo {author} {\bibfnamefont {N.}~\bibnamefont
  {Deppe}}, \bibinfo {author} {\bibfnamefont {W.}~\bibnamefont {Throwe}},
  \bibinfo {author} {\bibfnamefont {L.~E.}\ \bibnamefont {Kidder}}, \bibinfo
  {author} {\bibfnamefont {N.~L.}\ \bibnamefont {Vu}}, \bibinfo {author}
  {\bibfnamefont {F.}~\bibnamefont {H\'ebert}}, \bibinfo {author}
  {\bibfnamefont {J.}~\bibnamefont {Moxon}}, \bibinfo {author} {\bibfnamefont
  {C.}~\bibnamefont {Armaza}}, \bibinfo {author} {\bibfnamefont {G.~S.}\
  \bibnamefont {Bonilla}}, \bibinfo {author} {\bibfnamefont {Y.}~\bibnamefont
  {Kim}}, \bibinfo {author} {\bibfnamefont {P.}~\bibnamefont {Kumar}}, \bibinfo
  {author} {\bibfnamefont {G.}~\bibnamefont {Lovelace}}, \bibinfo {author}
  {\bibfnamefont {A.}~\bibnamefont {Macedo}}, \bibinfo {author} {\bibfnamefont
  {K.~C.}\ \bibnamefont {Nelli}}, \bibinfo {author} {\bibfnamefont
  {E.}~\bibnamefont {O'Shea}}, \bibinfo {author} {\bibfnamefont {H.~P.}\
  \bibnamefont {Pfeiffer}}, \bibinfo {author} {\bibfnamefont {M.~A.}\
  \bibnamefont {Scheel}}, \bibinfo {author} {\bibfnamefont {S.~A.}\
  \bibnamefont {Teukolsky}}, \bibinfo {author} {\bibfnamefont {N.~A.}\
  \bibnamefont {Wittek}},  \emph {et~al.},\ }\href {\doibase
  10.5281/zenodo.6949324} {\enquote {\bibinfo {title} {\texttt{SpECTRE
  v2022.08.01}},}\ }\bibinfo {howpublished}
  {\href{https://doi.org/10.5281/zenodo.6949324}{10.5281/zenodo.6949324}}
  (\bibinfo {year} {2022}{\natexlab{a}})\BibitemShut {NoStop}%
\bibitem [{\citenamefont {Kidder}\ \emph {et~al.}(2017)\citenamefont {Kidder}
  \emph {et~al.}}]{Kidder:2016hev}%
  \BibitemOpen
  \bibfield  {author} {\bibinfo {author} {\bibfnamefont {L.~E.}\ \bibnamefont
  {Kidder}} \emph {et~al.},\ }\href {\doibase 10.1016/j.jcp.2016.12.059}
  {\bibfield  {journal} {\bibinfo  {journal} {J. Comput. Phys.}\ }\textbf
  {\bibinfo {volume} {335}},\ \bibinfo {pages} {84} (\bibinfo {year} {2017})},\
  \Eprint {http://arxiv.org/abs/1609.00098} {arXiv:1609.00098 [astro-ph.HE]}
  \BibitemShut {NoStop}%
\bibitem [{\citenamefont {Kale}\ \emph {et~al.}(2020)\citenamefont {Kale},
  \citenamefont {Acun}, \citenamefont {Bak}, \citenamefont {Becker},
  \citenamefont {Bhandarkar}, \citenamefont {Bhat}, \citenamefont {Bhatele},
  \citenamefont {Bohm}, \citenamefont {Bordage}, \citenamefont {Brunner},
  \citenamefont {Buch}, \citenamefont {Chakravorty}, \citenamefont
  {Chandrasekar}, \citenamefont {Choi}, \citenamefont {Denardo}, \citenamefont
  {DeSouza}, \citenamefont {Diener}, \citenamefont {Dokania}, \citenamefont
  {Dooley}, \citenamefont {Fenton}, \citenamefont {Galvez}, \citenamefont
  {Gioachin}, \citenamefont {Gupta}, \citenamefont {Gupta}, \citenamefont
  {Gupta}, \citenamefont {Gursoy}, \citenamefont {Harsh}, \citenamefont {Hu},
  \citenamefont {Huang}, \citenamefont {Jagathesan}, \citenamefont {Jain},
  \citenamefont {Jetley}, \citenamefont {Jindal}, \citenamefont {Kanakagiri},
  \citenamefont {Koenig}, \citenamefont {Krishnan}, \citenamefont {Kumar},
  \citenamefont {Kunzman}, \citenamefont {Lang}, \citenamefont {Langer},
  \citenamefont {Lawlor}, \citenamefont {Lee}, \citenamefont {Lifflander},
  \citenamefont {Mahesh}, \citenamefont {Mendes}, \citenamefont {Menon},
  \citenamefont {Mei}, \citenamefont {Meneses}, \citenamefont {Mikida},
  \citenamefont {Miller}, \citenamefont {Mokos}, \citenamefont {Narayanan},
  \citenamefont {Ni}, \citenamefont {Nomura}, \citenamefont {Paranjpye},
  \citenamefont {Ramachandran}, \citenamefont {Ramkumar}, \citenamefont
  {Ramos}, \citenamefont {Robson}, \citenamefont {Saboo}, \citenamefont
  {Saletore}, \citenamefont {Sarood}, \citenamefont {Senthil}, \citenamefont
  {Shah}, \citenamefont {Shu}, \citenamefont {Sinha}, \citenamefont {Sun},
  \citenamefont {Sura}, \citenamefont {Totoni}, \citenamefont {Varadarajan},
  \citenamefont {Venkataraman}, \citenamefont {Wang}, \citenamefont
  {Wesolowski}, \citenamefont {White}, \citenamefont {Wilmarth}, \citenamefont
  {Wright}, \citenamefont {Yelon},\ and\ \citenamefont
  {Zheng}}]{laxmikant_kale_2020_3972617}%
  \BibitemOpen
  \bibfield  {author} {\bibinfo {author} {\bibfnamefont {L.}~\bibnamefont
  {Kale}}, \bibinfo {author} {\bibfnamefont {B.}~\bibnamefont {Acun}}, \bibinfo
  {author} {\bibfnamefont {S.}~\bibnamefont {Bak}}, \bibinfo {author}
  {\bibfnamefont {A.}~\bibnamefont {Becker}}, \bibinfo {author} {\bibfnamefont
  {M.}~\bibnamefont {Bhandarkar}}, \bibinfo {author} {\bibfnamefont
  {N.}~\bibnamefont {Bhat}}, \bibinfo {author} {\bibfnamefont {A.}~\bibnamefont
  {Bhatele}}, \bibinfo {author} {\bibfnamefont {E.}~\bibnamefont {Bohm}},
  \bibinfo {author} {\bibfnamefont {C.}~\bibnamefont {Bordage}}, \bibinfo
  {author} {\bibfnamefont {R.}~\bibnamefont {Brunner}}, \bibinfo {author}
  {\bibfnamefont {R.}~\bibnamefont {Buch}}, \bibinfo {author} {\bibfnamefont
  {S.}~\bibnamefont {Chakravorty}}, \bibinfo {author} {\bibfnamefont
  {K.}~\bibnamefont {Chandrasekar}}, \bibinfo {author} {\bibfnamefont
  {J.}~\bibnamefont {Choi}}, \bibinfo {author} {\bibfnamefont {M.}~\bibnamefont
  {Denardo}}, \bibinfo {author} {\bibfnamefont {J.}~\bibnamefont {DeSouza}},
  \bibinfo {author} {\bibfnamefont {M.}~\bibnamefont {Diener}}, \bibinfo
  {author} {\bibfnamefont {H.}~\bibnamefont {Dokania}}, \bibinfo {author}
  {\bibfnamefont {I.}~\bibnamefont {Dooley}}, \bibinfo {author} {\bibfnamefont
  {W.}~\bibnamefont {Fenton}}, \bibinfo {author} {\bibfnamefont
  {J.}~\bibnamefont {Galvez}}, \bibinfo {author} {\bibfnamefont
  {F.}~\bibnamefont {Gioachin}}, \bibinfo {author} {\bibfnamefont
  {A.}~\bibnamefont {Gupta}}, \bibinfo {author} {\bibfnamefont
  {G.}~\bibnamefont {Gupta}}, \bibinfo {author} {\bibfnamefont
  {M.}~\bibnamefont {Gupta}}, \bibinfo {author} {\bibfnamefont
  {A.}~\bibnamefont {Gursoy}}, \bibinfo {author} {\bibfnamefont
  {V.}~\bibnamefont {Harsh}}, \bibinfo {author} {\bibfnamefont
  {F.}~\bibnamefont {Hu}}, \bibinfo {author} {\bibfnamefont {C.}~\bibnamefont
  {Huang}}, \bibinfo {author} {\bibfnamefont {N.}~\bibnamefont {Jagathesan}},
  \bibinfo {author} {\bibfnamefont {N.}~\bibnamefont {Jain}}, \bibinfo {author}
  {\bibfnamefont {P.}~\bibnamefont {Jetley}}, \bibinfo {author} {\bibfnamefont
  {P.}~\bibnamefont {Jindal}}, \bibinfo {author} {\bibfnamefont
  {R.}~\bibnamefont {Kanakagiri}}, \bibinfo {author} {\bibfnamefont
  {G.}~\bibnamefont {Koenig}}, \bibinfo {author} {\bibfnamefont
  {S.}~\bibnamefont {Krishnan}}, \bibinfo {author} {\bibfnamefont
  {S.}~\bibnamefont {Kumar}}, \bibinfo {author} {\bibfnamefont
  {D.}~\bibnamefont {Kunzman}}, \bibinfo {author} {\bibfnamefont
  {M.}~\bibnamefont {Lang}}, \bibinfo {author} {\bibfnamefont {A.}~\bibnamefont
  {Langer}}, \bibinfo {author} {\bibfnamefont {O.}~\bibnamefont {Lawlor}},
  \bibinfo {author} {\bibfnamefont {C.~W.}\ \bibnamefont {Lee}}, \bibinfo
  {author} {\bibfnamefont {J.}~\bibnamefont {Lifflander}}, \bibinfo {author}
  {\bibfnamefont {K.}~\bibnamefont {Mahesh}}, \bibinfo {author} {\bibfnamefont
  {C.}~\bibnamefont {Mendes}}, \bibinfo {author} {\bibfnamefont
  {H.}~\bibnamefont {Menon}}, \bibinfo {author} {\bibfnamefont
  {C.}~\bibnamefont {Mei}}, \bibinfo {author} {\bibfnamefont {E.}~\bibnamefont
  {Meneses}}, \bibinfo {author} {\bibfnamefont {E.}~\bibnamefont {Mikida}},
  \bibinfo {author} {\bibfnamefont {P.}~\bibnamefont {Miller}}, \bibinfo
  {author} {\bibfnamefont {R.}~\bibnamefont {Mokos}}, \bibinfo {author}
  {\bibfnamefont {V.}~\bibnamefont {Narayanan}}, \bibinfo {author}
  {\bibfnamefont {X.}~\bibnamefont {Ni}}, \bibinfo {author} {\bibfnamefont
  {K.}~\bibnamefont {Nomura}}, \bibinfo {author} {\bibfnamefont
  {S.}~\bibnamefont {Paranjpye}}, \bibinfo {author} {\bibfnamefont
  {P.}~\bibnamefont {Ramachandran}}, \bibinfo {author} {\bibfnamefont
  {B.}~\bibnamefont {Ramkumar}}, \bibinfo {author} {\bibfnamefont
  {E.}~\bibnamefont {Ramos}}, \bibinfo {author} {\bibfnamefont
  {M.}~\bibnamefont {Robson}}, \bibinfo {author} {\bibfnamefont
  {N.}~\bibnamefont {Saboo}}, \bibinfo {author} {\bibfnamefont
  {V.}~\bibnamefont {Saletore}}, \bibinfo {author} {\bibfnamefont
  {O.}~\bibnamefont {Sarood}}, \bibinfo {author} {\bibfnamefont
  {K.}~\bibnamefont {Senthil}}, \bibinfo {author} {\bibfnamefont
  {N.}~\bibnamefont {Shah}}, \bibinfo {author} {\bibfnamefont {W.}~\bibnamefont
  {Shu}}, \bibinfo {author} {\bibfnamefont {A.~B.}\ \bibnamefont {Sinha}},
  \bibinfo {author} {\bibfnamefont {Y.}~\bibnamefont {Sun}}, \bibinfo {author}
  {\bibfnamefont {Z.}~\bibnamefont {Sura}}, \bibinfo {author} {\bibfnamefont
  {E.}~\bibnamefont {Totoni}}, \bibinfo {author} {\bibfnamefont
  {K.}~\bibnamefont {Varadarajan}}, \bibinfo {author} {\bibfnamefont
  {R.}~\bibnamefont {Venkataraman}}, \bibinfo {author} {\bibfnamefont
  {J.}~\bibnamefont {Wang}}, \bibinfo {author} {\bibfnamefont {L.}~\bibnamefont
  {Wesolowski}}, \bibinfo {author} {\bibfnamefont {S.}~\bibnamefont {White}},
  \bibinfo {author} {\bibfnamefont {T.}~\bibnamefont {Wilmarth}}, \bibinfo
  {author} {\bibfnamefont {J.}~\bibnamefont {Wright}}, \bibinfo {author}
  {\bibfnamefont {J.}~\bibnamefont {Yelon}}, \ and\ \bibinfo {author}
  {\bibfnamefont {G.}~\bibnamefont {Zheng}},\ }\href {\doibase
  10.5281/zenodo.3972617} {\enquote {\bibinfo {title} {Uiuc-ppl/charm: Charm++
  version 6.10.2},}\ } (\bibinfo {year} {2020})\BibitemShut {NoStop}%
\bibitem [{\citenamefont {{Cowling}}(1941)}]{Cowling:1941}%
  \BibitemOpen
  \bibfield  {author} {\bibinfo {author} {\bibfnamefont {T.~G.}\ \bibnamefont
  {{Cowling}}},\ }\href {\doibase 10.1093/mnras/101.8.367} {\bibfield
  {journal} {\bibinfo  {journal} {mnras}\ }\textbf {\bibinfo {volume} {101}},\
  \bibinfo {pages} {367} (\bibinfo {year} {1941})}\BibitemShut {NoStop}%
\bibitem [{\citenamefont {Deppe}\ \emph
  {et~al.}(2022{\natexlab{b}})\citenamefont {Deppe}, \citenamefont
  {H{\'e}bert}, \citenamefont {Kidder},\ and\ \citenamefont
  {Teukolsky}}]{Deppe2022method}%
  \BibitemOpen
  \bibfield  {author} {\bibinfo {author} {\bibfnamefont {N.}~\bibnamefont
  {Deppe}}, \bibinfo {author} {\bibfnamefont {F.}~\bibnamefont {H{\'e}bert}},
  \bibinfo {author} {\bibfnamefont {L.~E.}\ \bibnamefont {Kidder}}, \ and\
  \bibinfo {author} {\bibfnamefont {S.~A.}\ \bibnamefont {Teukolsky}},\ }\href
  {\doibase 10.1088/1361-6382/ac8864} {\bibfield  {journal} {\bibinfo
  {journal} {Classical and Quantum Gravity}\ }\textbf {\bibinfo {volume}
  {39}},\ \bibinfo {pages} {195001} (\bibinfo {year}
  {2022}{\natexlab{b}})}\BibitemShut {NoStop}%
\bibitem [{\citenamefont {Deppe}\ \emph
  {et~al.}(2022{\natexlab{c}})\citenamefont {Deppe} \emph
  {et~al.}}]{Deppe:2021bhi}%
  \BibitemOpen
  \bibfield  {author} {\bibinfo {author} {\bibfnamefont {N.}~\bibnamefont
  {Deppe}} \emph {et~al.},\ }\href {\doibase 10.1103/PhysRevD.105.123031}
  {\bibfield  {journal} {\bibinfo  {journal} {Phys. Rev. D}\ }\textbf {\bibinfo
  {volume} {105}},\ \bibinfo {pages} {123031} (\bibinfo {year}
  {2022}{\natexlab{c}})},\ \Eprint {http://arxiv.org/abs/2109.12033}
  {arXiv:2109.12033 [gr-qc]} \BibitemShut {NoStop}%
\bibitem [{\citenamefont {Lindblom}\ and\ \citenamefont
  {Indik}(2014)}]{Lindblom:2013kra}%
  \BibitemOpen
  \bibfield  {author} {\bibinfo {author} {\bibfnamefont {L.}~\bibnamefont
  {Lindblom}}\ and\ \bibinfo {author} {\bibfnamefont {N.~M.}\ \bibnamefont
  {Indik}},\ }\href {\doibase 10.1103/PhysRevD.89.064003,
  10.1103/PhysRevD.93.129903} {\bibfield  {journal} {\bibinfo  {journal} {Phys.
  Rev.}\ }\textbf {\bibinfo {volume} {D89}},\ \bibinfo {pages} {064003}
  (\bibinfo {year} {2014})},\ \bibinfo {note} {[Erratum: Phys.
  Rev.D93,no.12,129903(2016)]},\ \Eprint {http://arxiv.org/abs/1310.0803}
  {arXiv:1310.0803 [astro-ph.HE]} \BibitemShut {NoStop}%
\bibitem [{\citenamefont {Han}\ and\ \citenamefont
  {Steiner}(2019)}]{Han:2018mtj}%
  \BibitemOpen
  \bibfield  {author} {\bibinfo {author} {\bibfnamefont {S.}~\bibnamefont
  {Han}}\ and\ \bibinfo {author} {\bibfnamefont {A.~W.}\ \bibnamefont
  {Steiner}},\ }\href {\doibase 10.1103/PhysRevD.99.083014} {\bibfield
  {journal} {\bibinfo  {journal} {Phys. Rev. D}\ }\textbf {\bibinfo {volume}
  {99}},\ \bibinfo {pages} {083014} (\bibinfo {year} {2019})},\ \Eprint
  {http://arxiv.org/abs/1810.10967} {arXiv:1810.10967 [nucl-th]} \BibitemShut
  {NoStop}%
\bibitem [{\citenamefont {Pang}\ \emph {et~al.}(2020)\citenamefont {Pang},
  \citenamefont {Dietrich}, \citenamefont {Tews},\ and\ \citenamefont {Van
  Den~Broeck}}]{Pang:2020ilf}%
  \BibitemOpen
  \bibfield  {author} {\bibinfo {author} {\bibfnamefont {P.~T.~H.}\
  \bibnamefont {Pang}}, \bibinfo {author} {\bibfnamefont {T.}~\bibnamefont
  {Dietrich}}, \bibinfo {author} {\bibfnamefont {I.}~\bibnamefont {Tews}}, \
  and\ \bibinfo {author} {\bibfnamefont {C.}~\bibnamefont {Van Den~Broeck}},\
  }\href {\doibase 10.1103/PhysRevResearch.2.033514} {\bibfield  {journal}
  {\bibinfo  {journal} {Phys. Rev. Res.}\ }\textbf {\bibinfo {volume} {2}},\
  \bibinfo {pages} {033514} (\bibinfo {year} {2020})},\ \Eprint
  {http://arxiv.org/abs/2006.14936} {arXiv:2006.14936 [astro-ph.HE]}
  \BibitemShut {NoStop}%
\bibitem [{\citenamefont {Wysocki}\ \emph {et~al.}(2020)\citenamefont
  {Wysocki}, \citenamefont {O'Shaughnessy}, \citenamefont {Wade},\ and\
  \citenamefont {Lange}}]{Wysocki:2020myz}%
  \BibitemOpen
  \bibfield  {author} {\bibinfo {author} {\bibfnamefont {D.}~\bibnamefont
  {Wysocki}}, \bibinfo {author} {\bibfnamefont {R.}~\bibnamefont
  {O'Shaughnessy}}, \bibinfo {author} {\bibfnamefont {L.}~\bibnamefont {Wade}},
  \ and\ \bibinfo {author} {\bibfnamefont {J.}~\bibnamefont {Lange}},\
  }\href@noop {} {\  (\bibinfo {year} {2020})},\ \Eprint
  {http://arxiv.org/abs/2001.01747} {arXiv:2001.01747 [gr-qc]} \BibitemShut
  {NoStop}%
\bibitem [{\citenamefont {{{\"O}zel}}\ and\ \citenamefont
  {{Psaltis}}(2009)}]{Ozel:2009}%
  \BibitemOpen
  \bibfield  {author} {\bibinfo {author} {\bibfnamefont {F.}~\bibnamefont
  {{{\"O}zel}}}\ and\ \bibinfo {author} {\bibfnamefont {D.}~\bibnamefont
  {{Psaltis}}},\ }\href {\doibase 10.1103/PhysRevD.80.103003} {\bibfield
  {journal} {\bibinfo  {journal} {\prd}\ }\textbf {\bibinfo {volume} {80}},\
  \bibinfo {eid} {103003} (\bibinfo {year} {2009})},\ \Eprint
  {http://arxiv.org/abs/0905.1959} {arXiv:0905.1959 [astro-ph.HE]} \BibitemShut
  {NoStop}%
\bibitem [{\citenamefont {Blackman}\ \emph {et~al.}(2015)\citenamefont
  {Blackman}, \citenamefont {Field}, \citenamefont {Galley}, \citenamefont
  {Szil\'agyi}, \citenamefont {Scheel}, \citenamefont {Tiglio},\ and\
  \citenamefont {Hemberger}}]{Blackman:2015pia}%
  \BibitemOpen
  \bibfield  {author} {\bibinfo {author} {\bibfnamefont {J.}~\bibnamefont
  {Blackman}}, \bibinfo {author} {\bibfnamefont {S.~E.}\ \bibnamefont {Field}},
  \bibinfo {author} {\bibfnamefont {C.~R.}\ \bibnamefont {Galley}}, \bibinfo
  {author} {\bibfnamefont {B.}~\bibnamefont {Szil\'agyi}}, \bibinfo {author}
  {\bibfnamefont {M.~A.}\ \bibnamefont {Scheel}}, \bibinfo {author}
  {\bibfnamefont {M.}~\bibnamefont {Tiglio}}, \ and\ \bibinfo {author}
  {\bibfnamefont {D.~A.}\ \bibnamefont {Hemberger}},\ }\href {\doibase
  10.1103/PhysRevLett.115.121102} {\bibfield  {journal} {\bibinfo  {journal}
  {Phys. Rev. Lett.}\ }\textbf {\bibinfo {volume} {115}},\ \bibinfo {pages}
  {121102} (\bibinfo {year} {2015})},\ \Eprint
  {http://arxiv.org/abs/1502.07758} {arXiv:1502.07758 [gr-qc]} \BibitemShut
  {NoStop}%
\bibitem [{\citenamefont {Varma}\ \emph {et~al.}(2019)\citenamefont {Varma},
  \citenamefont {Field}, \citenamefont {Scheel}, \citenamefont {Blackman},
  \citenamefont {Kidder},\ and\ \citenamefont {Pfeiffer}}]{Varma:2018mmi}%
  \BibitemOpen
  \bibfield  {author} {\bibinfo {author} {\bibfnamefont {V.}~\bibnamefont
  {Varma}}, \bibinfo {author} {\bibfnamefont {S.~E.}\ \bibnamefont {Field}},
  \bibinfo {author} {\bibfnamefont {M.~A.}\ \bibnamefont {Scheel}}, \bibinfo
  {author} {\bibfnamefont {J.}~\bibnamefont {Blackman}}, \bibinfo {author}
  {\bibfnamefont {L.~E.}\ \bibnamefont {Kidder}}, \ and\ \bibinfo {author}
  {\bibfnamefont {H.~P.}\ \bibnamefont {Pfeiffer}},\ }\href {\doibase
  10.1103/PhysRevD.99.064045} {\bibfield  {journal} {\bibinfo  {journal} {Phys.
  Rev. D}\ }\textbf {\bibinfo {volume} {99}},\ \bibinfo {pages} {064045}
  (\bibinfo {year} {2019})},\ \Eprint {http://arxiv.org/abs/1812.07865}
  {arXiv:1812.07865 [gr-qc]} \BibitemShut {NoStop}%
\bibitem [{\citenamefont {Wijngaarden}\ \emph {et~al.}(2022)\citenamefont
  {Wijngaarden}, \citenamefont {Chatziioannou}, \citenamefont {Bauswein},
  \citenamefont {Clark},\ and\ \citenamefont {Cornish}}]{Wijngaarden:2022sah}%
  \BibitemOpen
  \bibfield  {author} {\bibinfo {author} {\bibfnamefont {M.}~\bibnamefont
  {Wijngaarden}}, \bibinfo {author} {\bibfnamefont {K.}~\bibnamefont
  {Chatziioannou}}, \bibinfo {author} {\bibfnamefont {A.}~\bibnamefont
  {Bauswein}}, \bibinfo {author} {\bibfnamefont {J.~A.}\ \bibnamefont {Clark}},
  \ and\ \bibinfo {author} {\bibfnamefont {N.~J.}\ \bibnamefont {Cornish}},\
  }\href {\doibase 10.1103/PhysRevD.105.104019} {\bibfield  {journal} {\bibinfo
   {journal} {Phys. Rev. D}\ }\textbf {\bibinfo {volume} {105}},\ \bibinfo
  {pages} {104019} (\bibinfo {year} {2022})},\ \Eprint
  {http://arxiv.org/abs/2202.09382} {arXiv:2202.09382 [gr-qc]} \BibitemShut
  {NoStop}%
\bibitem [{\citenamefont {Breschi}\ \emph {et~al.}(2022)\citenamefont
  {Breschi}, \citenamefont {Bernuzzi}, \citenamefont {Chakravarti},
  \citenamefont {Camilletti}, \citenamefont {Prakash},\ and\ \citenamefont
  {Perego}}]{Breschi:2022xnc}%
  \BibitemOpen
  \bibfield  {author} {\bibinfo {author} {\bibfnamefont {M.}~\bibnamefont
  {Breschi}}, \bibinfo {author} {\bibfnamefont {S.}~\bibnamefont {Bernuzzi}},
  \bibinfo {author} {\bibfnamefont {K.}~\bibnamefont {Chakravarti}}, \bibinfo
  {author} {\bibfnamefont {A.}~\bibnamefont {Camilletti}}, \bibinfo {author}
  {\bibfnamefont {A.}~\bibnamefont {Prakash}}, \ and\ \bibinfo {author}
  {\bibfnamefont {A.}~\bibnamefont {Perego}},\ }\href@noop {} {\  (\bibinfo
  {year} {2022})},\ \Eprint {http://arxiv.org/abs/2205.09112} {arXiv:2205.09112
  [gr-qc]} \BibitemShut {NoStop}%
\bibitem [{\citenamefont {Shibata}\ and\ \citenamefont
  {Uryu}(2000)}]{Shibata:1999wm}%
  \BibitemOpen
  \bibfield  {author} {\bibinfo {author} {\bibfnamefont {M.}~\bibnamefont
  {Shibata}}\ and\ \bibinfo {author} {\bibfnamefont {K.}~\bibnamefont {Uryu}},\
  }\href {\doibase 10.1103/PhysRevD.61.064001} {\bibfield  {journal} {\bibinfo
  {journal} {Phys. Rev. D}\ }\textbf {\bibinfo {volume} {61}},\ \bibinfo
  {pages} {064001} (\bibinfo {year} {2000})},\ \Eprint
  {http://arxiv.org/abs/gr-qc/9911058} {arXiv:gr-qc/9911058} \BibitemShut
  {NoStop}%
\bibitem [{\citenamefont {Etienne}\ \emph {et~al.}(2008)\citenamefont
  {Etienne}, \citenamefont {Faber}, \citenamefont {Liu}, \citenamefont
  {Shapiro}, \citenamefont {Taniguchi},\ and\ \citenamefont
  {Baumgarte}}]{Etienne:2007jg}%
  \BibitemOpen
  \bibfield  {author} {\bibinfo {author} {\bibfnamefont {Z.~B.}\ \bibnamefont
  {Etienne}}, \bibinfo {author} {\bibfnamefont {J.~A.}\ \bibnamefont {Faber}},
  \bibinfo {author} {\bibfnamefont {Y.~T.}\ \bibnamefont {Liu}}, \bibinfo
  {author} {\bibfnamefont {S.~L.}\ \bibnamefont {Shapiro}}, \bibinfo {author}
  {\bibfnamefont {K.}~\bibnamefont {Taniguchi}}, \ and\ \bibinfo {author}
  {\bibfnamefont {T.~W.}\ \bibnamefont {Baumgarte}},\ }\href {\doibase
  10.1103/PhysRevD.77.084002} {\bibfield  {journal} {\bibinfo  {journal} {Phys.
  Rev. D}\ }\textbf {\bibinfo {volume} {77}},\ \bibinfo {pages} {084002}
  (\bibinfo {year} {2008})},\ \Eprint {http://arxiv.org/abs/0712.2460}
  {arXiv:0712.2460 [astro-ph]} \BibitemShut {NoStop}%
\bibitem [{\citenamefont {Baiotti}\ \emph {et~al.}(2008)\citenamefont
  {Baiotti}, \citenamefont {Giacomazzo},\ and\ \citenamefont
  {Rezzolla}}]{Baiotti:2008ra}%
  \BibitemOpen
  \bibfield  {author} {\bibinfo {author} {\bibfnamefont {L.}~\bibnamefont
  {Baiotti}}, \bibinfo {author} {\bibfnamefont {B.}~\bibnamefont {Giacomazzo}},
  \ and\ \bibinfo {author} {\bibfnamefont {L.}~\bibnamefont {Rezzolla}},\
  }\href {\doibase 10.1103/PhysRevD.78.084033} {\bibfield  {journal} {\bibinfo
  {journal} {Phys. Rev. D}\ }\textbf {\bibinfo {volume} {78}},\ \bibinfo
  {pages} {084033} (\bibinfo {year} {2008})},\ \Eprint
  {http://arxiv.org/abs/0804.0594} {arXiv:0804.0594 [gr-qc]} \BibitemShut
  {NoStop}%
\bibitem [{\citenamefont {Duez}\ \emph {et~al.}(2008)\citenamefont {Duez},
  \citenamefont {Foucart}, \citenamefont {Kidder}, \citenamefont {Pfeiffer},
  \citenamefont {Scheel},\ and\ \citenamefont {Teukolsky}}]{Duez:2008rb}%
  \BibitemOpen
  \bibfield  {author} {\bibinfo {author} {\bibfnamefont {M.~D.}\ \bibnamefont
  {Duez}}, \bibinfo {author} {\bibfnamefont {F.}~\bibnamefont {Foucart}},
  \bibinfo {author} {\bibfnamefont {L.~E.}\ \bibnamefont {Kidder}}, \bibinfo
  {author} {\bibfnamefont {H.~P.}\ \bibnamefont {Pfeiffer}}, \bibinfo {author}
  {\bibfnamefont {M.~A.}\ \bibnamefont {Scheel}}, \ and\ \bibinfo {author}
  {\bibfnamefont {S.~A.}\ \bibnamefont {Teukolsky}},\ }\href {\doibase
  10.1103/PhysRevD.78.104015} {\bibfield  {journal} {\bibinfo  {journal} {Phys.
  Rev. D}\ }\textbf {\bibinfo {volume} {78}},\ \bibinfo {pages} {104015}
  (\bibinfo {year} {2008})},\ \Eprint {http://arxiv.org/abs/0809.0002}
  {arXiv:0809.0002 [gr-qc]} \BibitemShut {NoStop}%
\bibitem [{\citenamefont {Radice}\ \emph {et~al.}(2014)\citenamefont {Radice},
  \citenamefont {Rezzolla},\ and\ \citenamefont {Galeazzi}}]{Radice:2013hxh}%
  \BibitemOpen
  \bibfield  {author} {\bibinfo {author} {\bibfnamefont {D.}~\bibnamefont
  {Radice}}, \bibinfo {author} {\bibfnamefont {L.}~\bibnamefont {Rezzolla}}, \
  and\ \bibinfo {author} {\bibfnamefont {F.}~\bibnamefont {Galeazzi}},\ }\href
  {\doibase 10.1093/mnrasl/slt137} {\bibfield  {journal} {\bibinfo  {journal}
  {Mon. Not. Roy. Astron. Soc.}\ }\textbf {\bibinfo {volume} {437}},\ \bibinfo
  {pages} {L46} (\bibinfo {year} {2014})},\ \Eprint
  {http://arxiv.org/abs/1306.6052} {arXiv:1306.6052 [gr-qc]} \BibitemShut
  {NoStop}%
\bibitem [{\citenamefont {DeBuhr}\ \emph {et~al.}(2018)\citenamefont {DeBuhr},
  \citenamefont {Zhang}, \citenamefont {Anderson}, \citenamefont {Neilsen},
  \citenamefont {Hirschmann}, \citenamefont {Grenga},\ and\ \citenamefont
  {Paolucci}}]{DeBuhr:2015jqk}%
  \BibitemOpen
  \bibfield  {author} {\bibinfo {author} {\bibfnamefont {J.}~\bibnamefont
  {DeBuhr}}, \bibinfo {author} {\bibfnamefont {B.}~\bibnamefont {Zhang}},
  \bibinfo {author} {\bibfnamefont {M.}~\bibnamefont {Anderson}}, \bibinfo
  {author} {\bibfnamefont {D.}~\bibnamefont {Neilsen}}, \bibinfo {author}
  {\bibfnamefont {E.~W.}\ \bibnamefont {Hirschmann}}, \bibinfo {author}
  {\bibfnamefont {T.}~\bibnamefont {Grenga}}, \ and\ \bibinfo {author}
  {\bibfnamefont {S.}~\bibnamefont {Paolucci}},\ }\href {\doibase
  10.3847/1538-4357/aae5f9} {\bibfield  {journal} {\bibinfo  {journal}
  {Astrophys. J.}\ }\textbf {\bibinfo {volume} {867}},\ \bibinfo {pages} {112}
  (\bibinfo {year} {2018})},\ \Eprint {http://arxiv.org/abs/1512.00386}
  {arXiv:1512.00386 [astro-ph.IM]} \BibitemShut {NoStop}%
\bibitem [{\citenamefont {Yagi}\ and\ \citenamefont
  {Yunes}(2013)}]{Yagi:2013awa}%
  \BibitemOpen
  \bibfield  {author} {\bibinfo {author} {\bibfnamefont {K.}~\bibnamefont
  {Yagi}}\ and\ \bibinfo {author} {\bibfnamefont {N.}~\bibnamefont {Yunes}},\
  }\href {\doibase 10.1103/PhysRevD.88.023009} {\bibfield  {journal} {\bibinfo
  {journal} {Phys. Rev. D}\ }\textbf {\bibinfo {volume} {88}},\ \bibinfo
  {pages} {023009} (\bibinfo {year} {2013})},\ \Eprint
  {http://arxiv.org/abs/1303.1528} {arXiv:1303.1528 [gr-qc]} \BibitemShut
  {NoStop}%
\bibitem [{\citenamefont {Read}\ \emph
  {et~al.}(2009{\natexlab{b}})\citenamefont {Read}, \citenamefont {Markakis},
  \citenamefont {Shibata}, \citenamefont {Uryu}, \citenamefont {Creighton}
  \emph {et~al.}}]{Read:2009yp}%
  \BibitemOpen
  \bibfield  {author} {\bibinfo {author} {\bibfnamefont {J.~S.}\ \bibnamefont
  {Read}}, \bibinfo {author} {\bibfnamefont {C.}~\bibnamefont {Markakis}},
  \bibinfo {author} {\bibfnamefont {M.}~\bibnamefont {Shibata}}, \bibinfo
  {author} {\bibfnamefont {K.}~\bibnamefont {Uryu}}, \bibinfo {author}
  {\bibfnamefont {J.~D.}\ \bibnamefont {Creighton}},  \emph {et~al.},\ }\href
  {\doibase 10.1103/PhysRevD.79.124033} {\bibfield  {journal} {\bibinfo
  {journal} {Phys. Rev. D}\ }\textbf {\bibinfo {volume} {79}},\ \bibinfo
  {pages} {124033} (\bibinfo {year} {2009}{\natexlab{b}})},\ \Eprint
  {http://arxiv.org/abs/0901.3258} {arXiv:0901.3258 [gr-qc]} \BibitemShut
  {NoStop}%
\bibitem [{\citenamefont {Ujevic}\ \emph {et~al.}(2022)\citenamefont {Ujevic},
  \citenamefont {Gieg}, \citenamefont {Schianchi}, \citenamefont {Chaurasia},
  \citenamefont {Tews},\ and\ \citenamefont {Dietrich}}]{Ujevic:2022nkr}%
  \BibitemOpen
  \bibfield  {author} {\bibinfo {author} {\bibfnamefont {M.}~\bibnamefont
  {Ujevic}}, \bibinfo {author} {\bibfnamefont {H.}~\bibnamefont {Gieg}},
  \bibinfo {author} {\bibfnamefont {F.}~\bibnamefont {Schianchi}}, \bibinfo
  {author} {\bibfnamefont {S.~V.}\ \bibnamefont {Chaurasia}}, \bibinfo {author}
  {\bibfnamefont {I.}~\bibnamefont {Tews}}, \ and\ \bibinfo {author}
  {\bibfnamefont {T.}~\bibnamefont {Dietrich}},\ }\href@noop {} {\  (\bibinfo
  {year} {2022})},\ \Eprint {http://arxiv.org/abs/2211.04662} {arXiv:2211.04662
  [gr-qc]} \BibitemShut {NoStop}%
\bibitem [{\citenamefont {Hotokezaka}\ \emph {et~al.}(2011)\citenamefont
  {Hotokezaka}, \citenamefont {Kyutoku}, \citenamefont {Okawa}, \citenamefont
  {Shibata},\ and\ \citenamefont {Kiuchi}}]{Hotokezaka:2011dh}%
  \BibitemOpen
  \bibfield  {author} {\bibinfo {author} {\bibfnamefont {K.}~\bibnamefont
  {Hotokezaka}}, \bibinfo {author} {\bibfnamefont {K.}~\bibnamefont {Kyutoku}},
  \bibinfo {author} {\bibfnamefont {H.}~\bibnamefont {Okawa}}, \bibinfo
  {author} {\bibfnamefont {M.}~\bibnamefont {Shibata}}, \ and\ \bibinfo
  {author} {\bibfnamefont {K.}~\bibnamefont {Kiuchi}},\ }\href {\doibase
  10.1103/PhysRevD.83.124008} {\bibfield  {journal} {\bibinfo  {journal} {Phys.
  Rev. D}\ }\textbf {\bibinfo {volume} {83}},\ \bibinfo {pages} {124008}
  (\bibinfo {year} {2011})},\ \Eprint {http://arxiv.org/abs/1105.4370}
  {arXiv:1105.4370 [astro-ph.HE]} \BibitemShut {NoStop}%
\bibitem [{\citenamefont {Lackey}\ \emph {et~al.}(2014)\citenamefont {Lackey},
  \citenamefont {Kyutoku}, \citenamefont {Shibata}, \citenamefont {Brady},\
  and\ \citenamefont {Friedman}}]{Lackey:2013axa}%
  \BibitemOpen
  \bibfield  {author} {\bibinfo {author} {\bibfnamefont {B.~D.}\ \bibnamefont
  {Lackey}}, \bibinfo {author} {\bibfnamefont {K.}~\bibnamefont {Kyutoku}},
  \bibinfo {author} {\bibfnamefont {M.}~\bibnamefont {Shibata}}, \bibinfo
  {author} {\bibfnamefont {P.~R.}\ \bibnamefont {Brady}}, \ and\ \bibinfo
  {author} {\bibfnamefont {J.~L.}\ \bibnamefont {Friedman}},\ }\href {\doibase
  10.1103/PhysRevD.89.043009} {\bibfield  {journal} {\bibinfo  {journal} {Phys.
  Rev. D}\ }\textbf {\bibinfo {volume} {89}},\ \bibinfo {pages} {043009}
  (\bibinfo {year} {2014})},\ \Eprint {http://arxiv.org/abs/1303.6298}
  {arXiv:1303.6298 [gr-qc]} \BibitemShut {NoStop}%
\bibitem [{\citenamefont {Dietrich}\ \emph {et~al.}(2018)\citenamefont
  {Dietrich}, \citenamefont {Bernuzzi}, \citenamefont {Bruegmann},\ and\
  \citenamefont {Tichy}}]{Dietrich:2018upm}%
  \BibitemOpen
  \bibfield  {author} {\bibinfo {author} {\bibfnamefont {T.}~\bibnamefont
  {Dietrich}}, \bibinfo {author} {\bibfnamefont {S.}~\bibnamefont {Bernuzzi}},
  \bibinfo {author} {\bibfnamefont {B.}~\bibnamefont {Bruegmann}}, \ and\
  \bibinfo {author} {\bibfnamefont {W.}~\bibnamefont {Tichy}},\ }in\ \href
  {\doibase 10.1109/PDP2018.2018.00113} {\emph {\bibinfo {booktitle} {{26th
  Euromicro International Conference on Parallel, Distributed and Network-based
  Processing}}}}\ (\bibinfo {year} {2018})\ pp.\ \bibinfo {pages} {682--689},\
  \Eprint {http://arxiv.org/abs/1803.07965} {arXiv:1803.07965 [gr-qc]}
  \BibitemShut {NoStop}%
\bibitem [{\citenamefont {Dietrich}\ \emph {et~al.}(2017)\citenamefont
  {Dietrich}, \citenamefont {Bernuzzi},\ and\ \citenamefont
  {Tichy}}]{Dietrich:2017aum}%
  \BibitemOpen
  \bibfield  {author} {\bibinfo {author} {\bibfnamefont {T.}~\bibnamefont
  {Dietrich}}, \bibinfo {author} {\bibfnamefont {S.}~\bibnamefont {Bernuzzi}},
  \ and\ \bibinfo {author} {\bibfnamefont {W.}~\bibnamefont {Tichy}},\ }\href
  {\doibase 10.1103/PhysRevD.96.121501} {\bibfield  {journal} {\bibinfo
  {journal} {Phys. Rev.}\ }\textbf {\bibinfo {volume} {D96}},\ \bibinfo {pages}
  {121501} (\bibinfo {year} {2017})},\ \Eprint
  {http://arxiv.org/abs/1706.02969} {arXiv:1706.02969 [gr-qc]} \BibitemShut
  {NoStop}%
\bibitem [{\citenamefont {O'Boyle}\ \emph {et~al.}(2020)\citenamefont
  {O'Boyle}, \citenamefont {Markakis}, \citenamefont {Stergioulas},\ and\
  \citenamefont {Read}}]{OBoyle:2020qvf}%
  \BibitemOpen
  \bibfield  {author} {\bibinfo {author} {\bibfnamefont {M.~F.}\ \bibnamefont
  {O'Boyle}}, \bibinfo {author} {\bibfnamefont {C.}~\bibnamefont {Markakis}},
  \bibinfo {author} {\bibfnamefont {N.}~\bibnamefont {Stergioulas}}, \ and\
  \bibinfo {author} {\bibfnamefont {J.~S.}\ \bibnamefont {Read}},\ }\href
  {\doibase 10.1103/PhysRevD.102.083027} {\bibfield  {journal} {\bibinfo
  {journal} {Phys. Rev. D}\ }\textbf {\bibinfo {volume} {102}},\ \bibinfo
  {pages} {083027} (\bibinfo {year} {2020})},\ \Eprint
  {http://arxiv.org/abs/2008.03342} {arXiv:2008.03342 [astro-ph.HE]}
  \BibitemShut {NoStop}%
\bibitem [{\citenamefont {Press}\ \emph {et~al.}(2007)\citenamefont {Press},
  \citenamefont {Teukolsky}, \citenamefont {Vetterling},\ and\ \citenamefont
  {Flannery}}]{numericalrecipes}%
  \BibitemOpen
  \bibfield  {author} {\bibinfo {author} {\bibfnamefont {W.~H.}\ \bibnamefont
  {Press}}, \bibinfo {author} {\bibfnamefont {S.~A.}\ \bibnamefont
  {Teukolsky}}, \bibinfo {author} {\bibfnamefont {W.~T.}\ \bibnamefont
  {Vetterling}}, \ and\ \bibinfo {author} {\bibfnamefont {B.~P.}\ \bibnamefont
  {Flannery}},\ }\href@noop {} {\emph {\bibinfo {title} {Numerical Recipes 3rd
  Edition: The Art of Scientific Computing}}},\ \bibinfo {edition} {3rd}\ ed.\
  (\bibinfo  {publisher} {Cambridge University Press},\ \bibinfo {address}
  {USA},\ \bibinfo {year} {2007})\BibitemShut {NoStop}%
\bibitem [{\citenamefont {McLerran}\ and\ \citenamefont
  {Reddy}(2019)}]{McLerran:2018hbz}%
  \BibitemOpen
  \bibfield  {author} {\bibinfo {author} {\bibfnamefont {L.}~\bibnamefont
  {McLerran}}\ and\ \bibinfo {author} {\bibfnamefont {S.}~\bibnamefont
  {Reddy}},\ }\href {\doibase 10.1103/PhysRevLett.122.122701} {\bibfield
  {journal} {\bibinfo  {journal} {Phys. Rev. Lett.}\ }\textbf {\bibinfo
  {volume} {122}},\ \bibinfo {pages} {122701} (\bibinfo {year} {2019})},\
  \Eprint {http://arxiv.org/abs/1811.12503} {arXiv:1811.12503 [nucl-th]}
  \BibitemShut {NoStop}%
\bibitem [{\citenamefont {Tews}\ \emph
  {et~al.}(2018{\natexlab{a}})\citenamefont {Tews}, \citenamefont {Carlson},
  \citenamefont {Gandolfi},\ and\ \citenamefont {Reddy}}]{Tews:2018kmu}%
  \BibitemOpen
  \bibfield  {author} {\bibinfo {author} {\bibfnamefont {I.}~\bibnamefont
  {Tews}}, \bibinfo {author} {\bibfnamefont {J.}~\bibnamefont {Carlson}},
  \bibinfo {author} {\bibfnamefont {S.}~\bibnamefont {Gandolfi}}, \ and\
  \bibinfo {author} {\bibfnamefont {S.}~\bibnamefont {Reddy}},\ }\href
  {\doibase 10.3847/1538-4357/aac267} {\bibfield  {journal} {\bibinfo
  {journal} {Astrophys. J.}\ }\textbf {\bibinfo {volume} {860}},\ \bibinfo
  {pages} {149} (\bibinfo {year} {2018}{\natexlab{a}})},\ \Eprint
  {http://arxiv.org/abs/1801.01923} {arXiv:1801.01923 [nucl-th]} \BibitemShut
  {NoStop}%
\bibitem [{\citenamefont {Kapusta}\ and\ \citenamefont
  {Welle}(2021)}]{Kapusta:2021ney}%
  \BibitemOpen
  \bibfield  {author} {\bibinfo {author} {\bibfnamefont {J.~I.}\ \bibnamefont
  {Kapusta}}\ and\ \bibinfo {author} {\bibfnamefont {T.}~\bibnamefont
  {Welle}},\ }\href {\doibase 10.1103/PhysRevC.104.L012801} {\bibfield
  {journal} {\bibinfo  {journal} {Phys. Rev. C}\ }\textbf {\bibinfo {volume}
  {104}},\ \bibinfo {pages} {L012801} (\bibinfo {year} {2021})},\ \Eprint
  {http://arxiv.org/abs/2103.16633} {arXiv:2103.16633 [nucl-th]} \BibitemShut
  {NoStop}%
\bibitem [{\citenamefont {Landry}\ and\ \citenamefont
  {Kumar}(2018)}]{Landry:2018jyg}%
  \BibitemOpen
  \bibfield  {author} {\bibinfo {author} {\bibfnamefont {P.}~\bibnamefont
  {Landry}}\ and\ \bibinfo {author} {\bibfnamefont {B.}~\bibnamefont {Kumar}},\
  }\href {\doibase 10.3847/2041-8213/aaee76} {\bibfield  {journal} {\bibinfo
  {journal} {Astrophys. J.}\ }\textbf {\bibinfo {volume} {868}},\ \bibinfo
  {pages} {L22} (\bibinfo {year} {2018})},\ \Eprint
  {http://arxiv.org/abs/1807.04727} {arXiv:1807.04727 [gr-qc]} \BibitemShut
  {NoStop}%
\bibitem [{\citenamefont {Essick}\ \emph
  {et~al.}(2020{\natexlab{a}})\citenamefont {Essick}, \citenamefont {Landry},\
  and\ \citenamefont {Holz}}]{Essick:2019ldf}%
  \BibitemOpen
  \bibfield  {author} {\bibinfo {author} {\bibfnamefont {R.}~\bibnamefont
  {Essick}}, \bibinfo {author} {\bibfnamefont {P.}~\bibnamefont {Landry}}, \
  and\ \bibinfo {author} {\bibfnamefont {D.~E.}\ \bibnamefont {Holz}},\ }\href
  {\doibase 10.1103/PhysRevD.101.063007} {\bibfield  {journal} {\bibinfo
  {journal} {Phys. Rev. D}\ }\textbf {\bibinfo {volume} {101}},\ \bibinfo
  {pages} {063007} (\bibinfo {year} {2020}{\natexlab{a}})},\ \Eprint
  {http://arxiv.org/abs/1910.09740} {arXiv:1910.09740 [astro-ph.HE]}
  \BibitemShut {NoStop}%
\bibitem [{\citenamefont {M\"osta}\ \emph {et~al.}(2014)\citenamefont
  {M\"osta}, \citenamefont {Mundim}, \citenamefont {Faber}, \citenamefont
  {Haas}, \citenamefont {Noble}, \citenamefont {Bode}, \citenamefont
  {L\"offler}, \citenamefont {Ott}, \citenamefont {Reisswig},\ and\
  \citenamefont {Schnetter}}]{Mosta:2013gwu}%
  \BibitemOpen
  \bibfield  {author} {\bibinfo {author} {\bibfnamefont {P.}~\bibnamefont
  {M\"osta}}, \bibinfo {author} {\bibfnamefont {B.~C.}\ \bibnamefont {Mundim}},
  \bibinfo {author} {\bibfnamefont {J.~A.}\ \bibnamefont {Faber}}, \bibinfo
  {author} {\bibfnamefont {R.}~\bibnamefont {Haas}}, \bibinfo {author}
  {\bibfnamefont {S.~C.}\ \bibnamefont {Noble}}, \bibinfo {author}
  {\bibfnamefont {T.}~\bibnamefont {Bode}}, \bibinfo {author} {\bibfnamefont
  {F.}~\bibnamefont {L\"offler}}, \bibinfo {author} {\bibfnamefont {C.~D.}\
  \bibnamefont {Ott}}, \bibinfo {author} {\bibfnamefont {C.}~\bibnamefont
  {Reisswig}}, \ and\ \bibinfo {author} {\bibfnamefont {E.}~\bibnamefont
  {Schnetter}},\ }\href {\doibase 10.1088/0264-9381/31/1/015005} {\bibfield
  {journal} {\bibinfo  {journal} {Class. Quant. Grav.}\ }\textbf {\bibinfo
  {volume} {31}},\ \bibinfo {pages} {015005} (\bibinfo {year} {2014})},\
  \Eprint {http://arxiv.org/abs/1304.5544} {arXiv:1304.5544 [gr-qc]}
  \BibitemShut {NoStop}%
\bibitem [{\citenamefont {Tews}(2020)}]{Tews:2020}%
  \BibitemOpen
  \bibfield  {author} {\bibinfo {author} {\bibfnamefont {I.}~\bibnamefont
  {Tews}},\ }\href {\doibase 10.3389/fphy.2020.00153} {\bibfield  {journal}
  {\bibinfo  {journal} {Frontiers in Physics}\ }\textbf {\bibinfo {volume} {8}}
  (\bibinfo {year} {2020}),\ 10.3389/fphy.2020.00153}\BibitemShut {NoStop}%
\bibitem [{\citenamefont {Drischler}\ \emph {et~al.}(2016)\citenamefont
  {Drischler}, \citenamefont {Carbone}, \citenamefont {Hebeler},\ and\
  \citenamefont {Schwenk}}]{Drischler:2016djf}%
  \BibitemOpen
  \bibfield  {author} {\bibinfo {author} {\bibfnamefont {C.}~\bibnamefont
  {Drischler}}, \bibinfo {author} {\bibfnamefont {A.}~\bibnamefont {Carbone}},
  \bibinfo {author} {\bibfnamefont {K.}~\bibnamefont {Hebeler}}, \ and\
  \bibinfo {author} {\bibfnamefont {A.}~\bibnamefont {Schwenk}},\ }\href
  {\doibase 10.1103/PhysRevC.94.054307} {\bibfield  {journal} {\bibinfo
  {journal} {Phys. Rev. C}\ }\textbf {\bibinfo {volume} {94}},\ \bibinfo
  {pages} {054307} (\bibinfo {year} {2016})},\ \Eprint
  {http://arxiv.org/abs/1608.05615} {arXiv:1608.05615 [nucl-th]} \BibitemShut
  {NoStop}%
\bibitem [{\citenamefont {Essick}\ \emph
  {et~al.}(2020{\natexlab{b}})\citenamefont {Essick}, \citenamefont {Tews},
  \citenamefont {Landry}, \citenamefont {Reddy},\ and\ \citenamefont
  {Holz}}]{Essick:2020flb}%
  \BibitemOpen
  \bibfield  {author} {\bibinfo {author} {\bibfnamefont {R.}~\bibnamefont
  {Essick}}, \bibinfo {author} {\bibfnamefont {I.}~\bibnamefont {Tews}},
  \bibinfo {author} {\bibfnamefont {P.}~\bibnamefont {Landry}}, \bibinfo
  {author} {\bibfnamefont {S.}~\bibnamefont {Reddy}}, \ and\ \bibinfo {author}
  {\bibfnamefont {D.~E.}\ \bibnamefont {Holz}},\ }\href {\doibase
  10.1103/PhysRevC.102.055803} {\bibfield  {journal} {\bibinfo  {journal}
  {Phys. Rev. C}\ }\textbf {\bibinfo {volume} {102}},\ \bibinfo {pages}
  {055803} (\bibinfo {year} {2020}{\natexlab{b}})},\ \Eprint
  {http://arxiv.org/abs/2004.07744} {arXiv:2004.07744 [astro-ph.HE]}
  \BibitemShut {NoStop}%
\bibitem [{\citenamefont {Roca-Maza}\ \emph {et~al.}(2015)\citenamefont
  {Roca-Maza}, \citenamefont {Vi\~nas}, \citenamefont {Centelles},
  \citenamefont {Agrawal}, \citenamefont {Colo'}, \citenamefont {Paar},
  \citenamefont {Piekarewicz},\ and\ \citenamefont
  {Vretenar}}]{Roca-Maza:2015eza}%
  \BibitemOpen
  \bibfield  {author} {\bibinfo {author} {\bibfnamefont {X.}~\bibnamefont
  {Roca-Maza}}, \bibinfo {author} {\bibfnamefont {X.}~\bibnamefont {Vi\~nas}},
  \bibinfo {author} {\bibfnamefont {M.}~\bibnamefont {Centelles}}, \bibinfo
  {author} {\bibfnamefont {B.~K.}\ \bibnamefont {Agrawal}}, \bibinfo {author}
  {\bibfnamefont {G.}~\bibnamefont {Colo'}}, \bibinfo {author} {\bibfnamefont
  {N.}~\bibnamefont {Paar}}, \bibinfo {author} {\bibfnamefont {J.}~\bibnamefont
  {Piekarewicz}}, \ and\ \bibinfo {author} {\bibfnamefont {D.}~\bibnamefont
  {Vretenar}},\ }\href {\doibase 10.1103/PhysRevC.92.064304} {\bibfield
  {journal} {\bibinfo  {journal} {Phys. Rev. C}\ }\textbf {\bibinfo {volume}
  {92}},\ \bibinfo {pages} {064304} (\bibinfo {year} {2015})},\ \Eprint
  {http://arxiv.org/abs/1510.01874} {arXiv:1510.01874 [nucl-th]} \BibitemShut
  {NoStop}%
\bibitem [{\citenamefont {Essick}\ \emph
  {et~al.}(2021{\natexlab{a}})\citenamefont {Essick}, \citenamefont {Landry},
  \citenamefont {Schwenk},\ and\ \citenamefont {Tews}}]{Essick:2021ezp}%
  \BibitemOpen
  \bibfield  {author} {\bibinfo {author} {\bibfnamefont {R.}~\bibnamefont
  {Essick}}, \bibinfo {author} {\bibfnamefont {P.}~\bibnamefont {Landry}},
  \bibinfo {author} {\bibfnamefont {A.}~\bibnamefont {Schwenk}}, \ and\
  \bibinfo {author} {\bibfnamefont {I.}~\bibnamefont {Tews}},\ }\href {\doibase
  10.1103/PhysRevC.104.065804} {\bibfield  {journal} {\bibinfo  {journal}
  {Phys. Rev. C}\ }\textbf {\bibinfo {volume} {104}},\ \bibinfo {pages}
  {065804} (\bibinfo {year} {2021}{\natexlab{a}})}\BibitemShut {NoStop}%
\bibitem [{\citenamefont {Essick}\ \emph
  {et~al.}(2021{\natexlab{b}})\citenamefont {Essick}, \citenamefont {Tews},
  \citenamefont {Landry},\ and\ \citenamefont {Schwenk}}]{Essick:2021kjb}%
  \BibitemOpen
  \bibfield  {author} {\bibinfo {author} {\bibfnamefont {R.}~\bibnamefont
  {Essick}}, \bibinfo {author} {\bibfnamefont {I.}~\bibnamefont {Tews}},
  \bibinfo {author} {\bibfnamefont {P.}~\bibnamefont {Landry}}, \ and\ \bibinfo
  {author} {\bibfnamefont {A.}~\bibnamefont {Schwenk}},\ }\href {\doibase
  10.1103/PhysRevLett.127.192701} {\bibfield  {journal} {\bibinfo  {journal}
  {Phys. Rev. Lett.}\ }\textbf {\bibinfo {volume} {127}},\ \bibinfo {pages}
  {192701} (\bibinfo {year} {2021}{\natexlab{b}})},\ \Eprint
  {http://arxiv.org/abs/2102.10074} {arXiv:2102.10074 [nucl-th]} \BibitemShut
  {NoStop}%
\bibitem [{\citenamefont {Tews}\ \emph
  {et~al.}(2018{\natexlab{b}})\citenamefont {Tews}, \citenamefont {Margueron},\
  and\ \citenamefont {Reddy}}]{Tews:2018iwm}%
  \BibitemOpen
  \bibfield  {author} {\bibinfo {author} {\bibfnamefont {I.}~\bibnamefont
  {Tews}}, \bibinfo {author} {\bibfnamefont {J.}~\bibnamefont {Margueron}}, \
  and\ \bibinfo {author} {\bibfnamefont {S.}~\bibnamefont {Reddy}},\ }\href
  {\doibase 10.1103/PhysRevC.98.045804} {\bibfield  {journal} {\bibinfo
  {journal} {Phys. Rev. C}\ }\textbf {\bibinfo {volume} {98}},\ \bibinfo
  {pages} {045804} (\bibinfo {year} {2018}{\natexlab{b}})},\ \Eprint
  {http://arxiv.org/abs/1804.02783} {arXiv:1804.02783 [nucl-th]} \BibitemShut
  {NoStop}%
\bibitem [{\citenamefont {Landry}\ and\ \citenamefont
  {Essick}(2019)}]{Landry:2018prl}%
  \BibitemOpen
  \bibfield  {author} {\bibinfo {author} {\bibfnamefont {P.}~\bibnamefont
  {Landry}}\ and\ \bibinfo {author} {\bibfnamefont {R.}~\bibnamefont
  {Essick}},\ }\href {\doibase 10.1103/PhysRevD.99.084049} {\bibfield
  {journal} {\bibinfo  {journal} {Phys. Rev. D}\ }\textbf {\bibinfo {volume}
  {99}},\ \bibinfo {pages} {084049} (\bibinfo {year} {2019})},\ \Eprint
  {http://arxiv.org/abs/1811.12529} {arXiv:1811.12529 [gr-qc]} \BibitemShut
  {NoStop}%
\bibitem [{\citenamefont {Alford}\ \emph {et~al.}(2005)\citenamefont {Alford},
  \citenamefont {Braby}, \citenamefont {Paris},\ and\ \citenamefont
  {Reddy}}]{Alford:2004pf}%
  \BibitemOpen
  \bibfield  {author} {\bibinfo {author} {\bibfnamefont {M.}~\bibnamefont
  {Alford}}, \bibinfo {author} {\bibfnamefont {M.}~\bibnamefont {Braby}},
  \bibinfo {author} {\bibfnamefont {M.~W.}\ \bibnamefont {Paris}}, \ and\
  \bibinfo {author} {\bibfnamefont {S.}~\bibnamefont {Reddy}},\ }\href
  {\doibase 10.1086/430902} {\bibfield  {journal} {\bibinfo  {journal}
  {Astrophys. J.}\ }\textbf {\bibinfo {volume} {629}},\ \bibinfo {pages} {969}
  (\bibinfo {year} {2005})},\ \Eprint {http://arxiv.org/abs/nucl-th/0411016}
  {arXiv:nucl-th/0411016} \BibitemShut {NoStop}%
\bibitem [{\citenamefont {Potekhin}\ \emph {et~al.}(2013)\citenamefont
  {Potekhin}, \citenamefont {Fantina}, \citenamefont {Chamel}, \citenamefont
  {Pearson},\ and\ \citenamefont {Goriely}}]{Potekhin:2013qqa}%
  \BibitemOpen
  \bibfield  {author} {\bibinfo {author} {\bibfnamefont {A.~Y.}\ \bibnamefont
  {Potekhin}}, \bibinfo {author} {\bibfnamefont {A.~F.}\ \bibnamefont
  {Fantina}}, \bibinfo {author} {\bibfnamefont {N.}~\bibnamefont {Chamel}},
  \bibinfo {author} {\bibfnamefont {J.~M.}\ \bibnamefont {Pearson}}, \ and\
  \bibinfo {author} {\bibfnamefont {S.}~\bibnamefont {Goriely}},\ }\href
  {\doibase 10.1051/0004-6361/201321697} {\bibfield  {journal} {\bibinfo
  {journal} {Astron. Astrophys.}\ }\textbf {\bibinfo {volume} {560}},\ \bibinfo
  {pages} {A48} (\bibinfo {year} {2013})},\ \Eprint
  {http://arxiv.org/abs/1310.0049} {arXiv:1310.0049 [astro-ph.SR]} \BibitemShut
  {NoStop}%
\bibitem [{\citenamefont {Akmal}\ \emph {et~al.}(1998)\citenamefont {Akmal},
  \citenamefont {Pandharipande},\ and\ \citenamefont
  {Ravenhall}}]{Akmal:1998cf}%
  \BibitemOpen
  \bibfield  {author} {\bibinfo {author} {\bibfnamefont {A.}~\bibnamefont
  {Akmal}}, \bibinfo {author} {\bibfnamefont {V.~R.}\ \bibnamefont
  {Pandharipande}}, \ and\ \bibinfo {author} {\bibfnamefont {D.~G.}\
  \bibnamefont {Ravenhall}},\ }\href {\doibase 10.1103/PhysRevC.58.1804}
  {\bibfield  {journal} {\bibinfo  {journal} {Phys. Rev. C}\ }\textbf {\bibinfo
  {volume} {58}},\ \bibinfo {pages} {1804} (\bibinfo {year} {1998})},\ \Eprint
  {http://arxiv.org/abs/nucl-th/9804027} {arXiv:nucl-th/9804027} \BibitemShut
  {NoStop}%
\bibitem [{\citenamefont {Lackey}\ \emph {et~al.}(2006)\citenamefont {Lackey},
  \citenamefont {Nayyar},\ and\ \citenamefont {Owen}}]{Lackey:2005tk}%
  \BibitemOpen
  \bibfield  {author} {\bibinfo {author} {\bibfnamefont {B.~D.}\ \bibnamefont
  {Lackey}}, \bibinfo {author} {\bibfnamefont {M.}~\bibnamefont {Nayyar}}, \
  and\ \bibinfo {author} {\bibfnamefont {B.~J.}\ \bibnamefont {Owen}},\ }\href
  {\doibase 10.1103/PhysRevD.73.024021} {\bibfield  {journal} {\bibinfo
  {journal} {Phys. Rev. D}\ }\textbf {\bibinfo {volume} {73}},\ \bibinfo
  {pages} {024021} (\bibinfo {year} {2006})},\ \Eprint
  {http://arxiv.org/abs/astro-ph/0507312} {arXiv:astro-ph/0507312 [astro-ph]}
  \BibitemShut {NoStop}%
\bibitem [{\citenamefont {Baldo}\ \emph {et~al.}(1997)\citenamefont {Baldo},
  \citenamefont {Bombaci},\ and\ \citenamefont {Burgio}}]{Baldo:1997ag}%
  \BibitemOpen
  \bibfield  {author} {\bibinfo {author} {\bibfnamefont {M.}~\bibnamefont
  {Baldo}}, \bibinfo {author} {\bibfnamefont {I.}~\bibnamefont {Bombaci}}, \
  and\ \bibinfo {author} {\bibfnamefont {G.~F.}\ \bibnamefont {Burgio}},\
  }\href@noop {} {\bibfield  {journal} {\bibinfo  {journal} {Astron.
  Astrophys.}\ }\textbf {\bibinfo {volume} {328}},\ \bibinfo {pages} {274}
  (\bibinfo {year} {1997})},\ \Eprint {http://arxiv.org/abs/astro-ph/9707277}
  {arXiv:astro-ph/9707277} \BibitemShut {NoStop}%
\bibitem [{\citenamefont {Engvik}\ \emph {et~al.}(1996)\citenamefont {Engvik},
  \citenamefont {Bao}, \citenamefont {Hjorth-Jensen}, \citenamefont {Osnes},\
  and\ \citenamefont {Ostgaard}}]{Engvik:1995gn}%
  \BibitemOpen
  \bibfield  {author} {\bibinfo {author} {\bibfnamefont {L.}~\bibnamefont
  {Engvik}}, \bibinfo {author} {\bibfnamefont {G.}~\bibnamefont {Bao}},
  \bibinfo {author} {\bibfnamefont {M.}~\bibnamefont {Hjorth-Jensen}}, \bibinfo
  {author} {\bibfnamefont {E.}~\bibnamefont {Osnes}}, \ and\ \bibinfo {author}
  {\bibfnamefont {E.}~\bibnamefont {Ostgaard}},\ }\href {\doibase
  10.1086/177827} {\bibfield  {journal} {\bibinfo  {journal} {Astrophys. J.}\
  }\textbf {\bibinfo {volume} {469}},\ \bibinfo {pages} {794} (\bibinfo {year}
  {1996})},\ \Eprint {http://arxiv.org/abs/nucl-th/9509016}
  {arXiv:nucl-th/9509016} \BibitemShut {NoStop}%
\bibitem [{\citenamefont {M\"uther}\ \emph {et~al.}(1987)\citenamefont
  {M\"uther}, \citenamefont {Prakash},\ and\ \citenamefont
  {Ainsworth}}]{Muther:1987xaa}%
  \BibitemOpen
  \bibfield  {author} {\bibinfo {author} {\bibfnamefont {H.}~\bibnamefont
  {M\"uther}}, \bibinfo {author} {\bibfnamefont {M.}~\bibnamefont {Prakash}}, \
  and\ \bibinfo {author} {\bibfnamefont {T.~L.}\ \bibnamefont {Ainsworth}},\
  }\href {\doibase 10.1016/0370-2693(87)91611-X} {\bibfield  {journal}
  {\bibinfo  {journal} {Phys. Lett. B}\ }\textbf {\bibinfo {volume} {199}},\
  \bibinfo {pages} {469} (\bibinfo {year} {1987})}\BibitemShut {NoStop}%
\bibitem [{\citenamefont {{M{\"u}ller}}\ and\ \citenamefont
  {{Serot}}(1996)}]{Mueller:1996pm}%
  \BibitemOpen
  \bibfield  {author} {\bibinfo {author} {\bibfnamefont {H.}~\bibnamefont
  {{M{\"u}ller}}}\ and\ \bibinfo {author} {\bibfnamefont {B.~D.}\ \bibnamefont
  {{Serot}}},\ }\href {\doibase 10.1016/0375-9474(96)00187-X} {\bibfield
  {journal} {\bibinfo  {journal} {Nucl. Phys.}\ }\textbf {\bibinfo {volume}
  {A606}},\ \bibinfo {pages} {508} (\bibinfo {year} {1996})},\ \Eprint
  {http://arxiv.org/abs/nucl-th/9603037} {arXiv:nucl-th/9603037 [nucl-th]}
  \BibitemShut {NoStop}%
\bibitem [{\citenamefont {Rikovska-Stone}\ \emph {et~al.}(2007)\citenamefont
  {Rikovska-Stone}, \citenamefont {Guichon}, \citenamefont {Matevosyan},\ and\
  \citenamefont {Thomas}}]{Rikovska-Stone:2006gml}%
  \BibitemOpen
  \bibfield  {author} {\bibinfo {author} {\bibfnamefont {J.}~\bibnamefont
  {Rikovska-Stone}}, \bibinfo {author} {\bibfnamefont {P.~A.~M.}\ \bibnamefont
  {Guichon}}, \bibinfo {author} {\bibfnamefont {H.~H.}\ \bibnamefont
  {Matevosyan}}, \ and\ \bibinfo {author} {\bibfnamefont {A.~W.}\ \bibnamefont
  {Thomas}},\ }\href {\doibase 10.1016/j.nuclphysa.2007.05.011} {\bibfield
  {journal} {\bibinfo  {journal} {Nucl. Phys. A}\ }\textbf {\bibinfo {volume}
  {792}},\ \bibinfo {pages} {341} (\bibinfo {year} {2007})},\ \Eprint
  {http://arxiv.org/abs/nucl-th/0611030} {arXiv:nucl-th/0611030} \BibitemShut
  {NoStop}%
\bibitem [{\citenamefont {Douchin}\ and\ \citenamefont
  {Haensel}(2001)}]{Douchin:2001sv}%
  \BibitemOpen
  \bibfield  {author} {\bibinfo {author} {\bibfnamefont {F.}~\bibnamefont
  {Douchin}}\ and\ \bibinfo {author} {\bibfnamefont {P.}~\bibnamefont
  {Haensel}},\ }\href {\doibase 10.1051/0004-6361:20011402} {\bibfield
  {journal} {\bibinfo  {journal} {Astron. Astrophys.}\ }\textbf {\bibinfo
  {volume} {380}},\ \bibinfo {pages} {151} (\bibinfo {year} {2001})},\ \Eprint
  {http://arxiv.org/abs/astro-ph/0111092} {arXiv:astro-ph/0111092} \BibitemShut
  {NoStop}%
\bibitem [{\citenamefont {Wiringa}\ \emph {et~al.}(1988)\citenamefont
  {Wiringa}, \citenamefont {Fiks},\ and\ \citenamefont
  {Fabrocini}}]{Wiringa:1988tp}%
  \BibitemOpen
  \bibfield  {author} {\bibinfo {author} {\bibfnamefont {R.~B.}\ \bibnamefont
  {Wiringa}}, \bibinfo {author} {\bibfnamefont {V.}~\bibnamefont {Fiks}}, \
  and\ \bibinfo {author} {\bibfnamefont {A.}~\bibnamefont {Fabrocini}},\ }\href
  {\doibase 10.1103/PhysRevC.38.1010} {\bibfield  {journal} {\bibinfo
  {journal} {Phys. Rev. C}\ }\textbf {\bibinfo {volume} {38}},\ \bibinfo
  {pages} {1010} (\bibinfo {year} {1988})}\BibitemShut {NoStop}%
\bibitem [{\citenamefont {Carney}\ \emph {et~al.}(2018)\citenamefont {Carney},
  \citenamefont {Wade},\ and\ \citenamefont {Irwin}}]{Carney:2018sdv}%
  \BibitemOpen
  \bibfield  {author} {\bibinfo {author} {\bibfnamefont {M.~F.}\ \bibnamefont
  {Carney}}, \bibinfo {author} {\bibfnamefont {L.~E.}\ \bibnamefont {Wade}}, \
  and\ \bibinfo {author} {\bibfnamefont {B.~S.}\ \bibnamefont {Irwin}},\ }\href
  {\doibase 10.1103/PhysRevD.98.063004} {\bibfield  {journal} {\bibinfo
  {journal} {Phys. Rev.}\ }\textbf {\bibinfo {volume} {D98}},\ \bibinfo {pages}
  {063004} (\bibinfo {year} {2018})},\ \Eprint
  {http://arxiv.org/abs/1805.11217} {arXiv:1805.11217 [gr-qc]} \BibitemShut
  {NoStop}%
\bibitem [{\citenamefont {Legred}\ \emph {et~al.}(2022)\citenamefont {Legred},
  \citenamefont {Chatziioannou}, \citenamefont {Essick},\ and\ \citenamefont
  {Landry}}]{Legred:2022pyp}%
  \BibitemOpen
  \bibfield  {author} {\bibinfo {author} {\bibfnamefont {I.}~\bibnamefont
  {Legred}}, \bibinfo {author} {\bibfnamefont {K.}~\bibnamefont
  {Chatziioannou}}, \bibinfo {author} {\bibfnamefont {R.}~\bibnamefont
  {Essick}}, \ and\ \bibinfo {author} {\bibfnamefont {P.}~\bibnamefont
  {Landry}},\ }\href {\doibase 10.1103/PhysRevD.105.043016} {\bibfield
  {journal} {\bibinfo  {journal} {Phys. Rev. D}\ }\textbf {\bibinfo {volume}
  {105}},\ \bibinfo {pages} {043016} (\bibinfo {year} {2022})},\ \Eprint
  {http://arxiv.org/abs/2201.06791} {arXiv:2201.06791 [astro-ph.HE]}
  \BibitemShut {NoStop}%
\bibitem [{\citenamefont {Gross-Boelting}\ \emph {et~al.}(1999)\citenamefont
  {Gross-Boelting}, \citenamefont {Fuchs},\ and\ \citenamefont
  {Faessler}}]{Gross-Boelting:1998xsk}%
  \BibitemOpen
  \bibfield  {author} {\bibinfo {author} {\bibfnamefont {T.}~\bibnamefont
  {Gross-Boelting}}, \bibinfo {author} {\bibfnamefont {C.}~\bibnamefont
  {Fuchs}}, \ and\ \bibinfo {author} {\bibfnamefont {A.}~\bibnamefont
  {Faessler}},\ }\href {\doibase 10.1016/S0375-9474(99)00022-6} {\bibfield
  {journal} {\bibinfo  {journal} {Nucl. Phys. A}\ }\textbf {\bibinfo {volume}
  {648}},\ \bibinfo {pages} {105} (\bibinfo {year} {1999})},\ \Eprint
  {http://arxiv.org/abs/nucl-th/9810071} {arXiv:nucl-th/9810071} \BibitemShut
  {NoStop}%
\bibitem [{\citenamefont {Alford}\ \emph {et~al.}(2015)\citenamefont {Alford},
  \citenamefont {Burgio}, \citenamefont {Han}, \citenamefont {Taranto},\ and\
  \citenamefont {Zappal\`a}}]{Alford:2015dpa}%
  \BibitemOpen
  \bibfield  {author} {\bibinfo {author} {\bibfnamefont {M.~G.}\ \bibnamefont
  {Alford}}, \bibinfo {author} {\bibfnamefont {G.~F.}\ \bibnamefont {Burgio}},
  \bibinfo {author} {\bibfnamefont {S.}~\bibnamefont {Han}}, \bibinfo {author}
  {\bibfnamefont {G.}~\bibnamefont {Taranto}}, \ and\ \bibinfo {author}
  {\bibfnamefont {D.}~\bibnamefont {Zappal\`a}},\ }\href {\doibase
  10.1103/PhysRevD.92.083002} {\bibfield  {journal} {\bibinfo  {journal} {Phys.
  Rev. D}\ }\textbf {\bibinfo {volume} {92}},\ \bibinfo {pages} {083002}
  (\bibinfo {year} {2015})},\ \Eprint {http://arxiv.org/abs/1501.07902}
  {arXiv:1501.07902 [nucl-th]} \BibitemShut {NoStop}%
\bibitem [{\citenamefont {Courant}\ \emph {et~al.}(1967)\citenamefont
  {Courant}, \citenamefont {Friedrichs},\ and\ \citenamefont {Lewy}}]{CFL}%
  \BibitemOpen
  \bibfield  {author} {\bibinfo {author} {\bibfnamefont {R.}~\bibnamefont
  {Courant}}, \bibinfo {author} {\bibfnamefont {K.}~\bibnamefont {Friedrichs}},
  \ and\ \bibinfo {author} {\bibfnamefont {H.}~\bibnamefont {Lewy}},\ }\href
  {\doibase 10.1147/rd.112.0215} {\bibfield  {journal} {\bibinfo  {journal}
  {IBM Journal of Research and Development}\ }\textbf {\bibinfo {volume}
  {11}},\ \bibinfo {pages} {215} (\bibinfo {year} {1967})}\BibitemShut
  {NoStop}%
\bibitem [{\citenamefont {{Van Leer}}(1977)}]{VANLEER1977276}%
  \BibitemOpen
  \bibfield  {author} {\bibinfo {author} {\bibfnamefont {B.}~\bibnamefont {{Van
  Leer}}},\ }\href {\doibase 10.1016/0021-9991(77)90095-X} {\bibfield
  {journal} {\bibinfo  {journal} {Journal of Computational Physics}\ }\textbf
  {\bibinfo {volume} {23}},\ \bibinfo {pages} {276} (\bibinfo {year}
  {1977})}\BibitemShut {NoStop}%
\bibitem [{\citenamefont {Deppe}\ \emph {et~al.}(tion)\citenamefont {Deppe}
  \emph {et~al.}}]{PPAO-inprep}%
  \BibitemOpen
  \bibfield  {author} {\bibinfo {author} {\bibfnamefont {N.}~\bibnamefont
  {Deppe}} \emph {et~al.},\ }\href@noop {} {\  (\bibinfo {year} {in
  preparation})}\BibitemShut {NoStop}%
\bibitem [{\citenamefont {Oppenheimer}\ and\ \citenamefont
  {Volkoff}(1939)}]{Oppenheimer:1939ne}%
  \BibitemOpen
  \bibfield  {author} {\bibinfo {author} {\bibfnamefont {J.}~\bibnamefont
  {Oppenheimer}}\ and\ \bibinfo {author} {\bibfnamefont {G.}~\bibnamefont
  {Volkoff}},\ }\href {\doibase 10.1103/PhysRev.55.374} {\bibfield  {journal}
  {\bibinfo  {journal} {Phys. Rev.}\ }\textbf {\bibinfo {volume} {55}},\
  \bibinfo {pages} {374} (\bibinfo {year} {1939})}\BibitemShut {NoStop}%
\bibitem [{\citenamefont {Lattimer}\ and\ \citenamefont
  {Prakash}(2001)}]{Lattimer:2000nx}%
  \BibitemOpen
  \bibfield  {author} {\bibinfo {author} {\bibfnamefont {J.~M.}\ \bibnamefont
  {Lattimer}}\ and\ \bibinfo {author} {\bibfnamefont {M.}~\bibnamefont
  {Prakash}},\ }\href {\doibase 10.1086/319702} {\bibfield  {journal} {\bibinfo
   {journal} {Astrophys. J.}\ }\textbf {\bibinfo {volume} {550}},\ \bibinfo
  {pages} {426} (\bibinfo {year} {2001})},\ \Eprint
  {http://arxiv.org/abs/astro-ph/0002232} {arXiv:astro-ph/0002232} \BibitemShut
  {NoStop}%
\bibitem [{\citenamefont {Chatziioannou}\ and\ \citenamefont
  {Han}(2020)}]{Chatziioannou:2019yko}%
  \BibitemOpen
  \bibfield  {author} {\bibinfo {author} {\bibfnamefont {K.}~\bibnamefont
  {Chatziioannou}}\ and\ \bibinfo {author} {\bibfnamefont {S.}~\bibnamefont
  {Han}},\ }\href {\doibase 10.1103/PhysRevD.101.044019} {\bibfield  {journal}
  {\bibinfo  {journal} {Phys. Rev. D}\ }\textbf {\bibinfo {volume} {101}},\
  \bibinfo {pages} {044019} (\bibinfo {year} {2020})},\ \Eprint
  {http://arxiv.org/abs/1911.07091} {arXiv:1911.07091 [gr-qc]} \BibitemShut
  {NoStop}%
\bibitem [{\citenamefont {Gieg}\ \emph {et~al.}(2019)\citenamefont {Gieg},
  \citenamefont {Dietrich},\ and\ \citenamefont {Ujevic}}]{Gieg:2019yzq}%
  \BibitemOpen
  \bibfield  {author} {\bibinfo {author} {\bibfnamefont {H.}~\bibnamefont
  {Gieg}}, \bibinfo {author} {\bibfnamefont {T.}~\bibnamefont {Dietrich}}, \
  and\ \bibinfo {author} {\bibfnamefont {M.}~\bibnamefont {Ujevic}},\ }\href
  {\doibase 10.3390/particles2030023} {\bibfield  {journal} {\bibinfo
  {journal} {Particles}\ }\textbf {\bibinfo {volume} {2}},\ \bibinfo {pages}
  {365} (\bibinfo {year} {2019})},\ \Eprint {http://arxiv.org/abs/1908.03135}
  {arXiv:1908.03135 [gr-qc]} \BibitemShut {NoStop}%
\bibitem [{\citenamefont {Kurkela}\ \emph {et~al.}(2010)\citenamefont
  {Kurkela}, \citenamefont {Romatschke},\ and\ \citenamefont
  {Vuorinen}}]{Kurkela:2009gj}%
  \BibitemOpen
  \bibfield  {author} {\bibinfo {author} {\bibfnamefont {A.}~\bibnamefont
  {Kurkela}}, \bibinfo {author} {\bibfnamefont {P.}~\bibnamefont {Romatschke}},
  \ and\ \bibinfo {author} {\bibfnamefont {A.}~\bibnamefont {Vuorinen}},\
  }\href {\doibase 10.1103/PhysRevD.81.105021} {\bibfield  {journal} {\bibinfo
  {journal} {Phys. Rev. D}\ }\textbf {\bibinfo {volume} {81}},\ \bibinfo
  {pages} {105021} (\bibinfo {year} {2010})},\ \Eprint
  {http://arxiv.org/abs/0912.1856} {arXiv:0912.1856 [hep-ph]} \BibitemShut
  {NoStop}%
\bibitem [{\citenamefont {Kokkotas}\ and\ \citenamefont
  {Ruoff}(2001)}]{Kokkotas:2000up}%
  \BibitemOpen
  \bibfield  {author} {\bibinfo {author} {\bibfnamefont {K.~D.}\ \bibnamefont
  {Kokkotas}}\ and\ \bibinfo {author} {\bibfnamefont {J.}~\bibnamefont
  {Ruoff}},\ }\href {\doibase 10.1051/0004-6361:20000216} {\bibfield  {journal}
  {\bibinfo  {journal} {Astron. Astrophys.}\ }\textbf {\bibinfo {volume}
  {366}},\ \bibinfo {pages} {565} (\bibinfo {year} {2001})},\ \Eprint
  {http://arxiv.org/abs/gr-qc/0011093} {arXiv:gr-qc/0011093} \BibitemShut
  {NoStop}%
\bibitem [{\citenamefont {Sen}\ \emph {et~al.}(2022)\citenamefont {Sen},
  \citenamefont {Kumar}, \citenamefont {Kunjipurayi}, \citenamefont {Routaray},
  \citenamefont {Zhao},\ and\ \citenamefont {Kumar}}]{Sen:2022kva}%
  \BibitemOpen
  \bibfield  {author} {\bibinfo {author} {\bibfnamefont {S.}~\bibnamefont
  {Sen}}, \bibinfo {author} {\bibfnamefont {S.}~\bibnamefont {Kumar}}, \bibinfo
  {author} {\bibfnamefont {A.}~\bibnamefont {Kunjipurayi}}, \bibinfo {author}
  {\bibfnamefont {P.}~\bibnamefont {Routaray}}, \bibinfo {author}
  {\bibfnamefont {T.}~\bibnamefont {Zhao}}, \ and\ \bibinfo {author}
  {\bibfnamefont {B.}~\bibnamefont {Kumar}},\ }\href@noop {} {\  (\bibinfo
  {year} {2022})},\ \Eprint {http://arxiv.org/abs/2205.02076} {arXiv:2205.02076
  [nucl-th]} \BibitemShut {NoStop}%
\bibitem [{\citenamefont {Kastaun}\ \emph {et~al.}(2021)\citenamefont
  {Kastaun}, \citenamefont {Kalinani},\ and\ \citenamefont
  {Ciolfi}}]{Kastaun:2020uxr}%
  \BibitemOpen
  \bibfield  {author} {\bibinfo {author} {\bibfnamefont {W.}~\bibnamefont
  {Kastaun}}, \bibinfo {author} {\bibfnamefont {J.~V.}\ \bibnamefont
  {Kalinani}}, \ and\ \bibinfo {author} {\bibfnamefont {R.}~\bibnamefont
  {Ciolfi}},\ }\href {\doibase 10.1103/PhysRevD.103.023018} {\bibfield
  {journal} {\bibinfo  {journal} {Phys. Rev. D}\ }\textbf {\bibinfo {volume}
  {103}},\ \bibinfo {pages} {023018} (\bibinfo {year} {2021})},\ \Eprint
  {http://arxiv.org/abs/2005.01821} {arXiv:2005.01821 [gr-qc]} \BibitemShut
  {NoStop}%
\bibitem [{\citenamefont {Espino}\ and\ \citenamefont
  {Paschalidis}(2022)}]{Espino:2021adh}%
  \BibitemOpen
  \bibfield  {author} {\bibinfo {author} {\bibfnamefont {P.~L.}\ \bibnamefont
  {Espino}}\ and\ \bibinfo {author} {\bibfnamefont {V.}~\bibnamefont
  {Paschalidis}},\ }\href {\doibase 10.1103/PhysRevD.105.043014} {\bibfield
  {journal} {\bibinfo  {journal} {Phys. Rev. D}\ }\textbf {\bibinfo {volume}
  {105}},\ \bibinfo {pages} {043014} (\bibinfo {year} {2022})},\ \Eprint
  {http://arxiv.org/abs/2105.05269} {arXiv:2105.05269 [astro-ph.HE]}
  \BibitemShut {NoStop}%
\bibitem [{\citenamefont {Carbone}\ and\ \citenamefont
  {Schwenk}(2019)}]{Carbone:2019pkr}%
  \BibitemOpen
  \bibfield  {author} {\bibinfo {author} {\bibfnamefont {A.}~\bibnamefont
  {Carbone}}\ and\ \bibinfo {author} {\bibfnamefont {A.}~\bibnamefont
  {Schwenk}},\ }\href {\doibase 10.1103/PhysRevC.100.025805} {\bibfield
  {journal} {\bibinfo  {journal} {Phys. Rev. C}\ }\textbf {\bibinfo {volume}
  {100}},\ \bibinfo {pages} {025805} (\bibinfo {year} {2019})},\ \Eprint
  {http://arxiv.org/abs/1904.00924} {arXiv:1904.00924 [nucl-th]} \BibitemShut
  {NoStop}%
\bibitem [{\citenamefont {Raithel}\ \emph {et~al.}(2019)\citenamefont
  {Raithel}, \citenamefont {Ozel},\ and\ \citenamefont
  {Psaltis}}]{Raithel:2019gws}%
  \BibitemOpen
  \bibfield  {author} {\bibinfo {author} {\bibfnamefont {C.~A.}\ \bibnamefont
  {Raithel}}, \bibinfo {author} {\bibfnamefont {F.}~\bibnamefont {Ozel}}, \
  and\ \bibinfo {author} {\bibfnamefont {D.}~\bibnamefont {Psaltis}},\ }\href
  {\doibase 10.3847/1538-4357/ab08ea} {\bibfield  {journal} {\bibinfo
  {journal} {Astrophys. J.}\ }\textbf {\bibinfo {volume} {875}},\ \bibinfo
  {pages} {12} (\bibinfo {year} {2019})},\ \Eprint
  {http://arxiv.org/abs/1902.10735} {arXiv:1902.10735 [astro-ph.HE]}
  \BibitemShut {NoStop}%
\bibitem [{\citenamefont {Deppe}\ \emph {et~al.}(2023)\citenamefont {Deppe},
  \citenamefont {Throwe}, \citenamefont {Kidder}, \citenamefont {Vu},
  \citenamefont {Hébert}, \citenamefont {Moxon}, \citenamefont {Armaza},
  \citenamefont {Bonilla}, \citenamefont {Kim}, \citenamefont {Kumar},
  \citenamefont {Lovelace}, \citenamefont {Macedo}, \citenamefont {Nelli},
  \citenamefont {O'Shea}, \citenamefont {Pfeiffer}, \citenamefont {Scheel},
  \citenamefont {Teukolsky}, \citenamefont {Wittek}, \citenamefont
  {Anantpurkar}, \citenamefont {Boyle}, \citenamefont {Carpenter},
  \citenamefont {Ceja}, \citenamefont {Chaudhary}, \citenamefont {Foucart},
  \citenamefont {Ghadiri}, \citenamefont {Giesler}, \citenamefont {Guo},
  \citenamefont {Iozzo}, \citenamefont {Legred}, \citenamefont {Li},
  \citenamefont {Ma}, \citenamefont {Melchor}, \citenamefont {Morales},
  \citenamefont {Most}, \citenamefont {Pajkos}, \citenamefont {Ramirez},
  \citenamefont {Ring}, \citenamefont {Rüter}, \citenamefont {Sanchez},
  \citenamefont {Stein}, \citenamefont {Thomas}, \citenamefont {Vieira},
  \citenamefont {Wlodarczyk},\ and\ \citenamefont
  {Wu}}]{spectre_jan_23_release}%
  \BibitemOpen
  \bibfield  {author} {\bibinfo {author} {\bibfnamefont {N.}~\bibnamefont
  {Deppe}}, \bibinfo {author} {\bibfnamefont {W.}~\bibnamefont {Throwe}},
  \bibinfo {author} {\bibfnamefont {L.~E.}\ \bibnamefont {Kidder}}, \bibinfo
  {author} {\bibfnamefont {N.~L.}\ \bibnamefont {Vu}}, \bibinfo {author}
  {\bibfnamefont {F.}~\bibnamefont {Hébert}}, \bibinfo {author} {\bibfnamefont
  {J.}~\bibnamefont {Moxon}}, \bibinfo {author} {\bibfnamefont
  {C.}~\bibnamefont {Armaza}}, \bibinfo {author} {\bibfnamefont {G.~S.}\
  \bibnamefont {Bonilla}}, \bibinfo {author} {\bibfnamefont {Y.}~\bibnamefont
  {Kim}}, \bibinfo {author} {\bibfnamefont {P.}~\bibnamefont {Kumar}}, \bibinfo
  {author} {\bibfnamefont {G.}~\bibnamefont {Lovelace}}, \bibinfo {author}
  {\bibfnamefont {A.}~\bibnamefont {Macedo}}, \bibinfo {author} {\bibfnamefont
  {K.~C.}\ \bibnamefont {Nelli}}, \bibinfo {author} {\bibfnamefont
  {E.}~\bibnamefont {O'Shea}}, \bibinfo {author} {\bibfnamefont {H.~P.}\
  \bibnamefont {Pfeiffer}}, \bibinfo {author} {\bibfnamefont {M.~A.}\
  \bibnamefont {Scheel}}, \bibinfo {author} {\bibfnamefont {S.~A.}\
  \bibnamefont {Teukolsky}}, \bibinfo {author} {\bibfnamefont {N.~A.}\
  \bibnamefont {Wittek}}, \bibinfo {author} {\bibfnamefont {I.}~\bibnamefont
  {Anantpurkar}}, \bibinfo {author} {\bibfnamefont {M.}~\bibnamefont {Boyle}},
  \bibinfo {author} {\bibfnamefont {A.}~\bibnamefont {Carpenter}}, \bibinfo
  {author} {\bibfnamefont {A.}~\bibnamefont {Ceja}}, \bibinfo {author}
  {\bibfnamefont {H.}~\bibnamefont {Chaudhary}}, \bibinfo {author}
  {\bibfnamefont {F.}~\bibnamefont {Foucart}}, \bibinfo {author} {\bibfnamefont
  {N.}~\bibnamefont {Ghadiri}}, \bibinfo {author} {\bibfnamefont
  {M.}~\bibnamefont {Giesler}}, \bibinfo {author} {\bibfnamefont {J.~S.}\
  \bibnamefont {Guo}}, \bibinfo {author} {\bibfnamefont {D.~A.~B.}\
  \bibnamefont {Iozzo}}, \bibinfo {author} {\bibfnamefont {I.}~\bibnamefont
  {Legred}}, \bibinfo {author} {\bibfnamefont {D.}~\bibnamefont {Li}}, \bibinfo
  {author} {\bibfnamefont {S.}~\bibnamefont {Ma}}, \bibinfo {author}
  {\bibfnamefont {D.}~\bibnamefont {Melchor}}, \bibinfo {author} {\bibfnamefont
  {M.}~\bibnamefont {Morales}}, \bibinfo {author} {\bibfnamefont {E.~R.}\
  \bibnamefont {Most}}, \bibinfo {author} {\bibfnamefont {M.~A.}\ \bibnamefont
  {Pajkos}}, \bibinfo {author} {\bibfnamefont {T.}~\bibnamefont {Ramirez}},
  \bibinfo {author} {\bibfnamefont {N.}~\bibnamefont {Ring}}, \bibinfo {author}
  {\bibfnamefont {H.~R.}\ \bibnamefont {Rüter}}, \bibinfo {author}
  {\bibfnamefont {J.}~\bibnamefont {Sanchez}}, \bibinfo {author} {\bibfnamefont
  {L.~C.}\ \bibnamefont {Stein}}, \bibinfo {author} {\bibfnamefont
  {S.}~\bibnamefont {Thomas}}, \bibinfo {author} {\bibfnamefont
  {D.}~\bibnamefont {Vieira}}, \bibinfo {author} {\bibfnamefont
  {T.}~\bibnamefont {Wlodarczyk}}, \ and\ \bibinfo {author} {\bibfnamefont
  {D.}~\bibnamefont {Wu}},\ }\href {\doibase 10.5281/zenodo.7535144} {\enquote
  {\bibinfo {title} {Spectre},}\ } (\bibinfo {year} {2023})\BibitemShut
  {NoStop}%
\bibitem [{\citenamefont {SXS}(2023)}]{PaperGithub}%
  \BibitemOpen
  \bibfield  {author} {\bibinfo {author} {\bibnamefont {SXS}},\ }\href@noop {}
  {\enquote {\bibinfo {title} {paper-2023-spectre-enthalpy-eos},}\ }\bibinfo
  {howpublished}
  {\url{https://github.com/sxs-collaboration/paper-2023-spectre-enthalpy-eos}}
  (\bibinfo {year} {2023})\BibitemShut {NoStop}%
\bibitem [{\citenamefont {Virtanen}\ \emph {et~al.}(2020)\citenamefont
  {Virtanen}, \citenamefont {Gommers}, \citenamefont {Oliphant}, \citenamefont
  {Haberland}, \citenamefont {Reddy}, \citenamefont {Cournapeau}, \citenamefont
  {Burovski}, \citenamefont {Peterson}, \citenamefont {Weckesser},
  \citenamefont {Bright}, \citenamefont {{van der Walt}}, \citenamefont
  {Brett}, \citenamefont {Wilson}, \citenamefont {Millman}, \citenamefont
  {Mayorov}, \citenamefont {Nelson}, \citenamefont {Jones}, \citenamefont
  {Kern}, \citenamefont {Larson}, \citenamefont {Carey}, \citenamefont {Polat},
  \citenamefont {Feng}, \citenamefont {Moore}, \citenamefont {{VanderPlas}},
  \citenamefont {Laxalde}, \citenamefont {Perktold}, \citenamefont {Cimrman},
  \citenamefont {Henriksen}, \citenamefont {Quintero}, \citenamefont {Harris},
  \citenamefont {Archibald}, \citenamefont {Ribeiro}, \citenamefont
  {Pedregosa}, \citenamefont {{van Mulbregt}},\ and\ \citenamefont {{SciPy 1.0
  Contributors}}}]{scipy}%
  \BibitemOpen
  \bibfield  {author} {\bibinfo {author} {\bibfnamefont {P.}~\bibnamefont
  {Virtanen}}, \bibinfo {author} {\bibfnamefont {R.}~\bibnamefont {Gommers}},
  \bibinfo {author} {\bibfnamefont {T.~E.}\ \bibnamefont {Oliphant}}, \bibinfo
  {author} {\bibfnamefont {M.}~\bibnamefont {Haberland}}, \bibinfo {author}
  {\bibfnamefont {T.}~\bibnamefont {Reddy}}, \bibinfo {author} {\bibfnamefont
  {D.}~\bibnamefont {Cournapeau}}, \bibinfo {author} {\bibfnamefont
  {E.}~\bibnamefont {Burovski}}, \bibinfo {author} {\bibfnamefont
  {P.}~\bibnamefont {Peterson}}, \bibinfo {author} {\bibfnamefont
  {W.}~\bibnamefont {Weckesser}}, \bibinfo {author} {\bibfnamefont
  {J.}~\bibnamefont {Bright}}, \bibinfo {author} {\bibfnamefont {S.~J.}\
  \bibnamefont {{van der Walt}}}, \bibinfo {author} {\bibfnamefont
  {M.}~\bibnamefont {Brett}}, \bibinfo {author} {\bibfnamefont
  {J.}~\bibnamefont {Wilson}}, \bibinfo {author} {\bibfnamefont {K.~J.}\
  \bibnamefont {Millman}}, \bibinfo {author} {\bibfnamefont {N.}~\bibnamefont
  {Mayorov}}, \bibinfo {author} {\bibfnamefont {A.~R.~J.}\ \bibnamefont
  {Nelson}}, \bibinfo {author} {\bibfnamefont {E.}~\bibnamefont {Jones}},
  \bibinfo {author} {\bibfnamefont {R.}~\bibnamefont {Kern}}, \bibinfo {author}
  {\bibfnamefont {E.}~\bibnamefont {Larson}}, \bibinfo {author} {\bibfnamefont
  {C.~J.}\ \bibnamefont {Carey}}, \bibinfo {author} {\bibfnamefont
  {{\.I}.}~\bibnamefont {Polat}}, \bibinfo {author} {\bibfnamefont
  {Y.}~\bibnamefont {Feng}}, \bibinfo {author} {\bibfnamefont {E.~W.}\
  \bibnamefont {Moore}}, \bibinfo {author} {\bibfnamefont {J.}~\bibnamefont
  {{VanderPlas}}}, \bibinfo {author} {\bibfnamefont {D.}~\bibnamefont
  {Laxalde}}, \bibinfo {author} {\bibfnamefont {J.}~\bibnamefont {Perktold}},
  \bibinfo {author} {\bibfnamefont {R.}~\bibnamefont {Cimrman}}, \bibinfo
  {author} {\bibfnamefont {I.}~\bibnamefont {Henriksen}}, \bibinfo {author}
  {\bibfnamefont {E.~A.}\ \bibnamefont {Quintero}}, \bibinfo {author}
  {\bibfnamefont {C.~R.}\ \bibnamefont {Harris}}, \bibinfo {author}
  {\bibfnamefont {A.~M.}\ \bibnamefont {Archibald}}, \bibinfo {author}
  {\bibfnamefont {A.~H.}\ \bibnamefont {Ribeiro}}, \bibinfo {author}
  {\bibfnamefont {F.}~\bibnamefont {Pedregosa}}, \bibinfo {author}
  {\bibfnamefont {P.}~\bibnamefont {{van Mulbregt}}}, \ and\ \bibinfo {author}
  {\bibnamefont {{SciPy 1.0 Contributors}}},\ }\href {\doibase
  10.1038/s41592-019-0686-2} {\bibfield  {journal} {\bibinfo  {journal} {Nature
  Methods}\ }\textbf {\bibinfo {volume} {17}},\ \bibinfo {pages} {261}
  (\bibinfo {year} {2020})}\BibitemShut {NoStop}%
\bibitem [{\citenamefont {Oliphant}(06  )}]{numpy}%
  \BibitemOpen
  \bibfield  {author} {\bibinfo {author} {\bibfnamefont {T.}~\bibnamefont
  {Oliphant}},\ }\href {http://www.numpy.org/} {\enquote {\bibinfo {title}
  {{NumPy}: A guide to {NumPy}},}\ }\bibinfo {howpublished} {USA: Trelgol
  Publishing} (\bibinfo {year} {2006--}),\ \bibinfo {note} {[Online; accessed
  <today>]}\BibitemShut {NoStop}%
\bibitem [{\citenamefont {Hunter}(2007)}]{matplotlib}%
  \BibitemOpen
  \bibfield  {author} {\bibinfo {author} {\bibfnamefont {J.~D.}\ \bibnamefont
  {Hunter}},\ }\href {\doibase 10.1109/MCSE.2007.55} {\bibfield  {journal}
  {\bibinfo  {journal} {Computing in Science \& Engineering}\ }\textbf
  {\bibinfo {volume} {9}},\ \bibinfo {pages} {90} (\bibinfo {year}
  {2007})}\BibitemShut {NoStop}%
\bibitem [{\citenamefont {Ayachit}(2015)}]{paraview}%
  \BibitemOpen
  \bibfield  {author} {\bibinfo {author} {\bibfnamefont {U.}~\bibnamefont
  {Ayachit}},\ }\href@noop {} {\emph {\bibinfo {title} {The ParaView Guide: A
  Parallel Visualization Application}}}\ (\bibinfo  {publisher} {Kitware,
  Inc.},\ \bibinfo {address} {Clifton Park, NY, USA},\ \bibinfo {year}
  {2015})\BibitemShut {NoStop}%
\bibitem [{\citenamefont {Font}\ \emph {et~al.}(2002)\citenamefont {Font},
  \citenamefont {Goodale}, \citenamefont {Iyer}, \citenamefont {Miller},
  \citenamefont {Rezzolla}, \citenamefont {Seidel}, \citenamefont
  {Stergioulas}, \citenamefont {Suen},\ and\ \citenamefont
  {Tobias}}]{Font:2001ew}%
  \BibitemOpen
  \bibfield  {author} {\bibinfo {author} {\bibfnamefont {J.~A.}\ \bibnamefont
  {Font}}, \bibinfo {author} {\bibfnamefont {T.}~\bibnamefont {Goodale}},
  \bibinfo {author} {\bibfnamefont {S.}~\bibnamefont {Iyer}}, \bibinfo {author}
  {\bibfnamefont {M.~A.}\ \bibnamefont {Miller}}, \bibinfo {author}
  {\bibfnamefont {L.}~\bibnamefont {Rezzolla}}, \bibinfo {author}
  {\bibfnamefont {E.}~\bibnamefont {Seidel}}, \bibinfo {author} {\bibfnamefont
  {N.}~\bibnamefont {Stergioulas}}, \bibinfo {author} {\bibfnamefont {W.-M.}\
  \bibnamefont {Suen}}, \ and\ \bibinfo {author} {\bibfnamefont
  {M.}~\bibnamefont {Tobias}},\ }\href {\doibase 10.1103/PhysRevD.65.084024}
  {\bibfield  {journal} {\bibinfo  {journal} {Phys. Rev. D}\ }\textbf {\bibinfo
  {volume} {65}},\ \bibinfo {pages} {084024} (\bibinfo {year} {2002})},\
  \Eprint {http://arxiv.org/abs/gr-qc/0110047} {arXiv:gr-qc/0110047}
  \BibitemShut {NoStop}%
\end{thebibliography}%

\end{document}